\documentclass[12pt,a4paper,twoside]{article}

\usepackage{epsfig}

\begin{document}

\begin{center}
{\LARGE\bf Filtered overlap:\\[6pt]
speedup, locality, kernel non-normality and $Z_A\simeq1$}
\end{center}
\vspace{4pt}

\begin{center}
{\large
{\bf Stephan D\"urr}$\,{}^{a}$\hspace{2pt},\hspace{6pt}
{\bf Christian Hoelbling}$\,{}^{b}$\hspace{8pt}and\hspace{6pt}
{\bf Urs Wenger}$\,{}^{c}$
}\\[6pt]
${}^a\,$Institut f\"ur theoretische Physik, Universit\"at Bern,
Sidlerstr.\,5, CH-3012 Bern, Switzerland\\
${}^b\,$Bergische Universit\"at Wuppertal,
Gaussstr.\,20, D-42119 Wuppertal, Germany\\
${}^c\,$NIC/DESY Zeuthen,
Platanenallee 6, D-15738 Zeuthen, Germany
\end{center}
\vspace{4pt}

\begin{abstract}
\noindent
We investigate the overlap operator with a UV filtered Wilson kernel. The
filtering leads to a better localization of the operator even on coarse
lattices and with the untuned choice $\rho\!=\!1$. Furthermore, the
axial-vector renormalization constant $Z_A$ is much closer to 1, reducing the
mismatch with perturbation theory. We show that all these features persist over
a wide range of couplings and that the details of filtering prove immaterial.
We investigate the properties of the kernel spectrum and find that the kernel
non-normality is reduced. As a side effect we observe that for certain
applications of the filtered overlap a speed-up factor of 2-4 can be achieved.
\end{abstract}
\vspace{4pt}


\newcommand{\pad}{\partial}
\newcommand{\pas}{\partial\!\!\!/}
\newcommand{\Dsl}{D\!\!\!\!/\,}
\newcommand{\Psl}{P\!\!\!\!/\;\!}
\newcommand{\hqu}{\hbar}
\newcommand{\ovr}{\over}
\newcommand{\til}{\tilde}
\newcommand{\pri}{^\prime}
\renewcommand{\dag}{^\dagger}
\newcommand{\<}{\langle}
\renewcommand{\>}{\rangle}
\newcommand{\gaf}{\gamma_5}
\newcommand{\lap}{\triangle}
\newcommand{\trc}{{\rm tr}}
\newcommand{\nab}{\nabla}

\newcommand{\al}{\alpha}
\newcommand{\be}{\beta}
\newcommand{\ga}{\gamma}
\newcommand{\de}{\delta}
\newcommand{\ep}{\epsilon}
\newcommand{\ve}{\varepsilon}
\newcommand{\ze}{\zeta}
\newcommand{\et}{\eta}
\renewcommand{\th}{\theta}
\newcommand{\vt}{\vartheta}
\newcommand{\io}{\iota}
\newcommand{\ka}{\kappa}
\newcommand{\la}{\lambda}
\newcommand{\rh}{\rho}
\newcommand{\vr}{\varrho}
\newcommand{\si}{\sigma}
\newcommand{\ta}{\tau}
\newcommand{\ph}{\phi}
\newcommand{\vp}{\varphi}
\newcommand{\ch}{\chi}
\newcommand{\ps}{\psi}
\newcommand{\om}{\omega}
\newcommand{\psb}{\overline{\psi}}
\newcommand{\etb}{\overline{\eta}}
\newcommand{\psd}{\psi^{\dagger}}
\newcommand{\etd}{\eta^{\dagger}}
\newcommand{\beq}{\begin{equation}}

\newcommand{\eeq}{\end{equation}}
\newcommand{\bdm}{\begin{displaymath}}
\newcommand{\edm}{\end{displaymath}}
\newcommand{\bea}{\begin{eqnarray}}
\newcommand{\eea}{\end{eqnarray}}

\newcommand{\mr}{\mathrm}
\newcommand{\mb}{\mathbf}
\newcommand{\Nf}{{N_{\!f}}}
\newcommand{\Nc}{{N_{\!c}}}
\newcommand{\ri}{\mr{i}}
\newcommand{\HW}{H_\mr{W}}
\newcommand{\DW}{D_\mr{W}}

\newcommand{\Mpi}{M_\pi}
\newcommand{\Fpi}{F_\pi}
\newcommand{\MeV}{\,\mr{MeV}}
\newcommand{\GeV}{\,\mr{GeV}}
\newcommand{\fm}{\,\mr{fm}}

\hyphenation{topo-lo-gi-cal simu-la-tion theo-re-ti-cal mini-mum}




\section{Introduction}


From a theoretical viewpoint the ascent of ``overlap'' fermions
\cite{Kaplan:1992bt,Shamir:1993zy,overlap}, i.e.\ fermions which at zero quark
mass satisfy the Ginsparg Wilson (GW) relation \cite{Ginsparg:1981bj} ($\rh$
is a parameter that will be specified later)
\beq
\gaf D+D\hat\gaf=0,\qquad\hat\gaf=\gaf(1-{1\ovr\rh}D)
\label{GW}
\eeq
and thus realize a lattice version of the continuum chiral symmetry
\cite{Luscher:1998pq}
\beq
\de\ps=\hat\gaf\ps,\qquad\de\psb=\psb\gaf
\label{symm_luscher}
\eeq
together with an index theorem \cite{Hasenfratz:1998ri,Niedermayer:1998bi},
represents a major breakthrough in the field of non-perturbative studies of
QCD.
We know how to discretize fermions in a way that preserves the relevant
symmetries: $(i)$ gauge invariance, $(ii)$ flavor symmetry, and $(iii$) chiral
invariance.
Unfortunately, from a practical viewpoint the usefulness of this concept is
limited by the fact that the overlap tends to be one to two orders of magnitude
more expensive, in terms of CPU time, than a standard Wilson Dirac operator.

In this paper we study a variant of the overlap operator which makes use of a
UV filtered Wilson kernel.
Here, the ``filtering'' refers to replacing the original (``thin'') links of
the gauge configuration in the standard definition of the Wilson kernel by
``thick'' links obtained through APE \cite{ape} or HYP \cite{hyp} smearing.
This is a legal change of discretization as long as one keeps the iteration
level and smearing parameters fixed all the way down to the continuum, since
the ``thick'' links transform under a local gauge transformation in the same
way as the ``thin'' links; it should be seen as a modification of the operator
and not of the gauge background.
Such filtering has been used in the context of staggered quarks, where it has
been found to reduce UV fluctuations, in particular taste changing interactions
due to highly virtual gluons \cite{uvfilteredstag}.
In Ref.\,\cite{Durr:2004as} filtered staggered quarks were compared against
overlap quarks (where the filtered version was merely considered for
completeness), and it was observed that a single filtering step may speed up
the forward application of the overlap operator $D_\mr{ov}$ on a source vector
by a factor 2-4, depending on the gauge background.
This was seen to come through a reduction of the degree of the Chebychev
polynomial needed to approximate the inverse square root or sign function in
the definition of the massless overlap \cite{overlap}
\beq
aD_\mr{ov}=
\rh\Big[
1+D_{\mr{W},-\rh}(D_{\mr{W},-\rh}\dag D_{\mr{W},-\rh}^{})^{-1/2}
\Big]
=\rh
\Big[
1+\gaf\,\mr{sign}(a\gaf D_{\mr{W},-\rh})
\Big]
\label{over}
\eeq
with $D_{\mr{W},-\rh}\!=\!\DW-\!\rh/a$ the Wilson operator at negative mass
$-\rh/a$.
However, what matters in view of most phenomenological applications is
the performance of the massive operator (bare quark mass $m$)
\beq
D_{\mr{ov},m}=(1\!-{am\ovr 2\rh})D_\mr{ov}+m
\label{over_mass}
\eeq
in the process of calculating a given physical observable to a pre-defined
accuracy.
In other words the total CPU time spent depends on:
\begin{enumerate}
\itemsep
-4pt
\item The number of forward applications of the shifted Wilson operator
      $D_{\mr{W},-\rh}$ (or, generally speaking, of the kernel) needed to
      construct the massless overlap operator (\ref{over}).
\item The number of iterations spent on inverting the so-constructed massive
      operator (\ref{over_mass}) for a given renormalized quark mass (or a
      given $M_\pi^2$).
\item The number of gauge backgrounds needed to reach a pre-defined statistical
      accuracy of the desired observable at a given lattice spacing $a$.
\item The lattice spacing needed to enter the scaling window.
\end{enumerate}
The main emphasis of this paper will be on point 1; in particular we attempt
to give an understanding of the observed speedup in terms of the spectral
properties of the underlying hermitean (shifted) Wilson operator
$\HW\!=\!\gaf D_{\mr{W},-\rh}$.
At first sight it might seem that point 2 does not need to be considered at
all.
At fixed bare mass $m$ and fixed $\rh$ the filtered and the unfiltered overlap
do not differ on this point, since the number of forward applications of
$D_{\mr{ov},m}$ to get a column of the inverse depends only on its condition
number, and that is $2\rh/m$ for either variety.
As we shall see, the optimum $\rh$ (w.r.t.\ locality) gets reduced through
filtering whereas $Z_m\!=\!Z_S^{-1}\!=\!Z_P^{-1}$ increases and this means that
in the filtered case one has to use a smaller bare mass to work at a fixed
physical $m^\mr{ren}\!=\!Z_mm$.
These two aspects tend to compensate, and as a result there is little net
effect on point 2 from filtering.
Whether in points 3 and 4 filtering brings further savings is not clear, but we
plan to address this issue in the future.

\bigskip

\begin{figure}[!t]
\begin{center}
\epsfig{file=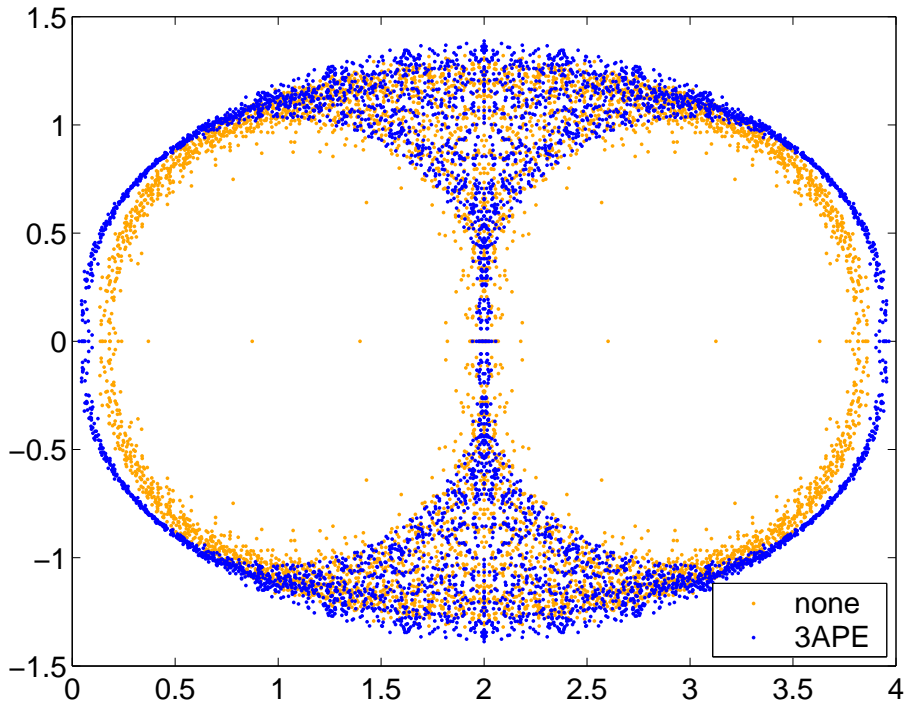,width=12cm}
\end{center}
\vspace{-6mm}
\caption{Eigenvalue spectra of $\DW$ with
$\ka\!=\!\ka_\mr{crit}^\mr{tree}\!=\!0.25$ in the quenched Schwinger model
($16^2$, $\be\!=\!3.2$, 10 configurations) without filtering and after 3 steps
with $\al\!=\!0.5$ (i.e.\ equal weight to the original link and the staple, see
\cite{DuHo_schwinger} for details; in 2D APE involves already the full
hypercube). Filtering depletes the ``bellies'', makes the physical (leftmost)
branch narrower and shifts it to the left.}
\label{fig:spec_schwinger}
\end{figure}

Let us try to obtain a first understanding of the effect of filtering in terms
of the spectrum of the underlying (non-hermitean) Wilson operator.
We are going to compute all eigenvalues, and to avoid spending too much CPU
time on this illustration, we shall do this in 2D, but it is clear that the
conceptual issue is --~mutatis mutandis~-- the same as in 4D.
In $d$ dimensions the Wilson Dirac operator has $d\!+\!1$ branches, and the
respective flavor multiplicities are
\beq
{d\choose 0}\;,\;
{d\choose 1}\;,\;
\;...\;     \;,\;
{d\choose d}\;.
\eeq
Thus in 2D the Wilson operator has 3 branches with multiplicities 1,2,1, while
in 4D it has 5 branches with multiplicities 1,4,6,4,1, respectively.
Fig.\,\ref{fig:spec_schwinger} shows the complete spectrum of $\DW$ with
the hopping parameter fixed at its tree-level critical value, $\ka\!=\!0.25$,
on 10 configurations of size $16^2$ at $\be\!=\!3.2$ in the quenched Schwinger
model.
Besides the ``thin'' link operator also its UV filtered descendent is shown.
In terms of the kernel spectrum the filtering is seen to have the following
effects:
\begin{enumerate}
\itemsep -4pt
\item[(a)] The two/four ``bellies'' are depleted --- in particular exactly real
           modes which cannot be assigned in a unique way to one of the
           three/five branches are severely suppressed.
\item[(b)] The horizontal scatter of any of the three/five branches diminishes.
\item[(c)] The additive mass renormalization of the physical (leftmost) branch
           is substantially reduced.
\end{enumerate}
If one were to ignore the kernel non-normality (we shall come back to this
point), the spectrum of $\DW$ could be linked, on a mode-by-mode basis, to the
one of $D_{\mr{W},-\rh}\dag D_{\mr{W},-\rh}^{}$,
$\HW\!=\!\gaf D_{\mr{W},-\rh}$ and $D_\mr{ov}$.
Then the first observation above (the reluctance of the filtered eigenvalues to
show up near the projection point $\rh$) simply means that the effect of
filtering on the spectrum of $\HW$ is to deplete the vicinity of the origin by
pushing the eigenvalues further towards the ends of the interval
$[-2d\!+\!1,2d\!-\!1]$.
In spite of the caveat mentioned, the thinning effect that (any kind of)
smearing has on the spectrum of $\HW$ near zero is indeed the reason for the
speedup in point 1 above.
A bigger interval $[0,\ep^2[$ or $]-\!\ep,\ep[$ that does not need to be
covered by the polynomial/rational approximation to the $1/\!\sqrt{.}$ or
$\mr{sign}(.)$ function translates into a lower degree and thus into fewer
forward applications of the kernel operator.

\bigskip

In the remainder of this article we shall address the spectral properties of
$\HW$ in more detail (Sec.\,2), and show that (a reasonable amount of)
filtering does not degrade the locality properties of $D_\mr{ov}$, but rather
makes the overlap operator \emph{more local} (Sec.\,3).
We continue with an explicit demonstration that the kernel non-normality gets
reduced by filtering (Sec.\,4).
We add some observations relevant to phenomenological applications of the
filtered overlap; in particular $Z_A$ is shown to be much closer to the
tree-level value 1 than for the unfiltered variety (Sec.\,5).
We rate this as a sign that perturbation theory might work far better for the
filtered overlap.
We make an attempt to compare our simple filtering recipe against other
approaches (Sec.\,6).
Finally, the appendix contains spectral data which suggest that the spectral
density of $\HW$ at the origin is non-zero for any $\be$ and any filtering
level.

We shall use pure gauge backgrounds and set the scale through the Sommer
parameter $r_0$ \cite{Sommer:1993ce}.
We choose the Wilson gauge action, and since $r_0(\be)$ is known
\cite{Guagnelli:1998ud} it is easy to select $\be$ values such that the
resulting lattices are matched, i.e.\ have fixed spatial size
$L\!\simeq\!1.5\fm$, with the resolution varying by a factor 3 from the
coarsest to the finest lattice -- see Tab.\,\ref{tab:parameters} for details.
Henceforth we set $a\!=\!1$.

\begin{table}
\begin{center}
\begin{tabular}{|l|cccc|}
\hline
$\be$    &$5.66$&$5.84$&$6.00$&$6.26$\\   
geometry &$ 8^4$&$12^4$&$16^4$&$24^4$\\   
geometry &$ 8^3\!\times\!16$&$12^3\!\times\!24$&$16^3\!\times\!32$&
          \hspace*{5mm}---\hspace*{5mm}\\ 
\hline
\end{tabular}
\end{center}
\vspace{-6mm}
\caption{Survey of matched 4D couplings and geometries with fixed
$L/r_0\!\approx\!3$, according to the interpolation formula of
Ref.\,\cite{Guagnelli:1998ud}. The first coupling is slightly out of bound
(see discussion in \cite{Guagnelli:1998ud}).}
\label{tab:parameters}
\end{table}


\section{Speedup and kernel spectrum}


In a quenched simulation the overhead, in terms of CPU time, of overlap versus
Wilson quarks comes in the first place from the polynomial or rational
approximation to the $1/\!\sqrt{.}$ or $\mr{sign}(.)$ function in (\ref{over}).
Let us assume%
\footnote{In fact, these values are rather close to the actual situation at
$\be\!\simeq\!6.0$, after projecting out the lowest 10-15 eigenvectors.}
that the lowest eigenvalue of the unfiltered $|\HW|$ is $0.14$ while the
highest eigenvalue takes the free field value 7.
This leads to the task to construct a polynomial/rational approximation of the
inverse square root over the range $[\ep^2,1]$ with $\ep\!=\!0.02$ the inverse
condition number of $|\HW|$.
Modest filtering will lift the lowest eigenvalue to something like $0.49$,
while the largest eigenvalue is almost invariant.
Then the task is to construct the approximation over the range $[\til\ep^2,1]$
with $\til\ep\!=\!0.07$ the filtered inverse condition number.
The lower bound increasing from $0.0004$ to $0.0049$ means that one gets away
with a smaller overall polynomial degree or an increased minimum root of the
denominator polynomial.
Therefore, the filtered overlap requires fewer forward applications of
$D_{\mr{W},-\rh}\dag D_{\mr{W},-\rh}^{}$ and this is how the savings on CPU
time in point 1 above come about.
In the remainder of this section we will elaborate on this statement, replace
the fictitious numbers by actual figures from real simulations and see that the
conclusion remains unchanged.

\bigskip

\begin{figure}[!b]
\vspace{-4mm}
\begin{center}
\epsfig{file=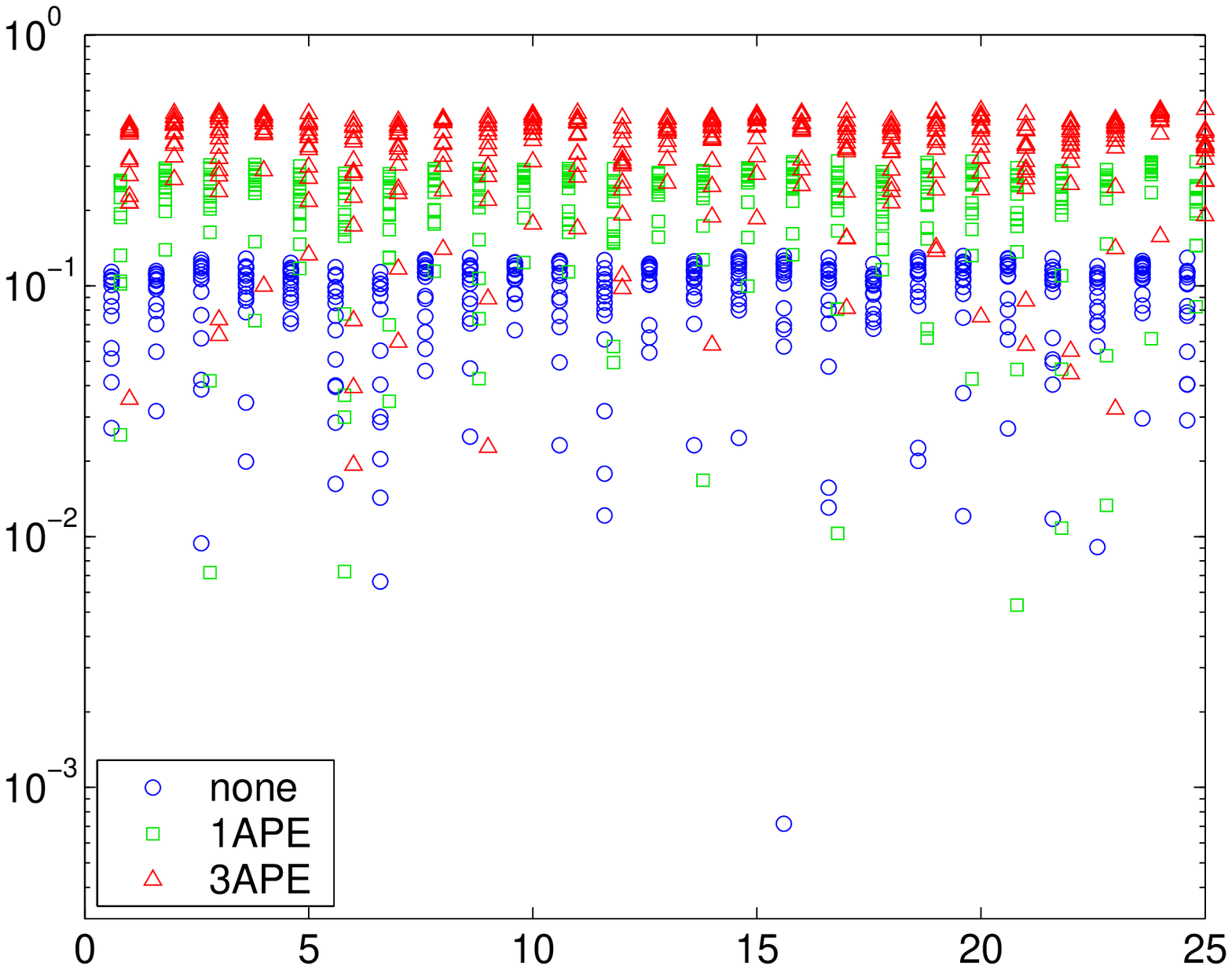,height=7.2cm}
\epsfig{file=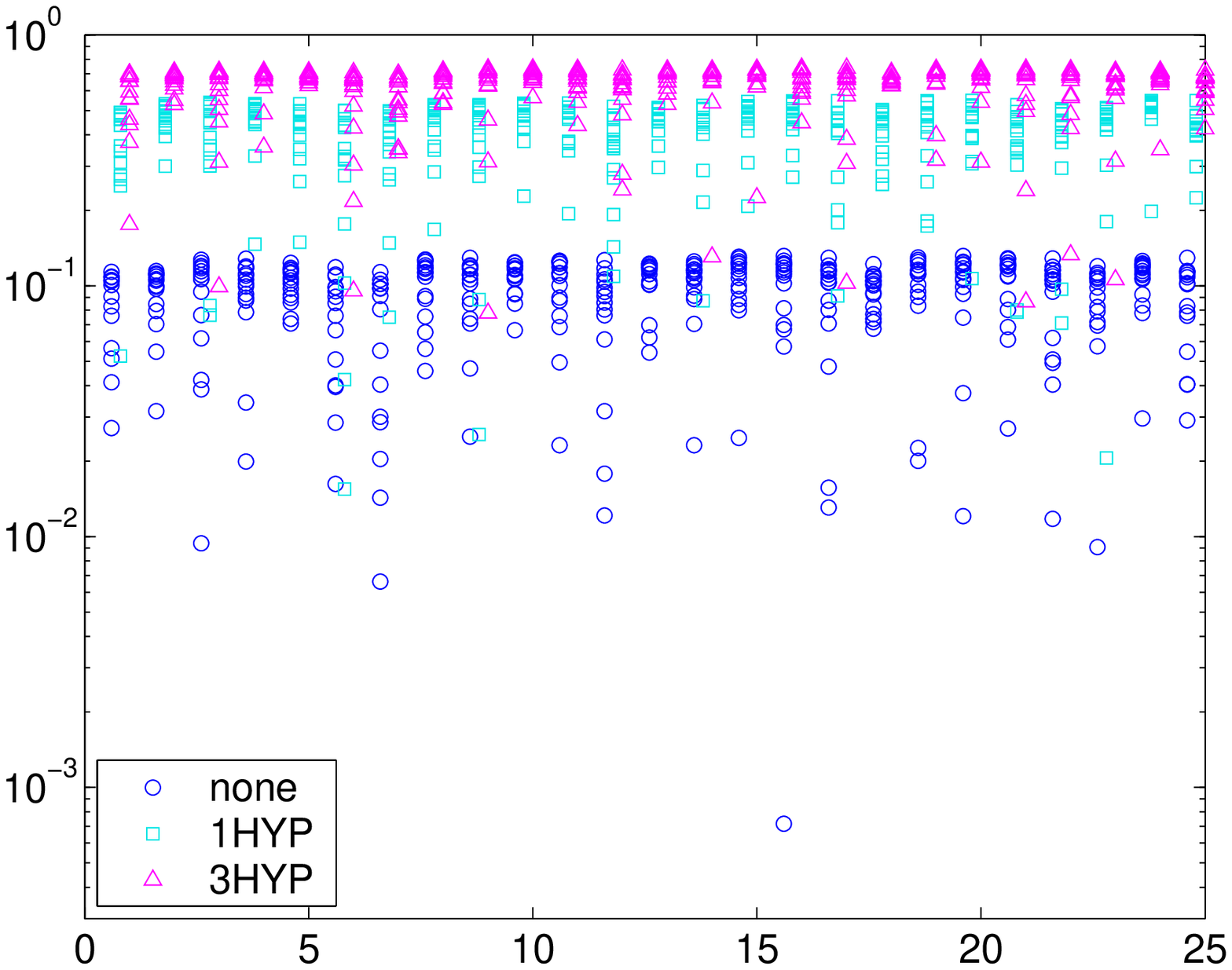,height=7.2cm}
\end{center}
\vspace{-6mm}
\caption{Sequence of the lowest 15 eigenvalues of $|\HW|$ on 25 configurations
at $\be\!=\!6.0$ without filtering and after 1,3 steps of APE (left) or HYP 
(right) filtering. Throughout $\rh\!=\!1$.}
\label{fig:low_sequence}
\end{figure}

\begin{figure}[!p]
\vspace{-4mm}
\begin{center}
\epsfig{file=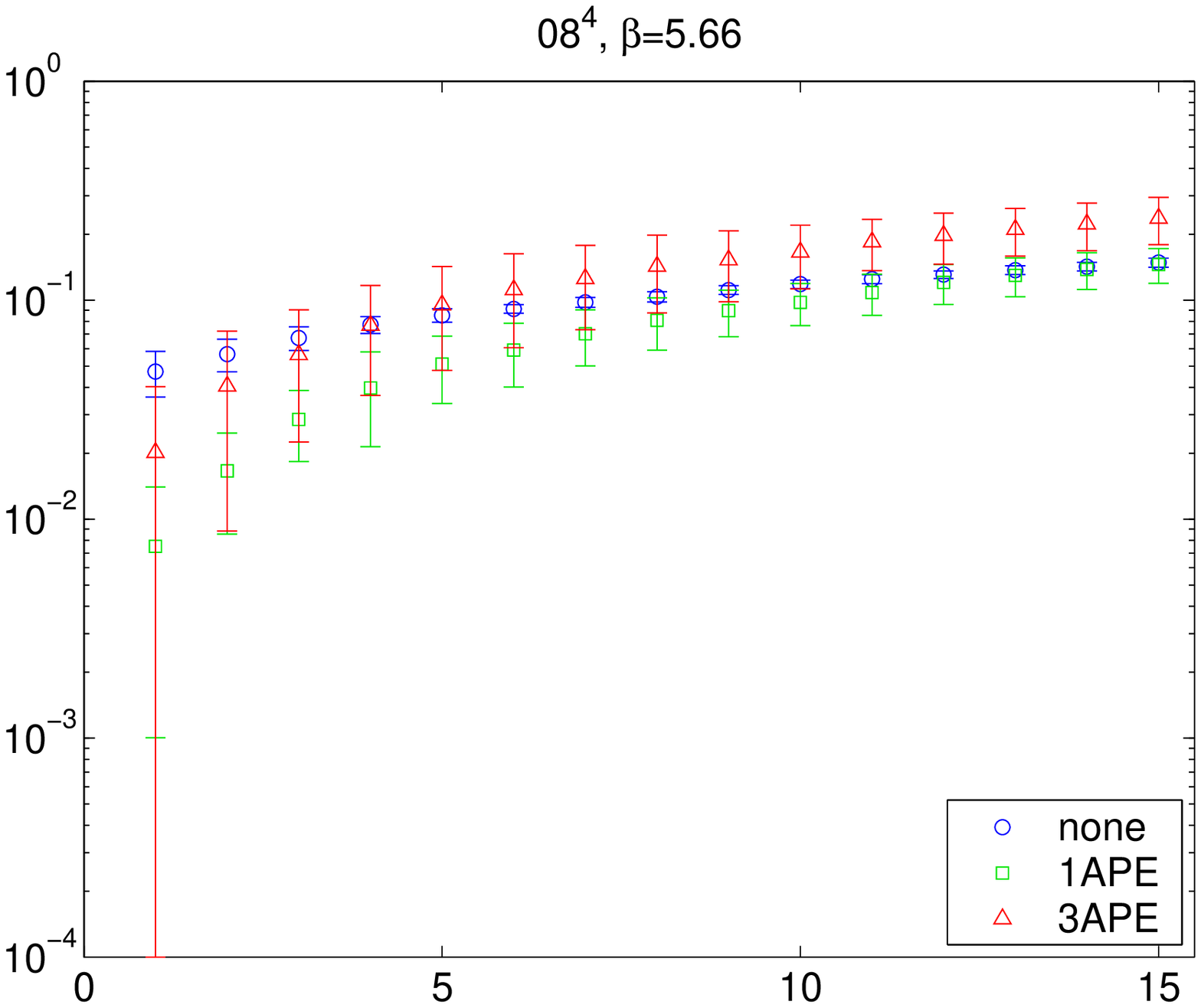,height=6cm}
\epsfig{file=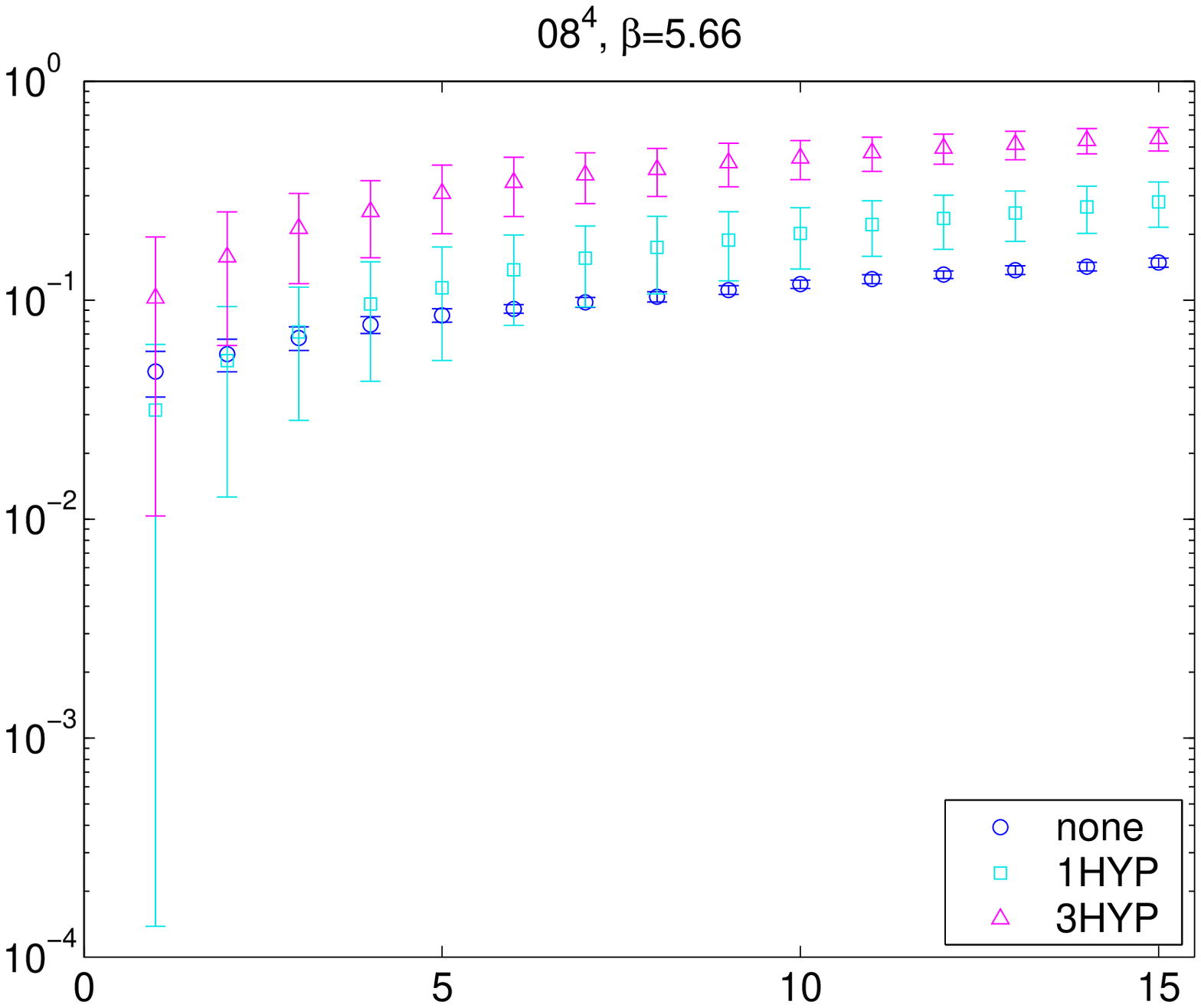,height=6cm}\\
\epsfig{file=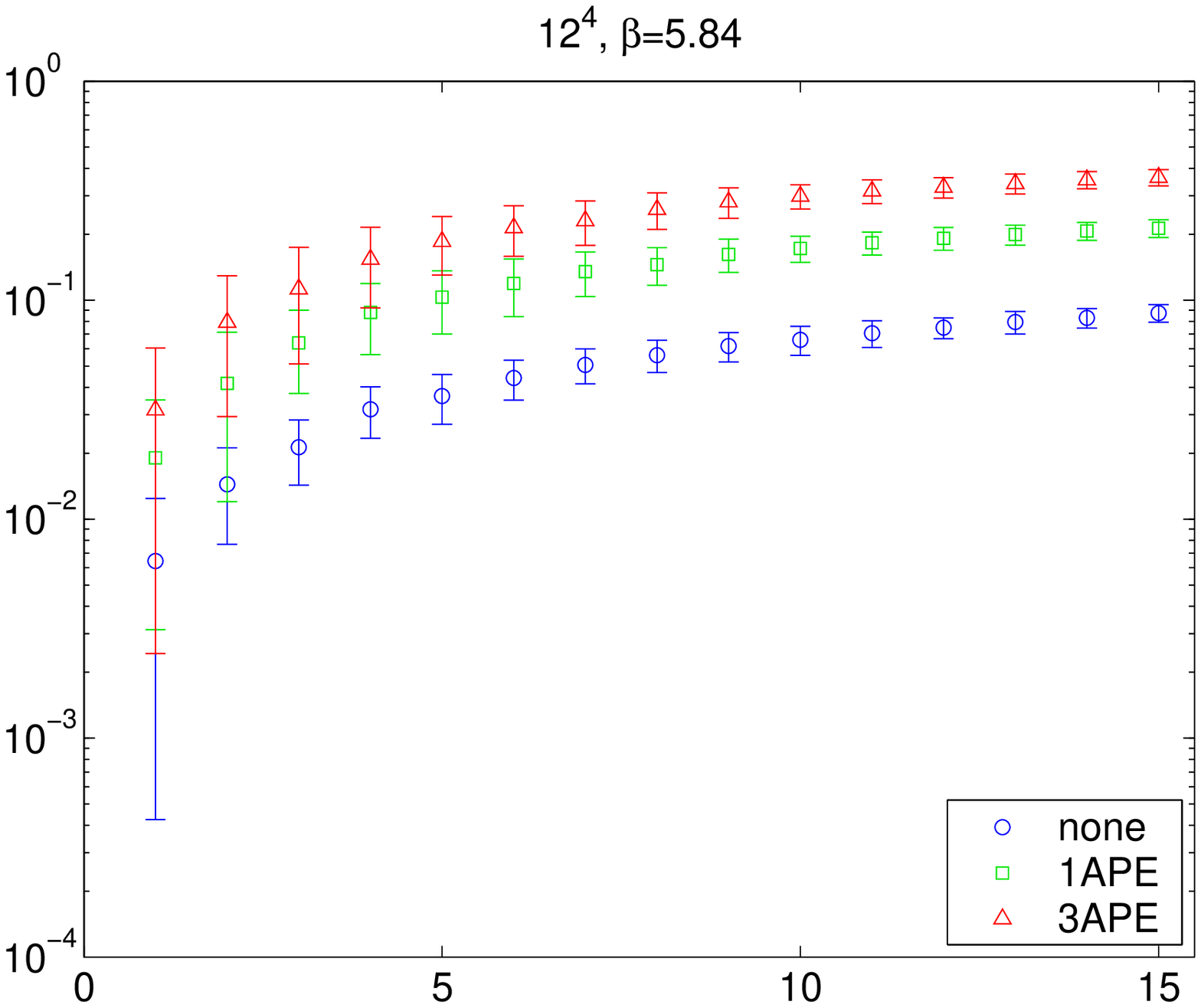,height=6cm}
\epsfig{file=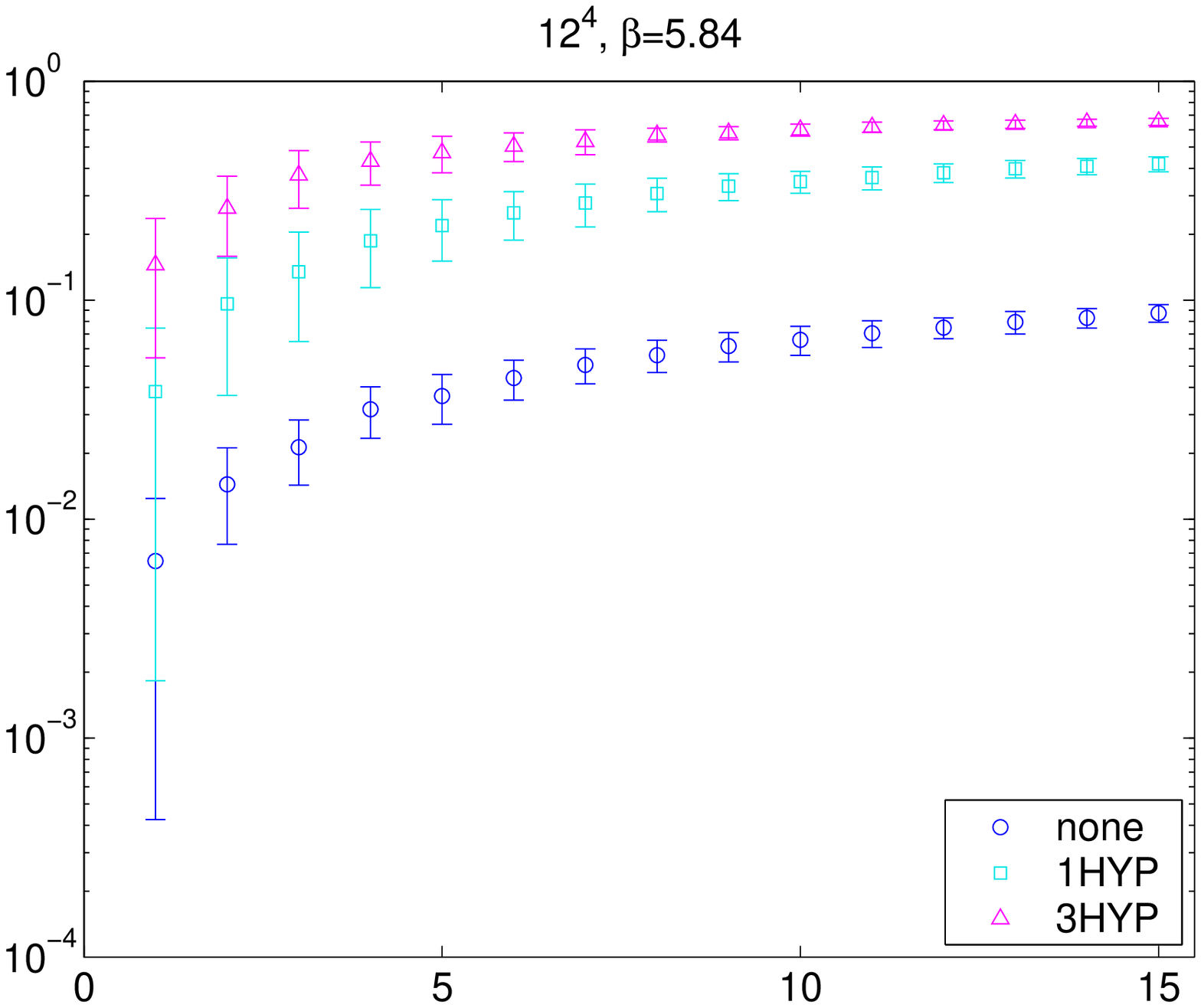,height=6cm}\\
\epsfig{file=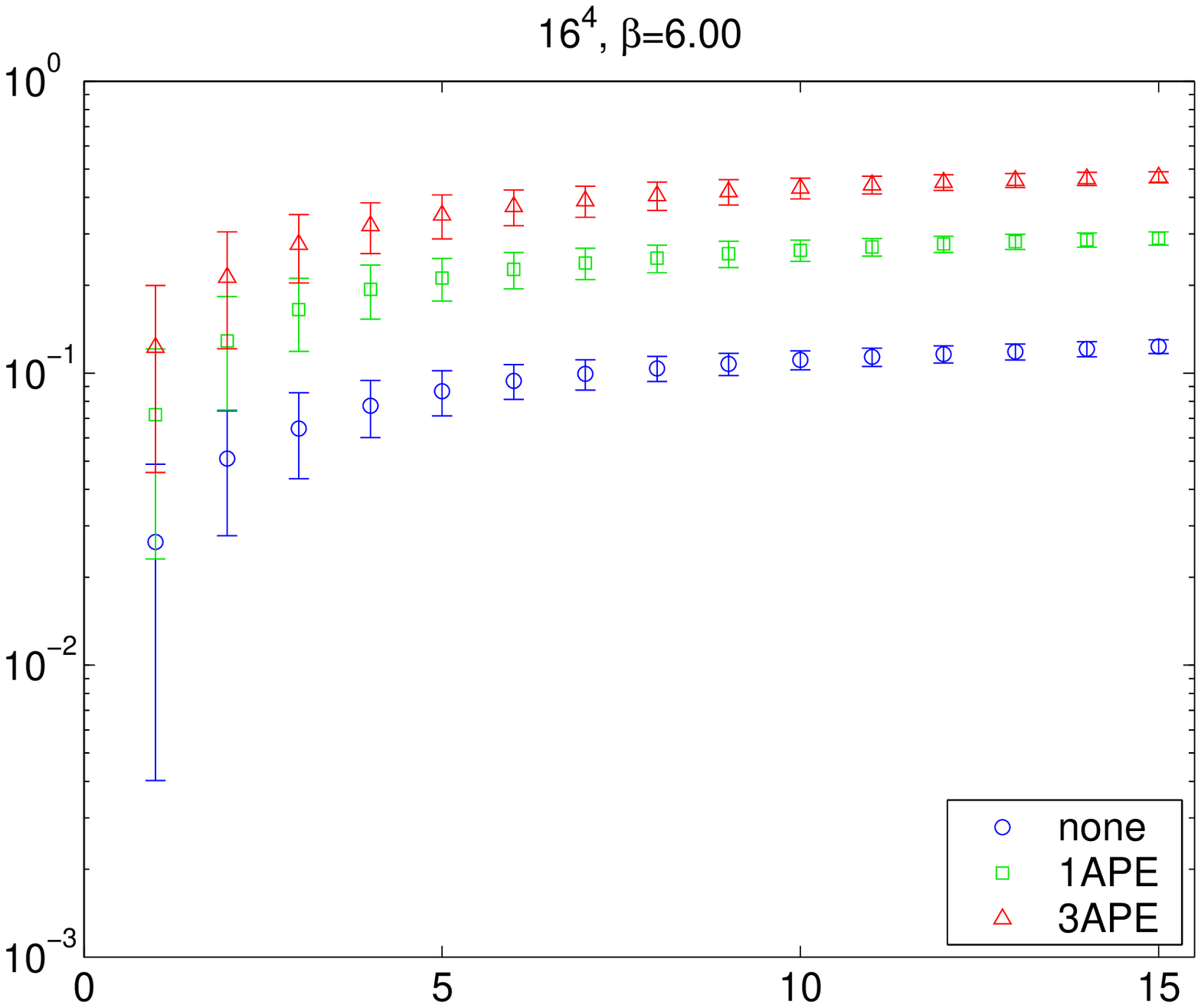,height=6cm}
\epsfig{file=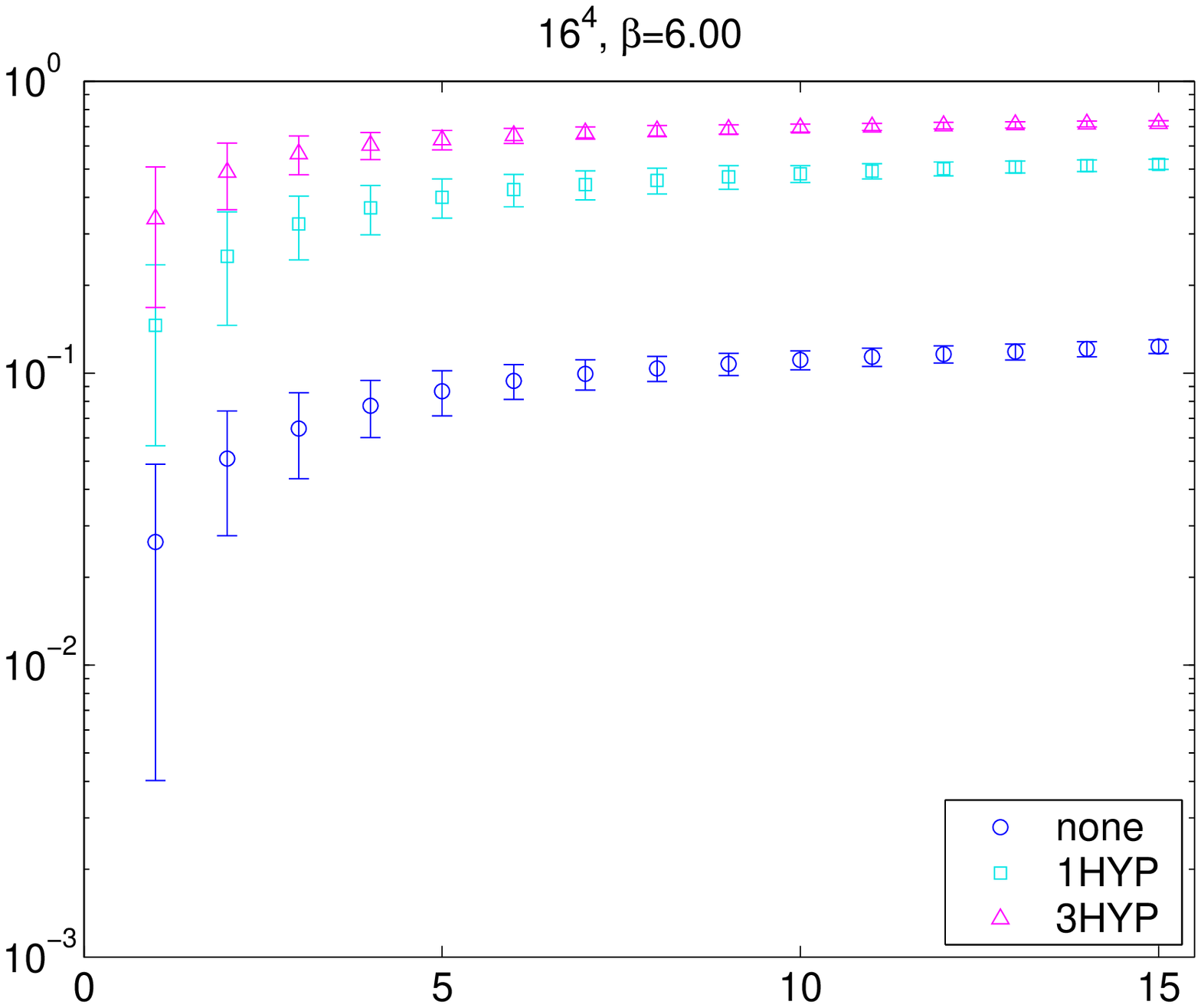,height=6cm}\\
\epsfig{file=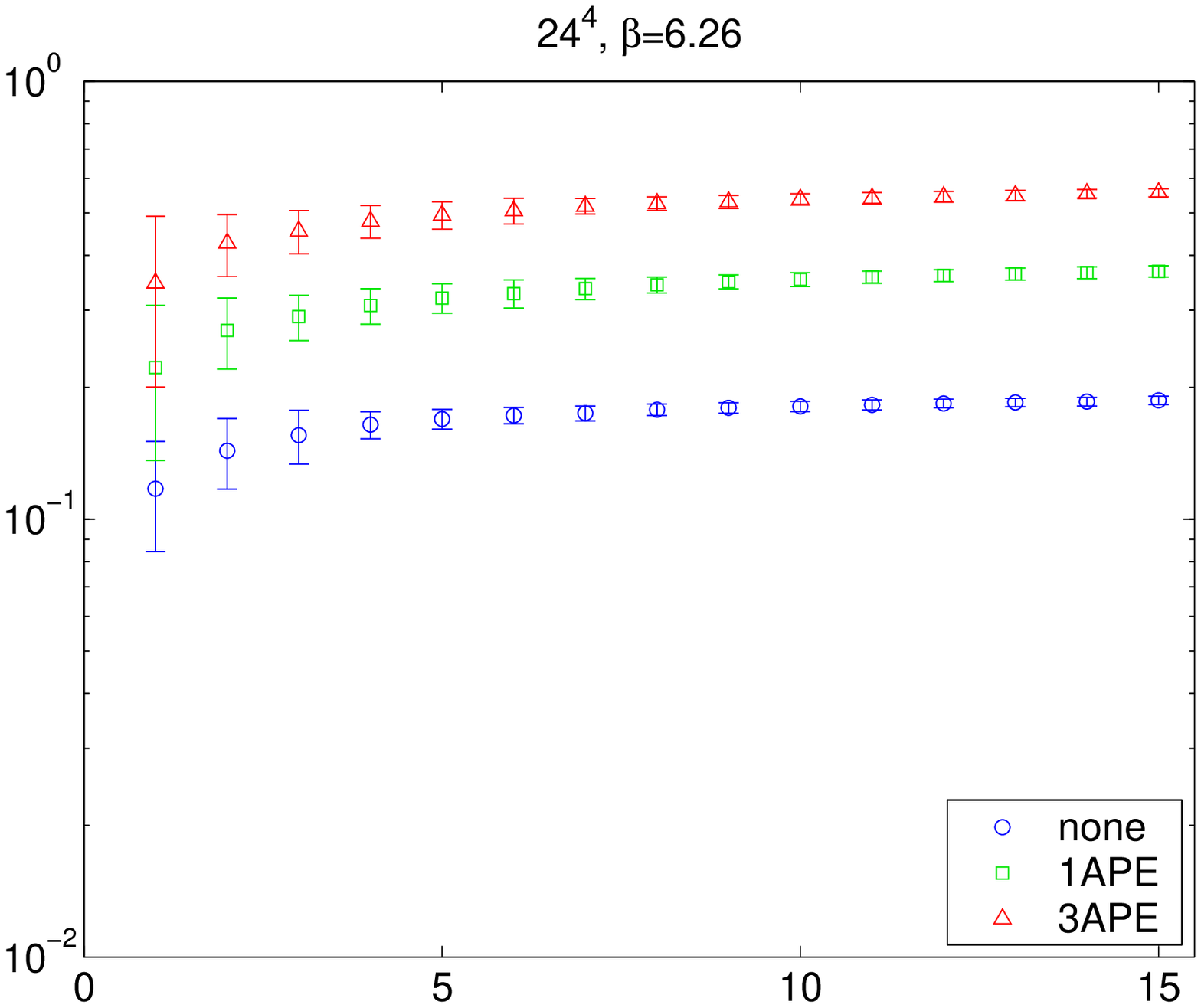,height=6cm}
\epsfig{file=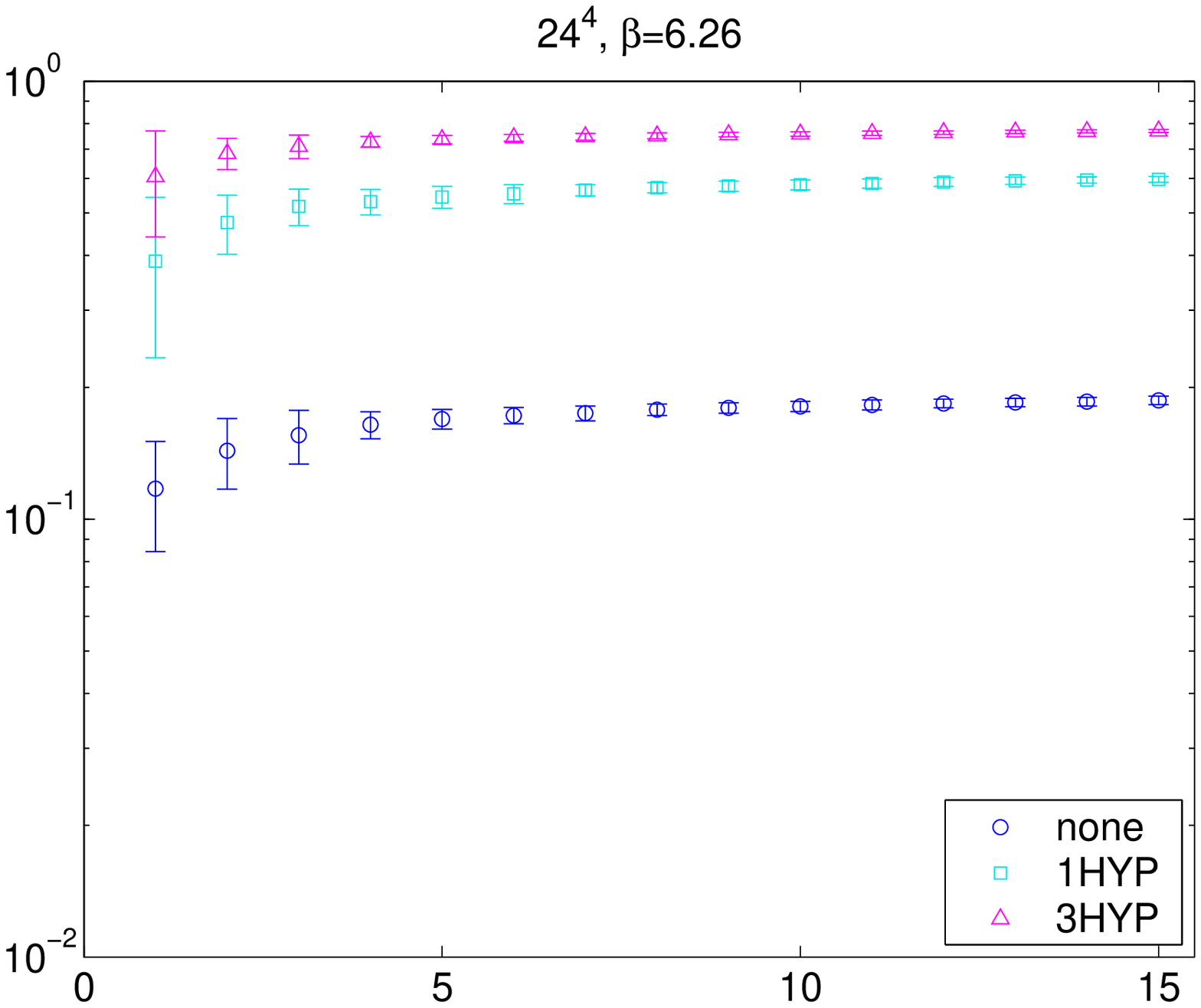,height=6cm}
\end{center}
\vspace{-7mm}
\caption{Mean and standard deviation ($\be\!=\!5.66,5.84,6.00,6.26$, from top
to bottom) of the 15 lowest eigenvalues of $|\HW|$ at $\rh\!=\!1$ in
semi-logarithmic form with 0,1,3 steps of APE or HYP filtering.}
\label{fig:low_average}
\end{figure}

Fig.\,\ref{fig:low_sequence} shows, as an illustration, the 15 lowest
eigenvalues \cite{Kalkreuter:1995mm} of $|\HW|$ on 25 configurations at
$\be\!=\!6.0$, without filtering and after 1,3 steps of APE or HYP smoothing.
The filtering increases the upper end of the band of eigenvalues shown.
In fact, just this upper end matters in terms of CPU time, since in practice
one projects out the lowest few modes \cite{Hernandez:2000sb} and constructs
the function in (\ref{over}) over the relevant spectral range of $|\HW|$ on the
subspace orthogonal to these modes.
Hence the sequence of the 15th eigenvalue represents the relevant quantity, if
14 modes are treated exactly, and this band gets lifted by filtering.
Evidently, a single APE step is less efficient than a single HYP step, and
adding two more steps lifts the 15th eigenvalue further, but the lifting factor
is no more as large as it was in the first step.
Here and below we use the parameters $\al_\mr{APE}\!=\!0.5$,
$\al_\mr{HYP}\!=\!(0.75,0.6,0.3)$ \cite{hyp} (for details of the $SU(3)$
projection see e.g.\ \cite{Kiskis:2003rd} or the appendix of
\cite{Durr:2004xu}) and, unless stated otherwise, $\rh\!=\!1$.

\begin{table}[!t]
\begin{center}
\begin{tabular}{|l|ccccc|}
\hline
$\be$  &  $5.66$ &  $5.84$ &  $6.00$ &$6.00\;(\rh=1.4)$&  $6.26$  \\
\hline
none   &0.149(02)&0.087(2)&0.123(1)&    0.222(3)    &0.187(1) \\
1\,APE &0.146(05)&0.213(4)&0.290(2)&    0.343(5)    &0.368(2) \\
3\,APE &0.237(12)&0.364(6)&0.469(2)&    0.547(7)    &0.557(2) \\
1\,HYP &0.281(14)&0.419(7)&0.519(2)&    0.618(5)    &0.597(2) \\
3\,HYP &0.548(14)&0.653(5)&0.717(1)&    0.639(1)    &0.770(1) \\
\hline
\end{tabular}
\end{center}
\vspace{-6mm}
\caption{Start of the ``bulk'' part of the eigenvalue spectrum of the shifted
hermitean Wilson operator $|\HW|$ without filtering and after one or three APE
or HYP steps. We use the mean of the 15th-smallest eigenvalue to define the
``bulk'' edge. Unless indicated otherwise, the numbers refer to the case
$\rh\!=\!1$.}
\label{tab:15th_average}
\end{table}

Fig.\,\ref{fig:low_average} shows the mean and the standard deviation of the 15
lowest eigenvalues of $|\HW|$, with our standard filtering options
(none, 1\,APE, 3\,APE, 1\,HYP, 3\,HYP).
In this logarithmic representation it is easy to see that (apart from the
coarsest lattice which represents a special case discussed in App.\,A) all
15 eigenvalues get lifted, at a given coupling, by virtually the same factor.
Specifically, the 15th%
\footnote{This number needs to be scaled with the physical box volume; working,
for any given $\be$, in a $(2.0\fm)^4$ box instead of $(1.5\fm)^4$, our
statement would most likely be adequate for the 47th mode.}
eigenvalue gets multiplied by $\la_\mr{1\,HYP}/\la_\mr{none}\!=\!4.8,4.2,3.2$
at $\be\!=\!5.84,6.00,6.26$.
Thus the lifting effect that filtering has on the ``bulk'' part of the $|\HW|$
spectrum diminishes somewhat towards the continuum, but for accessible
couplings it remains substantial.
Details of the ensemble average of the 15th eigenvalue are collected in
Tab.\,\ref{tab:15th_average}.
The second observation is that the bands become flatter at large $\be$, hence
the onset of the ``bulk'' becomes a less ambiguous concept at weaker coupling.
Had we chosen the 10th or 20th mode to
define the ``bulk edge'' instead of the 15th, this would cause a small change
at $\be\!=\!6.26$, but it would make a substantial difference at the smallest
$\be$ shown.

\begin{figure}
\begin{center}
\epsfig{file=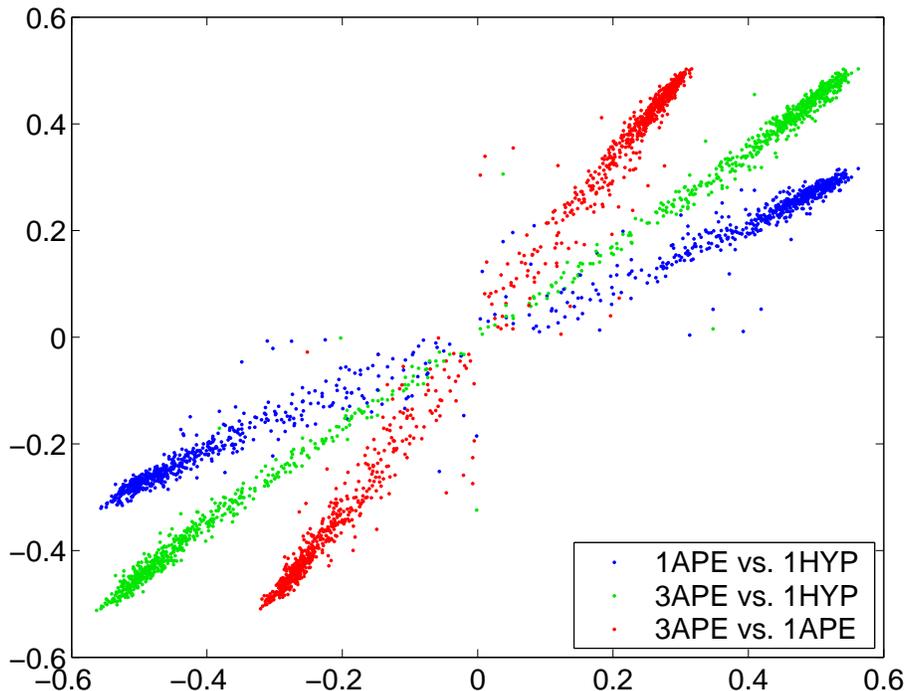,width=12cm}
\end{center}
\vspace{-6mm}
\caption{Correlation of the 15 eigenvalues closest to zero of $\HW$ with
various filtering options at $\be\!=\!6.0$.}
\label{fig:low_correlation}
\end{figure}

A point of theoretical interest is whether the low-lying eigenvalues of the
(shifted) hermitean Wilson operator $\HW$ are correlated, between different
smearing levels, just as the low-lying eigenvalues of the final $D_\mr{ov}$
were found to be correlated for large enough $\be$ \cite{Durr:2004as}.
Fig.\,\ref{fig:low_correlation} shows that this is almost true -- the
eigenvalues correlate if they are sufficiently large in absolute magnitude,
but the correlation weakens closer to the origin.
Here, a technical issue comes along.
Ideally, one would pair the eigenvalues by considering a smooth interpolation
between the two filtering recipes. Changes in topology (as seen by the overlap
operator) would then be evident as stray points in quadrants 2 or 4.
However, since we just know the eigenvalues
shown we decided to pair them starting from 0.
Now there are no points in quadrants 2 or 4 by definition and changes in
topology manifest themselves through a reduced correlation of the few lowest
eigenvalues in absolute magnitude.
Such topology changes are expected to occur with an $O(a^2)$ re-definition
of the overlap operator, e.g.\ by changing the filtering or $\rh$
\cite{Durr:2004as}.

\begin{figure}
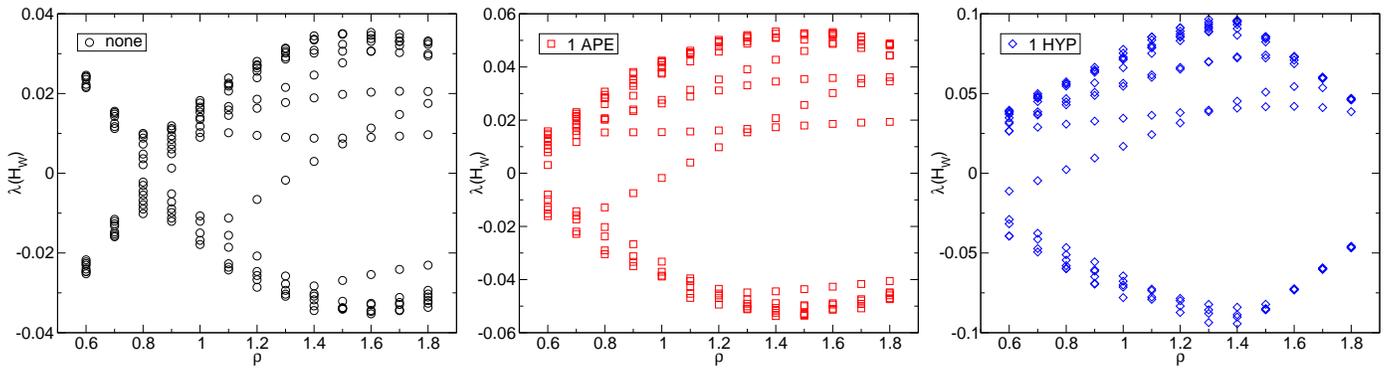

\begin{center}
\epsfig{file=nextstrike_v1.figs/ev_flow_L16_b6.00_w.19_Thin.eps,width=6cm}
\epsfig{file=nextstrike_v1.figs/ev_flow_L16_b6.00_w.19_APE1.eps,width=6cm}
\epsfig{file=nextstrike_v1.figs/ev_flow_L16_b6.00_w.19_HYP1.eps,width=6cm}\\
\end{center}
\vspace{-8mm}
\caption{Eigenvalue flows of $\HW$ with three filterings (none, 1\,APE, 1\,HYP)
on one $16^4$ configuration.}
\label{fig:flow}
\end{figure}

A similar conclusion is drawn from the flow of eigenvalues $\HW$ as shown in
Fig.\,\ref{fig:flow} for one $16^4$ configuration.
One effect of filtering is to stretch the whole scenery in the vertical
direction (note the vertical scale).
Filtering also shifts the entire eigenvalue flow to the left which is
consistent with the reduction of the additive mass renormalization of the
kernel operator as discussed in the introduction.
Note that there is, from a conceptual viewpoint, no reason to prefer one
filtering level over any other one; what we see is just a manifestation of the
$O(a^2)$ ambiguity of the overlap operator \cite{Durr:2004as}.

\enlargethispage{4mm}

\begin{figure}[!b]
\vspace*{-2mm}
\begin{center}
\epsfig{file=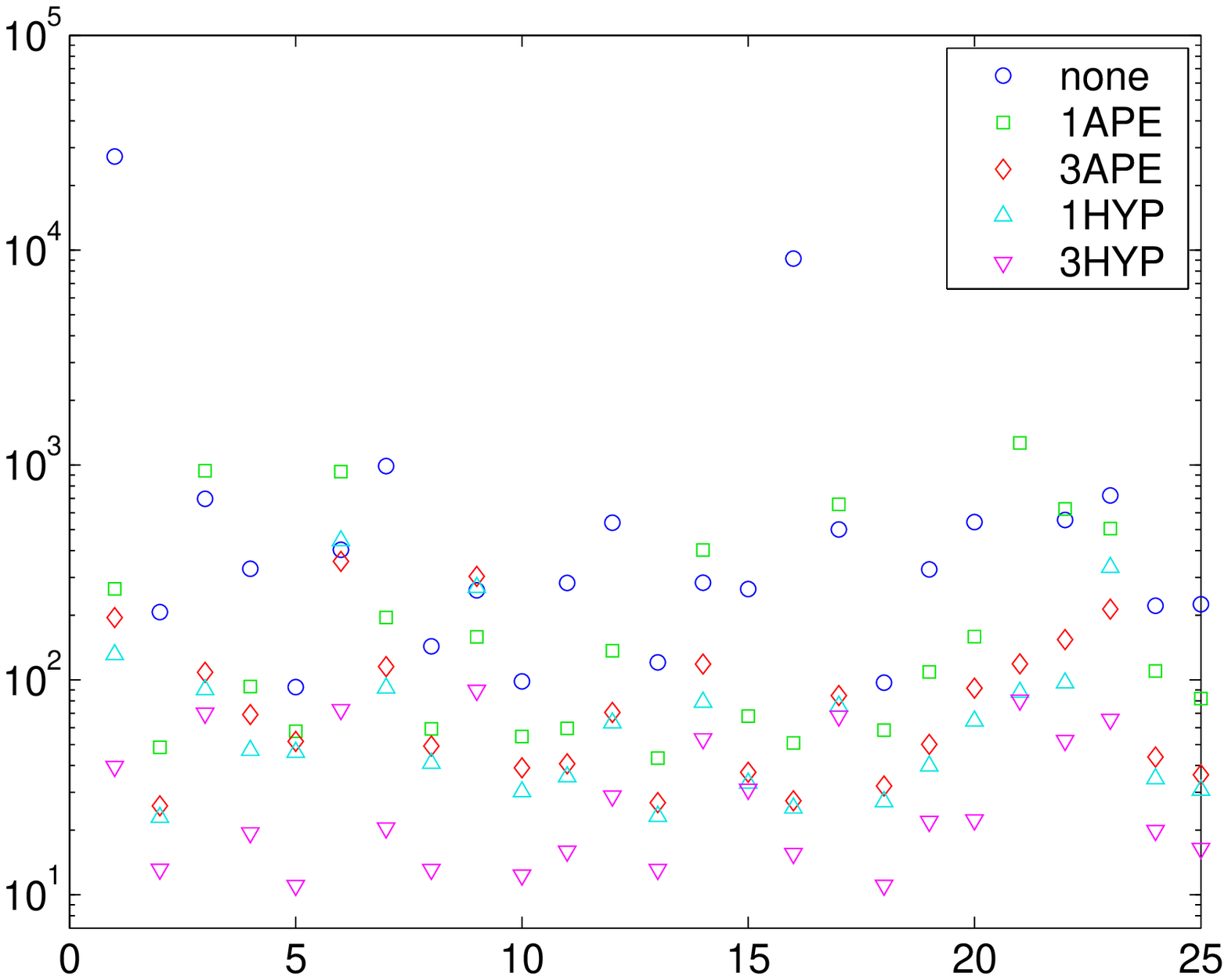,height=7.2cm}
\epsfig{file=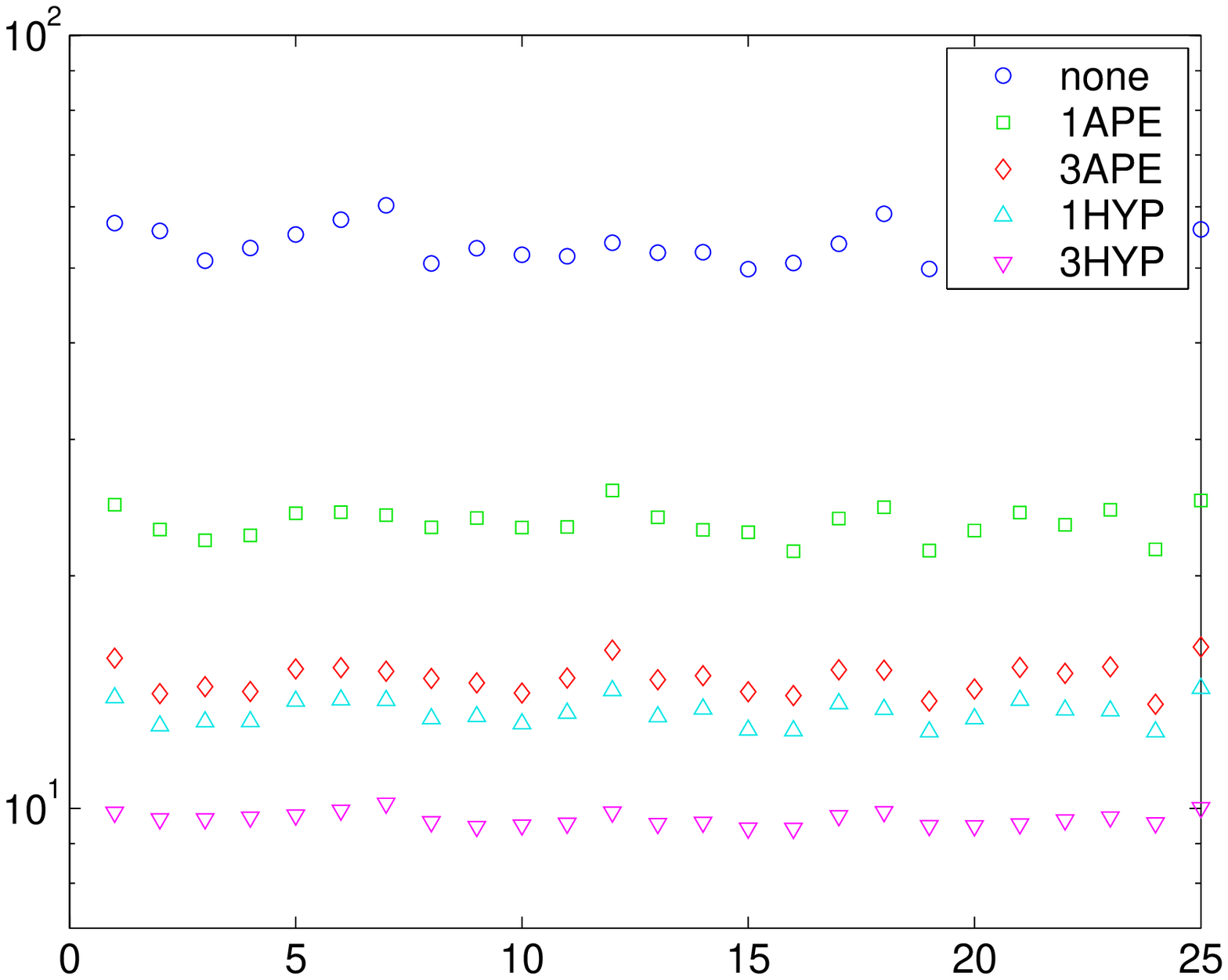,height=7.2cm}\\
\end{center}
\vspace{-6mm}
\caption{Condition number of $|\HW|$ on 25 configurations at $\be\!=\!6.0$
without projection (left) and with 14 modes handled exactly (right), after
0,1,3 steps of APE or HYP filtering.}
\label{fig:cond_sequence}
\end{figure}

\begin{table}
\begin{center}
\begin{tabular}{|l|ccccc|}
\hline
$\be$  &  $5.66$ &  $5.84$ &  $6.00$ &$6.00\;(\rh=1.4)$&  $6.26$ \\
\hline
none   & 43.5(04)& 75.2(15)& 53.2(4)&    27.9(4)     &35.3(2) \\
1\,APE & 47.2(18)& 31.8(06)& 23.4(2)&    18.6(3)     &18.5(1) \\
3\,APE & 30.5(17)& 19.0(03)& 14.7(1)&    11.9(2)     &12.4(1) \\
1\,HYP & 25.7(14)& 16.5(03)& 13.3(1)&    10.5(1)     &11.6(1) \\
3\,HYP & 12.8(04)& 10.6(01)& 9.69(3)&    10.2(1)     &9.04(2) \\
\hline
\end{tabular}
\end{center}
\vspace{-6mm}
\caption{Mean condition number $1/\ep$ of $|\HW|$ at $\rh\!=\!1$ after
projecting the 14 lowest eigenmodes.}
\label{tab:15th_cond}
\end{table}

To assess the CPU time needed for the massless overlap, the behavior of
the ``bulk edge'' of the $|\HW|$ spectrum is one ingredient.
What really matters is the condition number, thus we need to study the largest
eigenvalue, too.
From the naive discussion around Fig.\,\ref{fig:spec_schwinger} in the
introduction one expects that filtering barely affects the largest eigenvalue
of $|\HW|$.
It turns out that this is indeed true, for instance at $\be\!=\!6.0$ a single
HYP filtering step lifts it from $6.55(1)$ to $6.88(1)$.
Hence, filtering has an overall beneficial effect on the condition number as
illustrated in Fig.\,\ref{fig:cond_sequence}.
Without projection the condition number fluctuates wildly and occasionally it
may increase through filtering (i.e.\ the lowest eigenvalue decreases, cf.\
Fig.\,\ref{fig:low_sequence})
but after projecting 14 eigenmodes this never occurs.
The bottom line is that the combination of filtering and projection reduces
the condition number much more vigorously than either one alone could do.
Average condition numbers after projecting out 14 eigenmodes are collected in
Tab.\,\ref{tab:15th_cond} (regarding the first entry, cf.\ App.\,A).
As a side remark we note that the horizontal increase to the left explains why
in a fixed physical volume simulating unfiltered overlap quarks on a coarse
lattice is not so much cheaper than on a fine one; for the filtered version
this penalty is reduced.


We have also studied the condition number of $|\HW|$ as a function of the
parameter $\rh$.
With and without filtering the minimum is rather shallow and at a $\rh$ value
above $1$.
Since in the free case
\beq
\ep=\left\{
\begin{array}{ll}
\rh/(8\!-\!\rh)        &\mr{for}\;0\!<\!\rh\!\le\!1 \\
(2\!-\!\rh)/(8\!-\!\rh)&\mr{for}\;1\!\le\!\rh\!<\!2
\end{array}
\right.
\label{ep_vs_rho_free}
\eeq
we expect that larger $\be$ values will further drive the minimum location
towards $\rh\!=\!1$.

The last step is to convert the reduced condition number, brought by the
filtering, of $|\HW|$ on the subspace orthogonal to the lowest 14 modes into a
lower degree of the polynomial/rational approximation of the $1/\!\sqrt{.}$
function in (\ref{over}) and thus into actual savings of CPU time in step 1 of
the introduction.

\begin{figure}[!b]
\vspace{-2mm}
\begin{center}
\epsfig{file=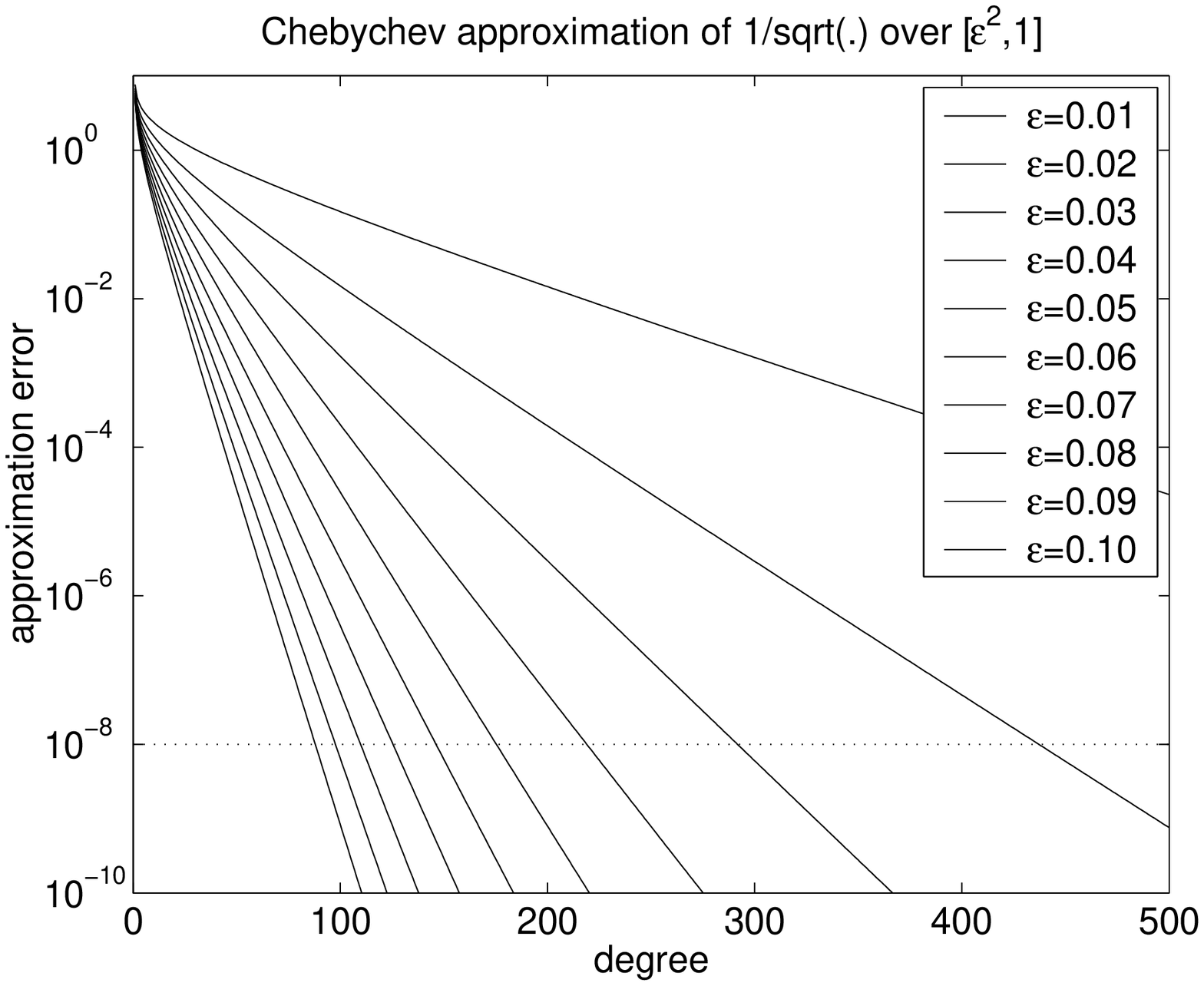,height=7.6cm}
\epsfig{file=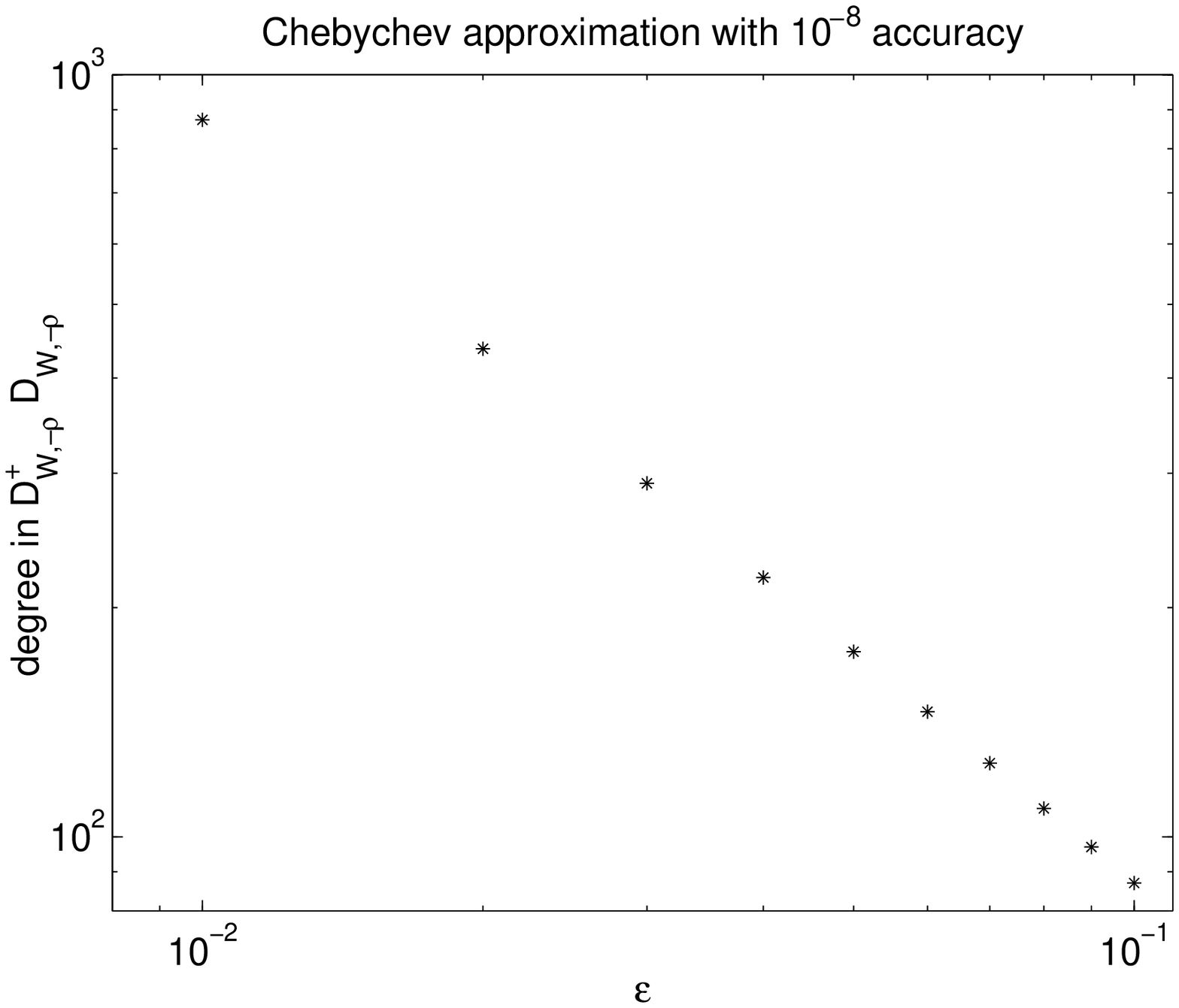,height=7.6cm}
\end{center}
\vspace{-6mm}
\caption{Accuracy of the Chebychev approximation of $(.)^{-1/2}$ over the
interval $[\ep^2,1]$ versus polynomial degree (left) and the degree (number
of forward applications of $D_{\mr{W},-\rh}\dag D_{\mr{W},-\rh}^{}$) needed to
reach absolute accuracy $10^{-8}$ versus the inverse condition number $\ep$
of the shifted hermitean Wilson operator $|\HW|$.}
\label{fig:chebychev_theo}
\end{figure}

\begin{figure}[!t]
\begin{center}
\epsfig{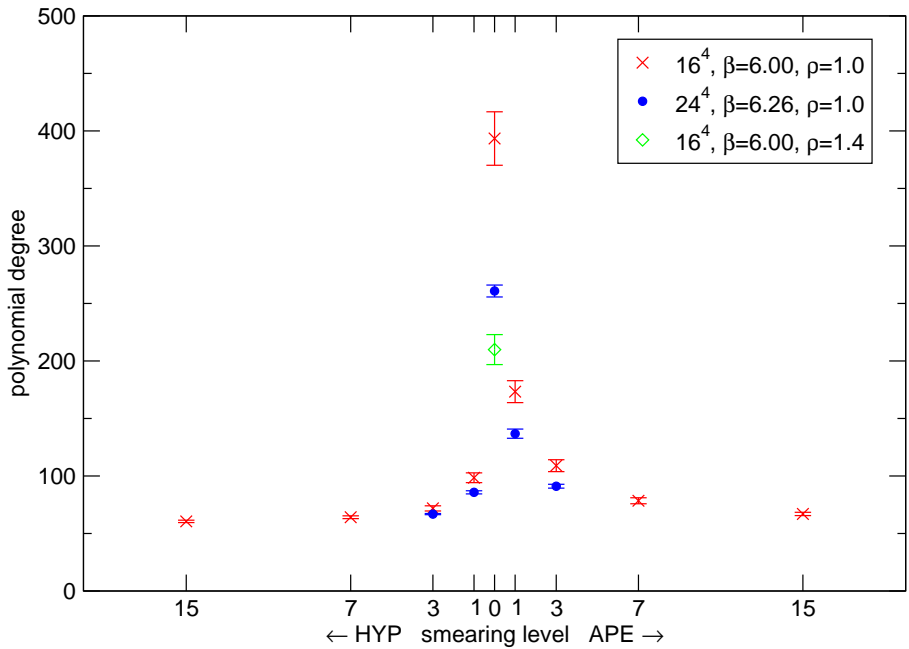}
\end{center}
\vspace{-8mm}
\caption{Mean and standard deviation of the Chebychev polynomial degree used to
achieve a minimax accuracy of $10^{-8}$. At $\be\!=\!6.0$ and $\rho\!=\!1$, a
single HYP step results in a speedup by a factor $\sim\!4$ for the massless
overlap operator. Comparing to the situation with $\rh\!=\!1.4$ and no
filtering the factor is $\sim\!2$.}
\label{fig:chebychev_prac}
\end{figure}

The precise speedup factor depends on the implementation of the massless
overlap operator (\ref{over}).
For definiteness let us consider the approximation of the inverse square root
over the range $[\ep^2,1]$ through Chebychev polynomials
\cite{Hernandez:2000sb}.
Fig.\,\ref{fig:chebychev_theo} shows on the l.h.s.\ for a few inverse condition
numbers $\ep$ of $|\HW|$ the well known exponential fall-off pattern of the
truncation error of the Chebychev approximation versus the number of
applications of $\HW^2\!=\!D_{\mr{W},-\rh}\dag D_{\mr{W},-\rh}^{}$.
What matters for our purpose is the dependence of the polynomial degree
required to reach a fixed minimax accuracy --~say $\de\!=\!10^{-8}$ over the
full approximation range~-- on $\ep$.
As is evident from the r.h.s.\ of that figure, the relation
\beq
\mr{degree}\propto{\ep^{-1}}
\label{deg_invprop_eps}
\eeq
holds in good approximation.
Thus, from (\ref{deg_invprop_eps}) and a look at Tab.\,\ref{tab:15th_cond} one
predicts that at $\be\!=\!6.0$ and $\rh\!=\!1$ a single HYP step will speed up
the construction of the overlap (on average) by a factor $53.2/13.3\!=\!4.00$,
and this is in good agreement with what we find in actual runs (see
Fig.\,\ref{fig:chebychev_prac}).
On a coarser lattice this factor would be somewhat larger
($4.56$ at $\be\!=\!5.84$) while on a finer lattice it tends to decrease
($3.04$ at $\be\!=\!6.26$), but it certainly remains substantial at all
accessible couplings.

To approximate the inverse square root or sign function over the relevant
range, two main strategies are found in the literature.
Polynomial \cite{Bunk:1997wj,Edwards:1998yw,Hernandez:2000sb} and rational
\cite{Edwards:1998yw,vandenEshof:2002ms} representations have been tried.
We have concentrated on the Chebychev variant, since this one is efficient and
easy to implement.
It goes without saying that the lifting effect on the bulk of the
$|\gaf D_\mr{kern}|$ eigenvalues translates into similar savings on CPU time in
step 1 of the introduction, if another representation is used.
For instance, in the rational approach it is the increase of the smallest zero
of the denominator polynomial that lets one get away with fewer iterations in
the inner multishift CG.

For five dimensional variants of the overlap operator \cite{Neuberger:1997bg,
borici_dwisover,Narayanan:2000qx,wenger_cf,Brower:2004xi}, in particular the
domain wall formulation, the computational gain comes from the reduction of the
extent of the fifth dimension needed to reach a given residual mass.
What we would like to stress here is simply that our proposal to replace the
``thin'' links by ``thick'' links is generically useful for any kind of
overlap variant.


\section{Locality}


It has been shown \cite{horvathbietenholz} that the overlap operator cannot be
ultralocal, as opposed to the Wilson operator where $\DW(x,y)\!=\!0$ for
$||x\!-\!y||_1\!>\!1$.
To guarantee the universality of the underlying field theory and hence to
obtain the correct continuum limit it is sufficient to have an operator with
\beq
D_\mr{ov}(x,y)\propto\exp(-\nu ||x-y||) \qquad\mr{for}\quad ||x-y||\gg1
\label{overlap_localization}
\eeq
where the localization $\nu$ is of the order of the cut-off, i.e.\
$\nu\!=\!O(1)$ [in lattice units].
In practice, for a given lattice spacing the condition
(\ref{overlap_localization}) gives an upper bound on any physical mass that one
can extract, and it is therefore crucial to have an operator as local as
possible, i.e.\ with a maximal $\nu$.
In \cite{Hernandez:1998et} it has been demonstrated that the standard overlap
operator indeed obeys (\ref{overlap_localization}).
It is clear that their proof goes through for our filtered variant, but it
is open in which way the localization $\nu$ is influenced.
Naively, one might think that the locality will deteriorate, since the original
links entering the covariant derivative of the filtered kernel spread over a
larger volume.
As first observed by Kovacs \cite{Kovacs}, the filtered overlap turns out to
be even more local than the standard one and this is achieved without tuning
$\rho$.

\begin{figure}[!t]
\begin{center}
\epsfig{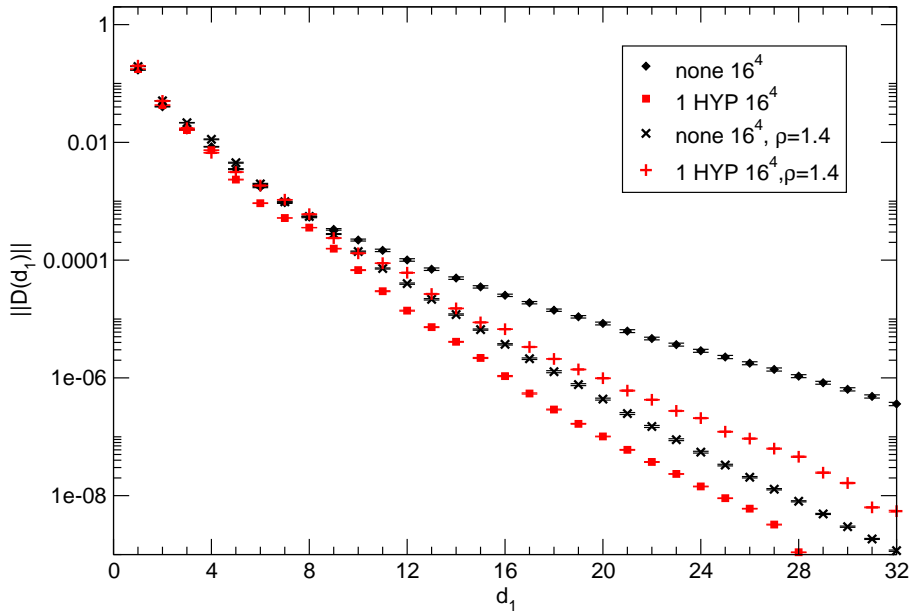}
\end{center}
\vspace{-8mm}
\caption{Localization of the overlap at $\beta\!=\!6.0$ without filtering and
after 1\,HYP step, for $\rho\!=\!1.0$ and $\rho\!=\!1.4$. A single HYP step
proves more efficient than optimizing $\rho$. Filtering and $\rho\!>\!1$ should
not be combined; for the filtered operator the untuned choice $\rho\!=\!1$ is
reasonable (but still not optimal).}
\label{fig:localization_6.0}
\end{figure}

In Fig.\,\ref{fig:localization_6.0} we plot the localization of $D_\mr{ov}$
at $\be\!=\!6.0$ with two projection parameters ($\rh\!=\!1.0,1.4$) and two
filtering options (none, 1\,HYP).
The ordinate is the maximum over the 2-norm of $D_\mr{ov}\et$ at $x$ with $\et$
a normalized $\de$-peak source vector at the point $y$ in the lattice, the
abscissa is the ``taxi driver'' distance $d_1\!=\!||x\!-\!y||_1$ to the
location of the $\de$-peak, i.e.\ we plot the function
\beq
f(d_1)=\mr{sup}\Big\{||(D_\mr{ov}\et)(x)||_2\;\Big|\;||x\!-\!y||_1\!=\!d_1\Big\}
\label{supremum}
\eeq
versus $d_1$, as first studied in \cite{Hernandez:1998et}.
Comparing the two unfiltered operators (black/dark diamonds and crosses) one
finds their result reproduced that (at this $\be$) adjusting $\rh$ to a value
around $1.4$ lets $f(d_1)$ fall off steeper than with the value $1.0$ which is
the canonical choice in view of the spectrum of the Wilson operator
sufficiently close to the continuum (cf.\ Fig.\,\ref{fig:spec_schwinger}).
The interesting observation is that a single HYP step together with
$\rh\!=\!1.0$ (red/light squares) results in an even steeper descent than the
unfiltered version with $\rh\!=\!1.4$ (which was chosen to nearly optimize
the locality of the unfiltered operator).
The last curve shown (red/light pluses) indicates that one should not attempt
to combine the filtering with a $\rh$ value that would be optimal for the
unfiltered operator.

\begin{figure}[!t]
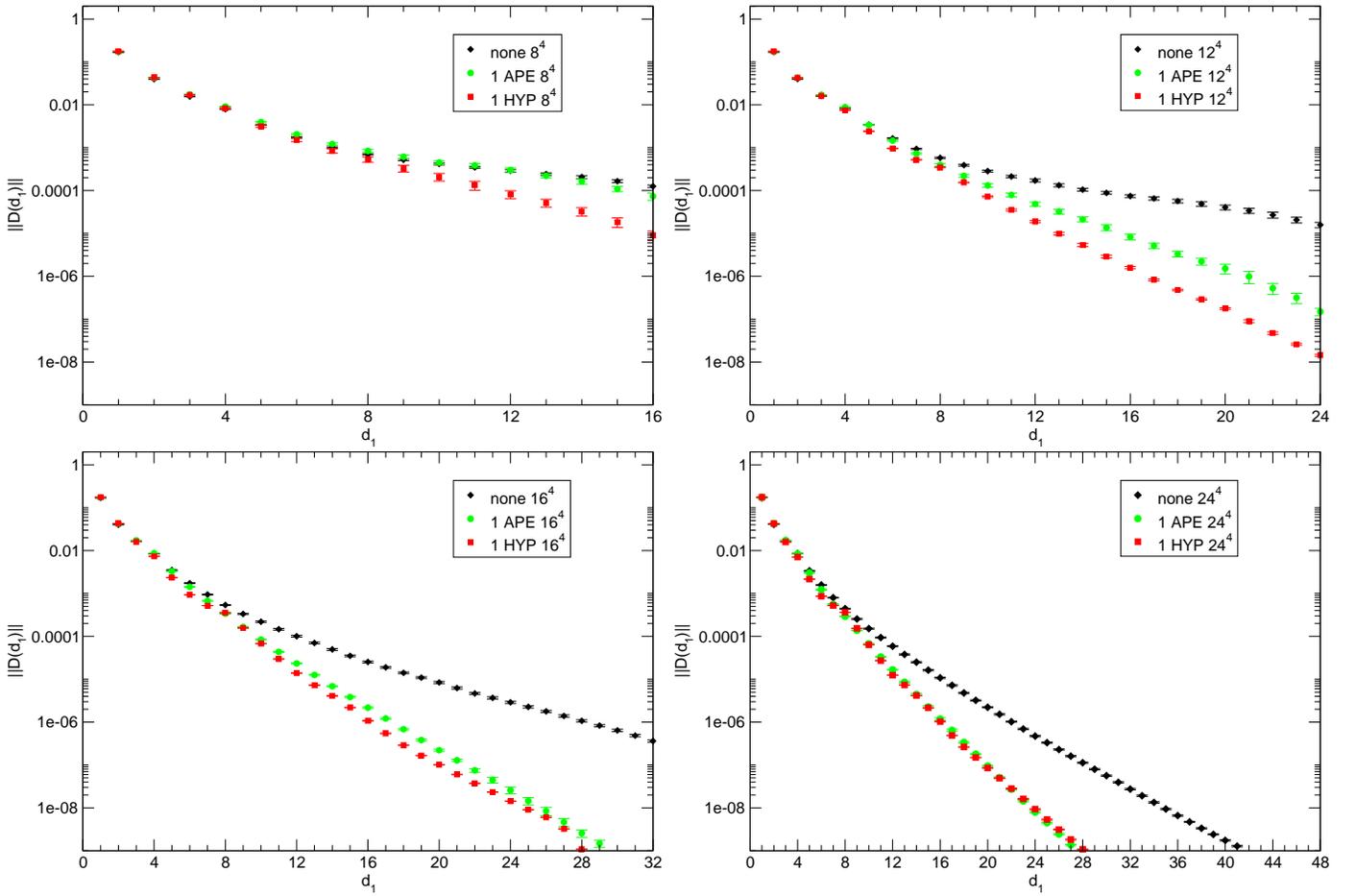

\begin{center}
\epsfig{file=nextstrike_v1.figs/loc_08.eps,width=9.1cm}
\epsfig{file=nextstrike_v1.figs/loc_12.eps,width=9.1cm}\\
\epsfig{file=nextstrike_v1.figs/loc_16.eps,width=9.1cm}
\epsfig{file=nextstrike_v1.figs/loc_24.eps,width=9.1cm}
\end{center}
\vspace{-6mm}
\caption{Localization of the unfiltered overlap and of the versions with one
APE or HYP step. The data are for the matched ensembles
($ 8^4$ at $\be\!=\!5.66$, $12^4$ at $\be\!=\!5.84$,
 $16^4$ at $\be\!=\!6.00$, $24^4$ at $\be\!=\!6.26$) and $\rho\!=\!1$
throughout. On sufficiently fine lattices the choice of smearing proves
irrelevant.}
\label{fig:localization_all}
\end{figure}

An obvious question is whether filtering remains useful on fine lattices.
Fig.\,\ref{fig:localization_all} shows the fall-off at four couplings with
no smearing, 1\,APE and 1\,HYP step, with $\rh\!=\!1$ fixed.
On the coarsest lattice smearing alters the locality just modestly, on the
two intermediate ones ($\be\!=\!5.84,6.00$) the locality gets substantially
improved, with HYP doing a better job than APE.
On the finest lattice, the improvement is still sizable, but there is almost
no difference among the two filtering recipes.
At this coupling further smearing steps would then diminish the locality.
The localization measured with the definition (\ref{def_nu}) [which we
use for technical reasons discussed below] is summarized in
Tab.\,\ref{tab:localization}.

\begin{table}[!t]
\begin{center}
\begin{tabular}{|l|ccccc|}
\hline
$\be$  &  $5.66$ &  $5.84$ &  $6.00$ &$6.00\;(\rh=1.4)$&  $6.26$ \\
\hline
none   &0.330(18)&0.236(17)&0.308(09)&0.571(10)&0.370(07)\\
1\,APE &0.344(30)&0.447(29)&0.577(13)&0.543(06)&0.586(18)\\
3\,APE &0.429(49)&0.634(30)&0.682(11)&0.485(04)&0.549(09)\\
1\,HYP &0.469(48)&0.642(32)&0.695(10)&0.480(05)&0.554(05)\\
3\,HYP &0.610(12)&0.630(32)&0.585(03)&0.476(08)&0.519(02)\\
\hline
\end{tabular}
\end{center}
\vspace{-6mm}
\caption{Localization $\nu$ of the overlap operator with an unfiltered Wilson
kernel and after 1 or 3 steps of APE or HYP filtering. At $\be\!=\!6.0$ we
compare to $\rh\!=\!1.4$ which is nearly optimal without filtering
\cite{Hernandez:1998et}. We use the definition (\ref{def_nu}); the error is
only statistical.}
\label{tab:localization}
\end{table}

\begin{figure}
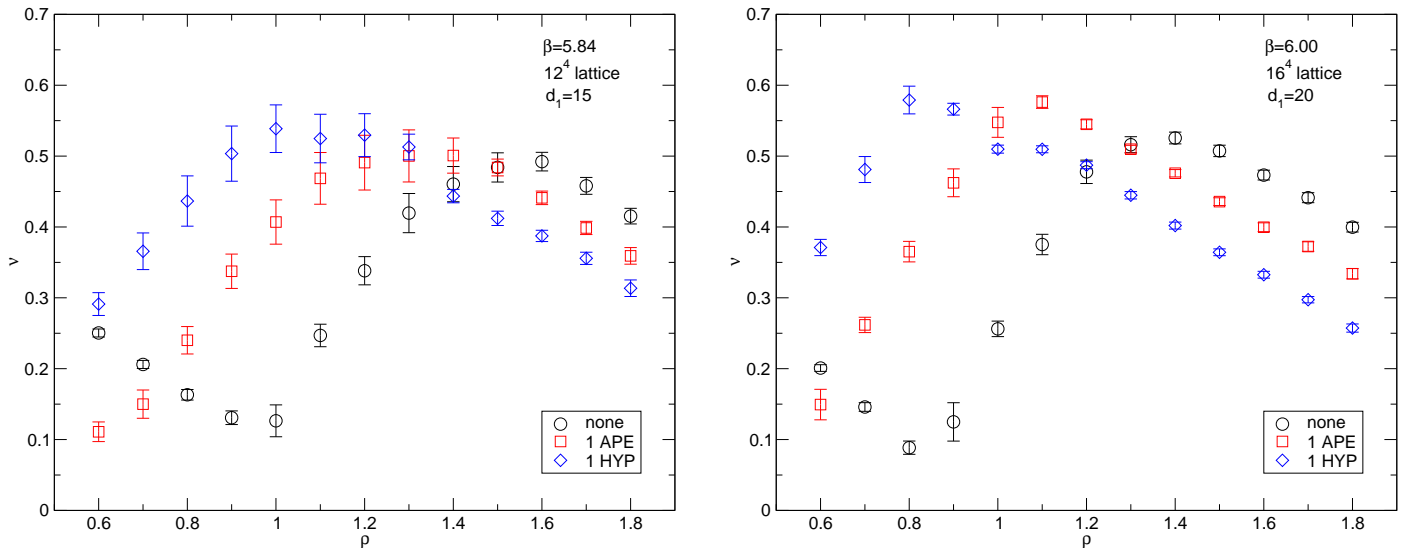

\begin{center}
\epsfig{file=nextstrike_v1.figs/eff_loc_L12_b5.84_r15.eps,width=89mm}\hfill
\epsfig{file=nextstrike_v1.figs/eff_loc_L16_b6.00_r20.eps,width=89mm}
\end{center}
\vspace{-8mm}
\caption{Localization $\nu$ vs.\ $\rh$ at $\be\!=\!5.84$ (left) and
$\be\!=\!6.00$ (right) with no filtering, 1\,APE or 1\,HYP step. This is the
only case where we deviate from our convention to define $\nu$ via
(\ref{def_nu}) and use $d_1\!=\!{5\ovr4}L$.}
\label{fig:optimumrho_L12L16}
\end{figure}

There is a loose connection between the localization of $D_\mr{ov}$ and the
spectrum of $\HW$, for instance
\beq
||D_\mr{ov}(x,y)|| \leq \mr{const}\times\exp(-{\th\ovr2}||x-y||_1)
\label{HJL1}
\eeq
is a bound found in \cite{Hernandez:1998et}, where $||.||$ is the matrix norm
in Dirac and color space.
The exponent $\th/2$ in (\ref{HJL1}) is defined via the largest and smallest
eigenvalue of $D_{\mr{W},-\rh}\dag D_{\mr{W},-\rh}^{}$ through
\beq
\cosh(\th)=
{\la_\mr{max}/\la_\mr{min}+1\ovr\la_\mr{max}/\la_\mr{min}-1}=
{1+\ep^2\ovr
 1-\ep^2}
\label{HJL2}
\eeq
where we like to express the r.h.s.\ in terms of the inverse condition number
$\ep$ of $|\HW|$.
Expanding either side to first order one obtains the simple relation (after
getting rid of the unphysical $\th\!<\!0$ solution)
\beq
{\th\ovr2}=\ep+O(\ep^2)
\;.
\label{HJL3}
\eeq
As already mentioned in \cite{Hernandez:1998et} the exponent $\th/2$,
defined via the spectral properties of the underlying $|\HW|$, is a rather bad
estimate for the actual localization $\nu$.
The situation is not much better for the filtered variety, as a brief
comparison of our Tabs.\,\ref{tab:15th_cond} and \ref{tab:localization}
reveals%
\footnote{Tab.\,\ref{tab:15th_cond} contains the condition number on the
subspace orthogonal to the 14 lowest modes, while $\ep$ in (\ref{HJL1},
\ref{HJL3}) refers to the full operator.
For two reasons we propose to re-interpret (\ref{HJL3}) as a prediction for the
locality of $D_\mr{ov}$ with $\ep$ the ratio of the lower to the upper end of
the \emph{bulk} of eigenvalues of $|\HW|$.
A practical hint is that the unprojected condition number fluctuates wildly
(see Fig.\,\ref{fig:cond_sequence}), whereas the localization is rather stable
for all configurations in an ensemble.
Furthermore, in \cite{Hernandez:1998et} it is shown that an isolated near-zero
mode of $|\HW|$ does normally not affect the locality of $D_\mr{ov}$.
Of course, one cannot repeat that argument indefinitely, but still a test
whether a modified $\ep$ helps is interesting.}.
Though quantitatively unsuccessful, this connection still gives a qualitative
hint that the overlap operator with a filtered Wilson kernel might enjoy better
localization properties due to the reduced condition number of $H_W$.
There are more detailed bounds in the literature \cite{Neuberger:1999pz,
Kikukawa:1999dk,Fujikawa:2001av,adams_bounds,Chiu:2002kj}, but it seems fair to
say that a quantitative understanding of the localization of $D_\mr{ov}$ in
terms of the spectral properties of $H_W$ is a challenge.

The localization $\nu$ as a function of the projection parameter $\rh$ is
presented in Fig.\,\ref{fig:optimumrho_L12L16}.
For $\be\!=\!5.84,6.00$ the optimum parameter for the unfiltered operator
is around $\rh\!=\!1.6,1.4$, respectively.
For the 1\,HYP operator the localization at $\rh\!=\!1.0$ does not fall short
of the maximal one by a large amount; this is why we restrict much of our
investigation with a filtered $D_\mr{ov}$ to the case $\rh\!=\!1.0$.
Still, the figure suggests that an optimal $\rh$ for the 1\,HYP filtered
operator may be \emph{smaller} than 1, and it decreases with increasing $\be$;
at $\be\!=\!5.84$ we find $\rh_\mr{opt}^\mr{1\,HYP}\!\simeq\!1.0$ and
at $\be\!=\!6.00$ we find $\rh_\mr{opt}^\mr{1\,HYP}\!\simeq\!0.8$.

\begin{figure}[!t]
\begin{center}
\epsfig{file=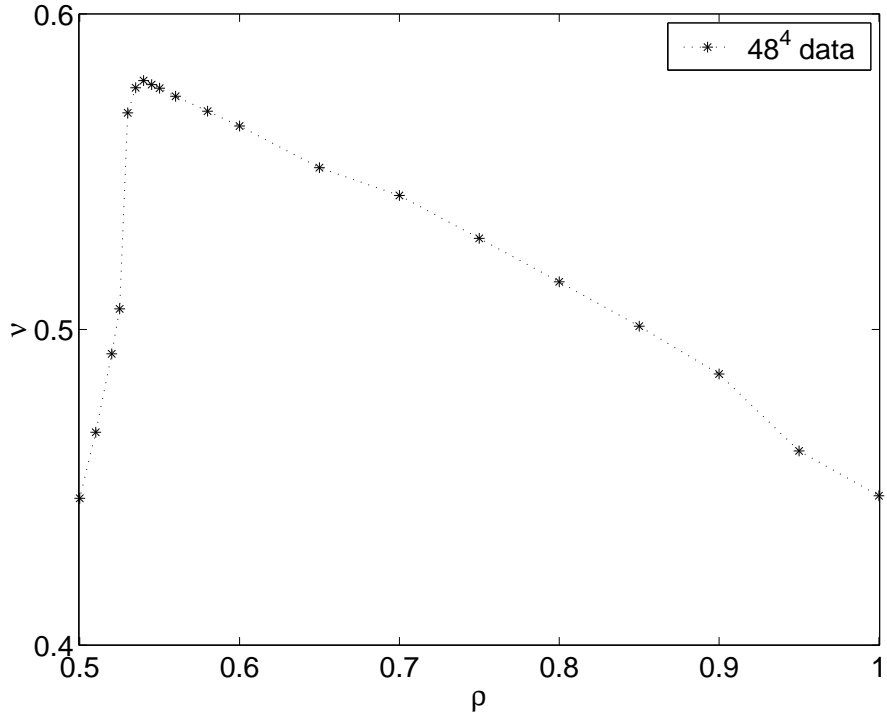,width=12cm}\\
\end{center}
\vspace{-8mm}
\caption{Localization (\ref{def_nu}) in the free field case with
an extremum at $\rh_\mr{opt}^\mr{free}\!\simeq\!0.54$ and a steep descent to
the left. Note that in the $\rh$ range shown the inverse condition number
$\ep\!=\!\rh/(8\!-\!\rh)$ is monotonic.}
\label{fig:localization_free_summary48}
\end{figure}

After dealing with some technical issues to make sure that an $48^4$ lattice
is large enough (see App.\,B), we have studied $\nu$ defined via (\ref{def_nu})
as function of $\rh$ in the free case; the result is shown in
Fig.\,\ref{fig:localization_free_summary48}.
The pattern observed in Fig.\,\ref{fig:optimumrho_L12L16} should thus not come
as a surprise, filtering simply drives the locality properties of the
overlap operator towards the free field case.
In fact, Fig.\,\ref{fig:localization_free_summary48} offers a simple
explanation why it is so difficult to predict the localization $\nu$ from
spectral properties of the underlying $|\HW|$ operator -- in the free case the
inverse condition number (\ref{ep_vs_rho_free}) in the range $0\!<\!\rh\!<\!1$
is monotonic, while $\nu$ has a non-trivial extremum at
$\rh_\mr{opt}^\mr{free}\!\simeq\!0.54$.


\section{Kernel non-normality}


An operator $A$ is called normal, if it commutes with its adjoint
\beq
[A,A\dag]=0
\label{def_nonnormal}
\eeq
which implies that its left and right eigenbasis coincide.
Normality has special implications for lattice Dirac operators.
For a normal Dirac operator $D=\sum_k\la_k |k\>\<k|$ which, in addition, is
$\gaf$-hermitean
\beq
\gaf D\gaf=D\dag
\label{gafh}
\eeq
we immediately obtain
\beq
D\dag=\sum_k\la_k^* |k\>\<k|=\sum_k\la_k\gaf|k\>\<k|\gaf
\eeq
and this implies that eigenmodes with real $\la_k$ are chiral (or may be
linearly combined to chiral modes in case of degeneracies).
Furthermore, for such a $D$ the eigenvectors of the hermitean Dirac operator
\beq
H=\gaf D=\sum_k\la_k \gaf|k\>\<k|
\eeq
are given by $\sqrt{\la_k^*}|k\>\pm\sqrt{\la_k^{}}\gaf|k\>$ with the
corresponding eigenvalues $\pm|\la_k|$.

The continuum Dirac operator is normal, and so are the naive and staggered
discretizations (but the latter two yield more than one flavor in the continuum
limit).
The GW relation (\ref{GW}) together with $\gaf$-hermiticity (\ref{gafh})
also implies normality of the operator, hence $D_\mr{ov}$ is normal.
In fact, the overlap construction can be described as extracting the unique
unitary part of $D_\mr{kern}/\rh$ \cite{Giusti:1999be}, and for a normal
kernel it reduces to a simple radial projection of the $D_\mr{kern}/\rh$
eigenvalues onto the unit circle.

\begin{figure}[!t]
\begin{center}
\epsfig{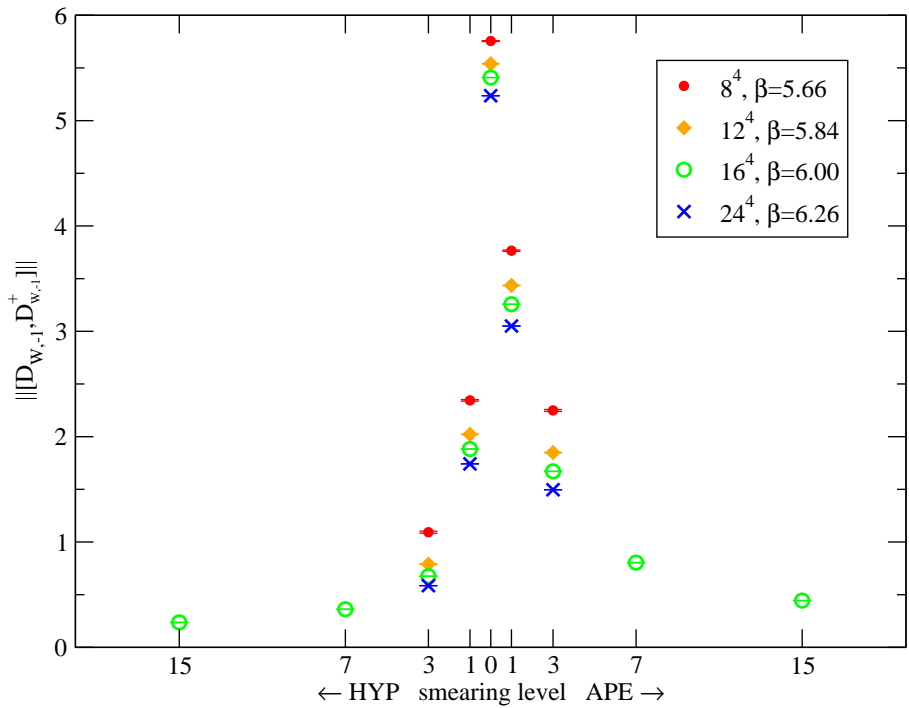}
\end{center}
\vspace{-6mm}
\caption{Non-normality (in lattice units) of the Wilson kernel, as defined in
(\ref{comm2}), as a function of smearing. The improvement of the kernel
normality does not seem to degrade towards the continuum.}
\label{fig:nonnormality}
\end{figure}

The shifted Wilson operator, which we use as a kernel, is not normal.
Some consequences of this non-normality have been explored in other contexts
\cite{kerler_hip_nonnormality}.
Here, it suffices to point out that the relations between the eigenmodes of the
overlap operator, its kernel and the hermitean Dirac operator are not as simple
as above for the case of a normal operator.
Typically, an eigenvector of the (hermitean) kernel will mix into every mode
of the overlap operator, which we expect to have a detrimental effect on the
efficiency of overlap construction algorithms.
Thus, a practically relevant question is whether UV filtering can reduce the
amount of non-normality of the overlap kernel.

To quantify the non-normality of $D_{\mr{W}}$ we measure the 2-norm of
the commutator; technically
\beq
|| [D_{\mr{W},-1}^{},D_{\mr{W},-1}\dag]|\et\> ||
\label{comm2}
\eeq
is averaged over a number of normalized random vectors $|\et\>$.
In Fig.\,\ref{fig:nonnormality} the commutator (\ref{comm2}) is shown for all
$\be$ and smearing levels (since this is not a physical observable, we use
lattice units).
Evidently, any kind of filtering reduces it -- the filtered kernel is thus
closer to normality and has left- and right-eigenvectors that are better
aligned than for the unfiltered version.
Whether ``smart'' overlap construction algorithms can be written which exploit
this property is an open question.


\section{Physics perspectives}


To explore the physics potential of filtered overlap quarks a quenched
spectroscopy study would be highly desirable.
Physical results should reproduce --~after a continuum extrapolation~--
results in the traditional ``thin link'' formulation.
It would be interesting to see whether the speedup in point 1 of the
introduction gets enhanced in points 2-4; in particular if scaling and/or
asymptotic scaling set in earlier, this would make a real difference.
Unfortunately, a detailed scaling study requires substantial computational
resources, but as a first step in this direction we want to investigate the
renormalization of the axial-vector current with filtered overlap quarks.

\begin{figure}[!b]
\begin{center}
\epsfig{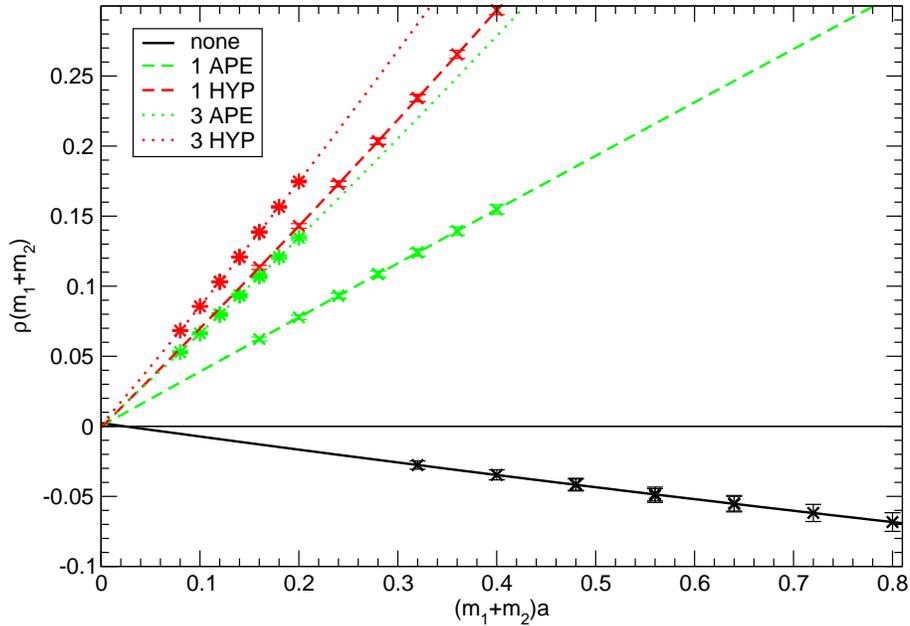}
\end{center}
\vspace{-8mm}
\caption{$m_1^\mr{AWI}\!+\!m_2^\mr{AWI}$ vs.\ $m_1^\mr{bare}\!+\!m_2^\mr{bare}$
[$\mr{slope}\!=\!Z_A^{-1}$] at $\be\!=\!5.66$ . Without filtering $Z_A\!<\!0$,
while any filtering prescription gives $Z_A\!>\!1$, with higher filtering
levels resulting in a value closer to 1.}
\label{fig:ZA_08}
\end{figure}

\begin{figure}[!b]
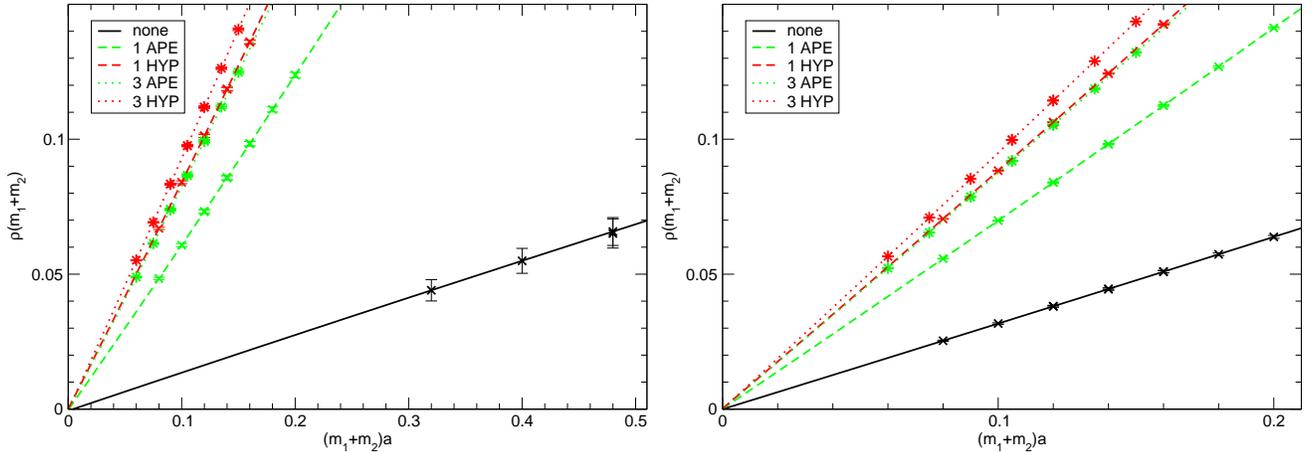

\begin{center}
\epsfig{file=nextstrike_v1.figs/za12_large.eps,height=6cm}
\epsfig{file=nextstrike_v1.figs/za16_large.eps,height=6cm}
\end{center}
\vspace{-8mm}
\caption{$m_1^\mr{AWI}\!+\!m_2^\mr{AWI}$ vs.\ $m_1^\mr{bare}\!+\!m_2^\mr{bare}$
at $\be\!=\!5.84,6.00$. Higher filtering levels shift $Z_A$ closer to 1.}
\label{fig:ZA_1216}
\end{figure}

We follow the method of \cite{Luscher:1996sc,Giusti:2001pk,Berruto:2003rt},
where one starts from the usual (chirally rotated) densities
\bea
    P(x)&=&\ps_1(x)       \gaf[(1-{1\ovr2\rh}D_\mr{ov})\ps_2](x)
\label{def_P}
\\
A_\mu(x)&=&\ps_1(x)\ga_\mu\gaf[(1-{1\ovr2\rh}D_\mr{ov})\ps_2](x)
\label{def_Amu}
\eea
with $\ps_1\!\neq\!\psi_2$ (flavor non-singlet) and defines the correlators
[$x\!=\!(\mb{x},t)$]
\bea
G_{PP}      (t)&=&\sum_\mb{x}P(\mb{x},t) P^c(\mb{0},0)
\label{G_PP}
\\
G_{\nab\!AP}(t)&=&\sum_\mb{x}\bar\nab_{\!4}\!A_4(\mb{x},t) P^c(\mb{0},0)
\label{G_nabAP}
\eea
where $\bar\nab_{\!4}$ is the symmetric derivative in the time direction and
$P^c$ is the conjugate of (\ref{def_P}), i.e.\ with the flavor indices
$1\!\leftrightarrow\!2$ interchanged.
With these correlators at hand one forms the ratio
\beq
\rh(t,m_1,m_2)={G_{\bar\nab\!AP}(t) \ovr G_{PP}(t)}
\label{rho_def}
\eeq
where the second and third argument indicate that the spinors $\ps_1$ and
$\ps_2$ in the densities (\ref{def_P}, \ref{def_Amu}) are solutions to the
massive operators $D_{\mr{ov},m_1}$ and $D_{\mr{ov},m_2}$, respectively.
On account of the axial Ward identity (AWI) the ratio $\rh$ should be constant 
in time, and for light enough quarks (\ref{rho_def}) tends indeed to plateau
rather nicely (see e.g.\ Fig.\,1 in \cite{Berruto:2003rt}).
In a slightly sloppy but transparent notation the plateau value is
$\rh(m_1,m_2)$.
This quantity will --~to the extent to which the AWI is respected at finite
lattice spacing~-- only depend on the sum%
\footnote{In principle, we might use the covariant conserved current for
overlap quarks (see \cite{Kikukawa:1998bg} and the 2nd work in
\cite{Hasenfratz}) with the ``thin'' links replaced by ``thick'' links. Then
the last term on the r.h.s.\ of (\ref{rho_mod}, \ref{rho_fit}) would be absent,
and the AWI would be an exact identity. However, there is a practical problem
with APE or HYP filtering, due to the $SU(3)$ projection involved. The
solution via stout/EXP links is in exact analogy to the dynamical case
discussed in the last section.}
of the quark masses, and thus defines the $m^\mr{AWI}$ quark masses
\beq
\rh(m_1,m_2)=\rh(m_1\!+\!m_2)\;+\;O(a^2)=m_1^\mr{AWI}+m_2^\mr{AWI}\;+\;O(a^2)
\;.
\label{rho_mod}
\eeq

The actual data for our $Z_A$ determination for quenched filtered and
unfiltered overlap quarks are generated with couplings and geometries as
given in the last line of Tab.\,\ref{tab:parameters}.
We restrict ourselves to the canonical choice $\rh\!=\!1$.
We plot $\rh(m_1,m_2)$ versus $m_1\!+\!m_2$ for various quark mass combinations
and filtering levels in Fig.\,\ref{fig:ZA_08} for $\be\!=\!5.66$ and in
Fig.\,\ref{fig:ZA_1216} for $\be\!=\!5.84,6.00$, respectively.
They form one universal band, i.e.\ different $m_1$ and $m_2$ combinations
with a fixed sum $m_1\!+\!m_2$ always give the same $\rh(m_1\!+\!m_2)$
[within errors].
Furthermore, the relationship is in good approximation linear, but there is
an anomaly without filtering at our strongest coupling (Fig.\,\ref{fig:ZA_08}).
Here, the slope is \emph{negative}, and this supports the view established in
App.\,A that with $\be\!=\!5.66$ and $\rh\!=\!1.0$ the projection point is
``in'' or ``to the left'' of the physical branch of the underlying Wilson
operator, and we effectively operate in the ``zero fermion'' sector.
Also at $\be\!=\!5.84$ the unfiltered plateau was not very pronounced either,
resulting in a large systematic uncertainty beyond the statistical error quoted
below. 
We use the ansatz
\beq
\rh(m_1\!+\!m_2)=
\mr{const}+{1\ovr Z_A}(m_1\!+\!m_2)+\mr{const}\,(m_1\!+\!m_2)^2
\label{rho_fit}
\eeq
and see whether we obtain acceptable fits and whether the first constant is
consistent with zero.
It turns out that this is the case, and the associate $Z_A$ values are
summarized in Tab.\,\ref{tab:Z_A}.

\begin{table}[!t]
\begin{center}
\begin{tabular}{|l|cccc|}
\hline
$\be$  &  $5.66$ &  $5.84$ &  $6.00$ &$6.00\;(\rh=1.4)$\\
\hline
$Z_A^\mr{none}  $&ill-def.&7.06(73)&3.145(94)&1.554(1)\cite{Berruto:2003rt}\\
$Z_A^\mr{1\,APE}$&2.57(7) &1.66(02)&1.452(04)&   ---   \\
$Z_A^\mr{3\,APE}$&1.55(3) &1.23(01)&1.160(06)&   ---   \\
$Z_A^\mr{1\,HYP}$&1.44(2) &1.22(01)&1.153(03)&   ---   \\
$Z_A^\mr{3\,HYP}$&1.21(1) &1.10(01)&1.072(02)&   ---   \\
\hline
$Z_A^\mr{none}|_\mr{1-loop}$ \cite{Alexandrou:2000kj}&1.280&1.272&1.264&1.120\\
\hline
\end{tabular}
\end{center}
\vspace{-6mm}
\caption{$Z_A$ determined via (\ref{rho_fit}) with various filtering
prescriptions. On the last line the 1-loop result \cite{Alexandrou:2000kj} for
the thin-link overlap
$Z_{V,A}=1+C_F\,0.198206g_0^2+O(g_0^4)=1+1.585648/\be+O(1/\be^2)$ at
$\rh\!=\!1.0$ and
$Z_{V,A}=1+C_F\,0.090301g_0^2+O(g_0^4)=1+0.722408/\be+O(1/\be^2)$ at
$\rh\!=\!1.4$ is added for comparison.}
\label{tab:Z_A}
\end{table}

\begin{figure}[!b]
\begin{center}
\epsfig{file=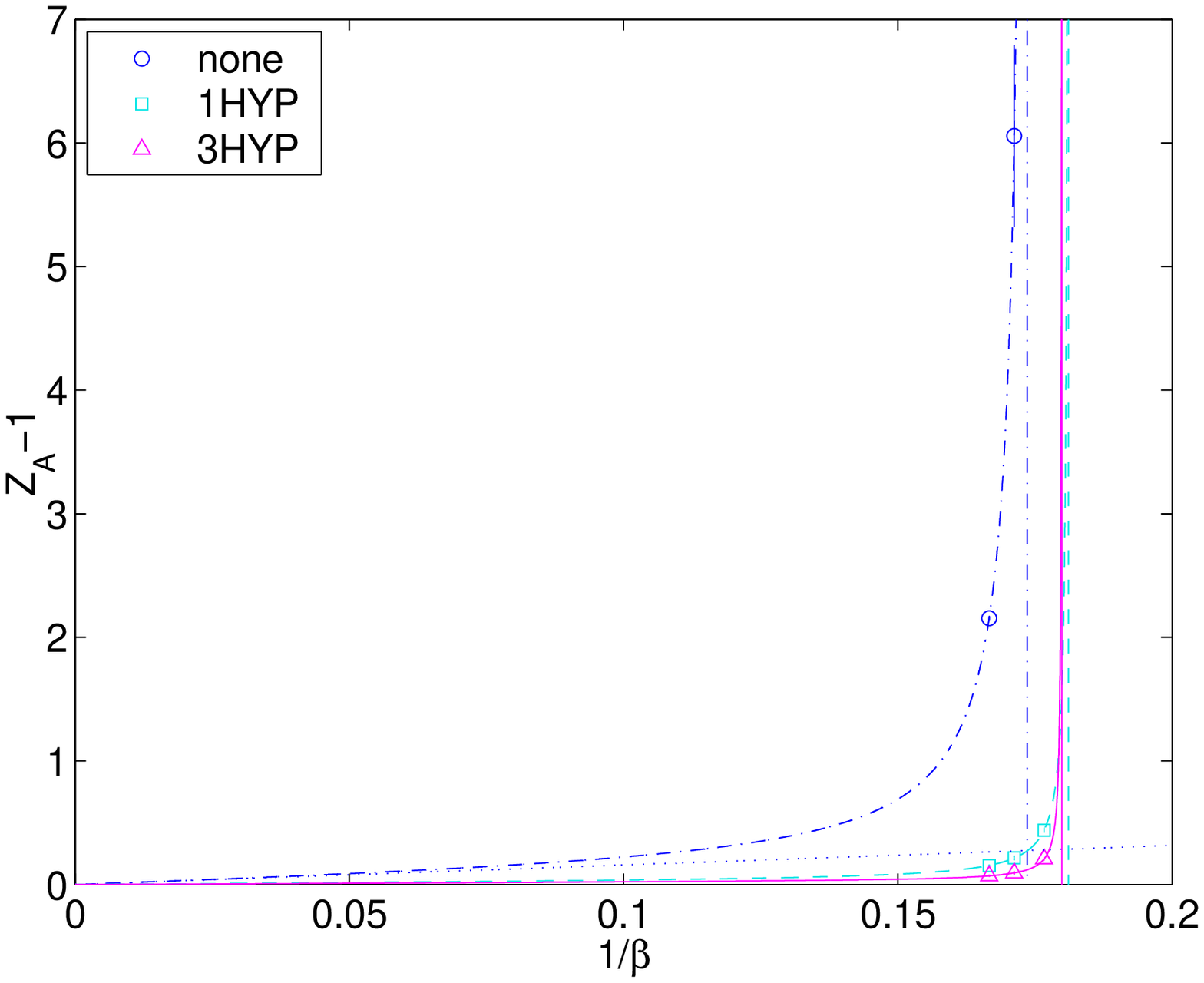,height=7.4cm}
\epsfig{file=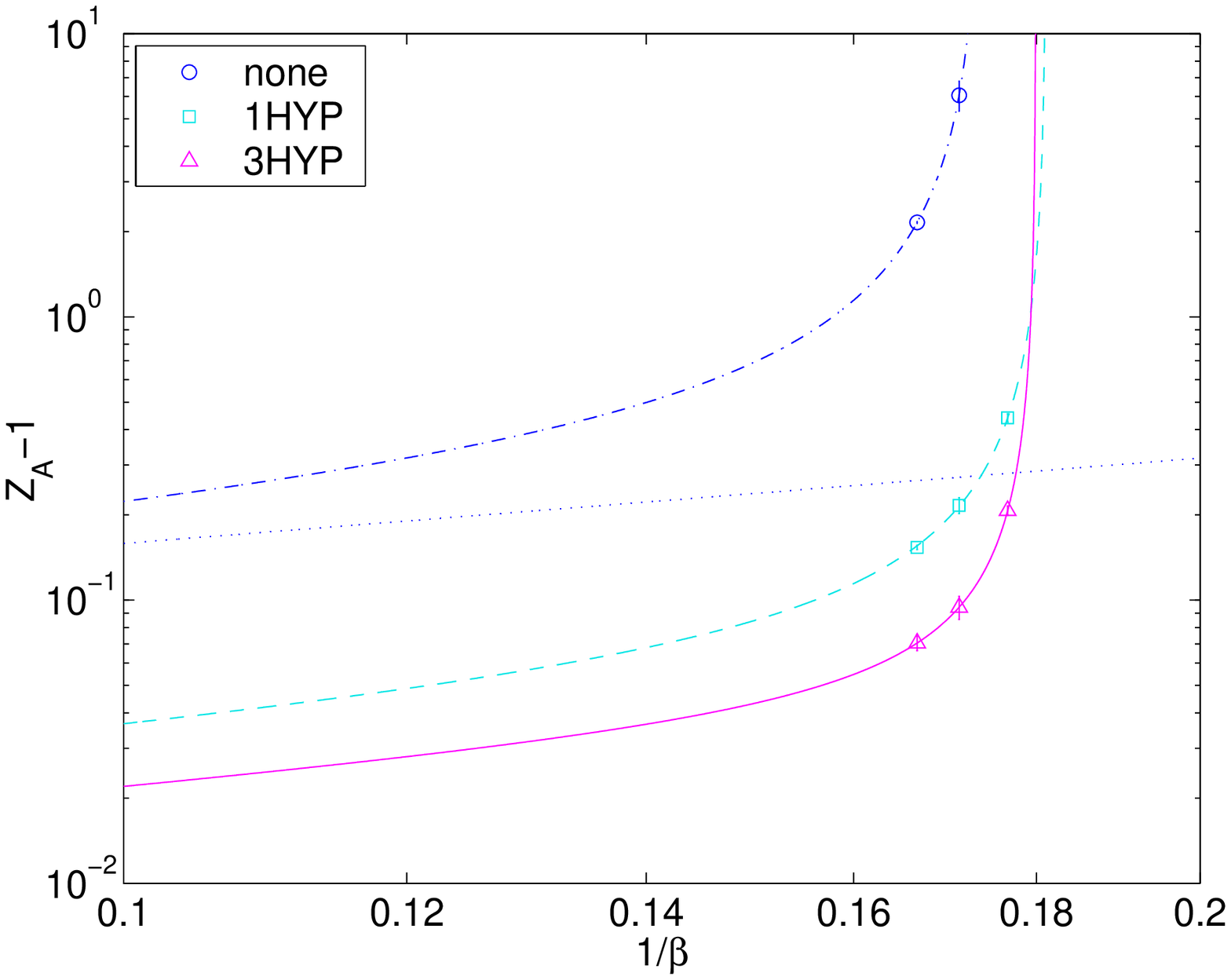,height=7.4cm}
\end{center}
\vspace{-8mm}
\caption{Pade functional forms through our $\rh\!=\!1$ data in standard (left)
or double logarithmic form (right). In the unfiltered case the constraint to
reproduce the known 1-loop behavior is built in and the latter is indicated
with a dotted line. It seems the 1-loop coefficients of the filtered $Z_A$ are
dramatically reduced, while the perturbative range (blow-up point of the Pade
ansatz) gets barely enlarged.}
\label{fig:pade}
\end{figure}

It is interesting to discuss both the general pattern of these $Z_A$ values and
the relation to 1-loop perturbation theory.
Evidently, at fixed $\be$ and $\rh$ the filtered $Z_A$ is much closer to the
tree-level value 1.
We recover the relative strength ordering of Sect.\,2, i.e.\ one APE step is
less efficient than 3\,APE or a single HYP step, but the latter is topped by
3\,HYP steps.
At $\be\!=\!6.0$ we compare to $\rh\!=\!1.4$ which is the the standard choice
for the ``thin link'' overlap.
Without filtering, $Z_{A}^\mr{none}(\rh\!=\!1.4)\!\simeq\!1.554$ is about half
of $Z_{A}^\mr{none}(\rh\!=\!1.0)\!\simeq\!3.145$, and this means that the
choice $\rh\!=\!1.4$ is not just near-optimal w.r.t.\ locality, but also
beneficial to tame (one particular) renormalization.
Once the filtering recipe is specified, $Z_A$ seems to be monotonic in
$6/\be\!=\!g_0^2$, as expected from perturbation theory.
In the unfiltered case the 1-loop value is included in the last line of
Tab.\,\ref{tab:Z_A} for comparison.
Assuming that in perturbation theory $1\!<\!Z_A^\mr{1\,HYP}\!<\!Z_A^\mr{none}$
holds for $\rh\!=\!1$, one may compare the deviation of the unfiltered
$\be\!=\!6.0$ operator $3.145\!-\!1.264\!=\!1.881$ to $1.153\!-\!1\!=\!0.153$
which then amounts to an upper bound in the 1\,HYP case.
Evidently, the discrepancy is dramatically reduced, which in view of the
perturbative results in \cite{Bernard:1999kc,DeGrand:2002va}, should not come
as a surprise.
To get a slightly more quantitative view, we consider it useful to fit our
data without filtering at $\rh\!=\!1$ to a Pade-type ansatz of the form
\beq
Z_A^\mr{none}={1+c_1x+c_2x^2\ovr1+(c_1-1.585648)x}
\label{ZA_none}
\eeq
with $x\!=\!1/\be$, where the perturbative knowledge \cite{Alexandrou:2000kj}
(cf.\ caption of Tab.\,\ref{tab:Z_A}) is built-in as a constraint.
In the same spirit a Pade ansatz for any of the filtered operators reads
\beq
Z_A^\mr{1HYP\,/\,3HYP}={1+c_1x+c_2x^2\ovr1+c_3x}
\label{ZA_filt}
\eeq
with --~as of now~-- no constraint on $c_1\!-\!c_3$ yet.
There is a problem with the functional forms (\ref{ZA_none}, \ref{ZA_filt}),
since our data sets contain 2 and 3 entries, respectively, and there is zero
degree of freedom.
Still, for an illustration such a ``fit'' might be worth while, and the result
is shown in Fig.\,\ref{fig:pade}.
With sufficient data the curves would contain two pieces of information.
The asymptotic slope for $x\!\to\!0$ would predict the perturbative 1-loop
coefficients for $Z_A^\mr{1\,HYP}$, $Z_A^\mr{3\,HYP}$.
And the pole in (\ref{ZA_none}, \ref{ZA_filt}), i.e.\ the values $c_1-1.585648$
or $c_3$, respectively, would predict the coupling where the perturbative
description breaks down.
Hence, if the curves in Fig.\,\ref{fig:pade} are indicative at all, it seems
that filtering renders the perturbative 1-loop coefficient of $Z_A$ much
smaller, but the perturbative range gets barely enhanced.


\section{Discussion}


In this paper we have studied the massless overlap operator constructed from
a filtered Wilson kernel where the original ``thin'' links were replaced
by ``thick'' links which behave in the same manner under local gauge
transformations.
This is a legal change of the fermion discretization as long as one particular
filtering recipe [e.g.\ 1\,HYP step with $\al_\mr{HYP}\!=\!(0.75,0.6,0.3)$]
is maintained at all couplings.
It amounts to an $O(a^2)$ re-definition of $D_\mr{ov}$ at fixed $\rh$,
as does a change of $\rh$ at fixed filtering level.

Our key observations are the following.
First, the onset of the ``bulk'' part of the spectrum of the underlying
shifted hermitean Wilson operator $\HW\!=\!\gaf(\DW\!-\!\rh)$ gets lifted.
This leads to an increased inverse condition number $\ep$ (after projection
typically by a factor 2-4 through a single HYP step) and the latter reflects
itself in a reduction (by the same factor) of the polynomial degree (and thus
the number of forward applications of $\HW^2$) needed to construct the inverse
square root over the relevant range.
What is the precise impact on CPU requirements to invert the massive operator
is a topic for future research.
Second, at standard couplings the filtered massless overlap is --~even with the
untuned canonical choice $\rh\!=\!1$~-- better localized than the unfiltered
version with an optimally tuned $\rh$ could ever be.
Our finding is backed by the observation that in the free case the optimum
$\rh$ (w.r.t.\ locality) is around $0.54$ and thus substantially smaller than
the typical $\rh\!\simeq\!1.4$ used in the past.
Our third observation is that the filtered kernel is much closer to being a
normal operator.
In other words the left- and right-eigenvectors of $\DW$ are better aligned
with higher filtering level, and in this respect the effect of the ``thick''
links is the same as a shift much closer towards the continuum under which the
overlap construction (\ref{over}) tends to be a simple radial projection of the
$\DW$ eigenvalues.
Finally, our fourth observation is that the renormalization constant of the
axial-vector current is much closer to 1 with filtering than without.
We rate this as a sign that lattice perturbation theory for the filtered
overlap might work much better than for the unfiltered variety.
If this is indeed so, and if it goes through for 4-fermion operators, it is
likely to be the most important consequence of our work, since it offers the
perspective of considerably reduced theoretical uncertainties in electroweak
precision studies.

\bigskip

Let us finally discuss a variety of proposals in the literature that are
similar in spirit to the one put forth in the present paper.

There is a top-level version deriving from ``parametrized fixed-point
fermions''.
The idea behind this approach pursued by Hasenfratz and Niedermayer is that
true fixed-point fermions would satisfy the GW relation exactly
\cite{Hasenfratz:1998ri}, but a practical implementation is always ultralocal.
Hence, sticking such an ansatz into the overlap formula (\ref{over}) yields
fermions with exact chiral symmetry and otherwise properties that are at least
as good (but typically better) than the version with a plain Wilson kernel
\cite{Hasenfratz}.

Bietenholz has considered a variety of actions, originally based on RG concepts
\cite{Bietenholz}.
The idea was that an action with a spectrum close to the GW circle could be
iteratively improved in its chiral properties.
Over time the focus has shifted towards using the overlap formula (\ref{over})
to have exact chiral symmetry, but it is clear that the kernel of his
``hypercubic overlap'' benefits from a larger inverse condition number $\ep$
of $|\gaf D_\mr{kern}|$ just as we do.

Gattringer and collaborators construct a ``chirally improved'' Dirac operator
that involves the full Dirac Clifford algebra with links restricted to the
hypercube.
The coefficients are adjusted such that (for a given coupling) the violation of
the GW relation is minimized \cite{Gattringer}.
The problem is the same as in the Bietenholz approach: a single forward
application with such a kernel is so expensive that the improvement, if it is
not ``perfect'', does not really pay off.

DeGrand has considered --~both perturbatively and non-perturbatively~-- Wilson
and clover action varieties that involve smeared gauge links
\cite{DeGrand:1999gp,Bernard:1999kc}.
Based on this experience he went on to construct a ``variant overlap'' which
starts from a kernel with only scalar/vector terms and smoothed links, and
is thus sufficiently cheap as to allow for sticking it into the overlap
formula \cite{DeGrand,DeGrand:2002va}.

The closest to what we do is found in the work of Kovacs \cite{Kovacs}.
He uses a ``fat-link clover'' overlap in which all links are smeared, together
with the tree-level value $c_\mr{SW}\!=\!1$.
As far as we know, he was the first author to notice that such a filtered
kernel allows for the untuned choice $\rh\!=\!1$, and still the resulting
overlap shows good localization properties.

A related approach has been pursued by the Adelaide group \cite{Kamleh:2001ff}.
Their ``fat link irrelevant clover'' overlap quarks are built from a clover
action in which only the irrelevant pieces (i.e.\ the Wilson and the
Sheikoleslami-Wohlert terms) use smeared links, but not the covariant
derivative.
They found a similar speedup factor in the construction of the overlap operator
(cf.\ ``step 1'' in the introduction) and tied it to the reduced spectral
density of $|\gaf D_\mr{kern}|$ near the origin.

Finally, ``overlap'' quarks with smeared gauge links have been used by several
lattice collaborations.
RBC has found that the residual mass of domain-wall fermions at fixed $N_5$
gets reduced \cite{Lin:2004xf}, though they miss out an important ingredient,
the projection to $SU(3)$.
UKQCD has used overlap valence quarks with 3-fold HYP smeared links on
staggered sea as supplied by the MILC collaboration, finding a surprisingly
good signal on as few as 10 configurations \cite{Bowler:2004hs}.
Similarly, LHP and NPLQCD have used filtered domain-wall valence quarks on
staggered sea to compute the pion form factor \cite{Bonnet:2004fr} and the
$I\!=\!2$ $\pi\pi$-scattering length \cite{Beane:2005rj}, respectively.

There is another idea that should not be confused with filtering.
Using an improved gauge action has been found to reduce $\rh_{|\HW|}(0)$ by up
to an order of magnitude \cite{AliKhan:2000kd,Aoki:2002vt,Jansen:2003jq}.
There is, however, an important practical difference to the filtering concept,
which is a modification of the fermion action.
As already discussed in \cite{Durr:2004as}, a better choice of the gauge action
improves, in the first place, the very low end of the $|\HW|$ eigenvalue
distribution.
After projecting out the lowest $O(15)$ eigenvectors (which nowadays is a
standard thing to do \cite{Hernandez:2000sb}) much of the advantage is lost (in
Fig.\,3 of \cite{Jansen:2003jq} the lifting factor diminishes to the right).
By contrast, filtering lifts the complete low-energy end of the $|\HW|$
eigenvalues (in our Fig.\,\ref{fig:low_average} one finds an almost-universal
lifting factor) and the usefulness of filtering is not vitiated by the
projection.
Still, it might be interesting to see whether the two ideas can be fruitfully
combined.

\bigskip

An extension of the filtering concept to full QCD is straightforward, albeit
hampered by a technical problem.
These days, most dynamical fermion simulations are set up with a HMC algorithm,
and the latter requires the fermion action to be differentiable w.r.t.\ the
gauge links.
The kernel of our filtered overlap quarks is differentiable w.r.t.\ the
``thick'' links, but not w.r.t\ the elements of the original set, due to the
projection involved in the APE \cite{ape} or HYP \cite{hyp} procedure.
A convenient way out is offered by the stout/EXP links introduced in
\cite{Morningstar:2003gk}, involving a differentiable mapping between the
``thick'' and ``thin'' links.
In pure gauge observables the usefulness of this smearing recipe was found to
be restricted to small parameter values \cite{Morningstar:2003gk,Durr:2004xu},
and one may fear that this feature persists in stout/EXP overlap quarks, since
in perturbation theory they are equivalent to APE filtered overlap fermions
with $\al_\mr{APE}=1/(1-6\al_\mr{EXP})$ \cite{DeGrand_private}.
Thus, due to the pole at $\al_\mr{EXP}\!=\!1/6$ we expect them to have a
``narrow therapeutic range'' in parameter space, but it is clear that there
is no conceptual issue in simulating full QCD with filtered overlap quarks
beyond the difficulties met in the unfiltered case \cite{full_qcd_overlap}.

To summarize, our suggestion is to use the overlap recipe (\ref{over}) with an
unimproved ($c_\mr{SW}\!=\!0$) Wilson kernel in which all links are replaced by
some smeared descendents of the actual gauge background.
We recommend to stay with a moderate amount of link ``fattening'', e.g.\ with
a single step of standard HYP smearing \cite{hyp}.
The projection parameter $\rh$ may be fixed at its canonical value 1, and in
this sense the filtered overlap involves \emph{less tuning} than the unfiltered
version%
\footnote{Of course, there are parameters in the filtering recipe, but our
results show that they hardly matter. Thus filtering allows one to trade a
parameter that needs to be tuned for parameters on which the lattice data
show very little sensitivity.}.
An important restriction is that the choice of iteration level and smearing
parameter must be the same for all couplings considered in a scaling study.
This is one point on which our proposal differs from some of the attempts
reviewed above which involve coefficients (e.g.\ in the extended $\ga$-algebra)
that are adjusted ``by hand'' to yield a GW-type spectrum at one standard
value of the gauge coupling.
The other difference is that our kernel remains cheap and still requires fewer
$D_\mr{kern}\dag D_\mr{kern}^{}$ forward applications.
From a practical viewpoint, a clear advantage is the ease of implementation of
the ``filtered overlap'' --- everyone with a running overlap code has it (in
disguise).

\subsection*{Acknowledgments}

It is a pleasure to thank Ferenc Niedermayer and Tom DeGrand for useful
conversation or correspondence.
This paper was supported by the Swiss NSF.


\section*{App.\,A: Cumulative eigenvalue distributions in 4D and 2D}


In this appendix we discuss what can be learned from the cumulative eigenvalue
distribution (CED).
We consider both the eigenvalues in 4D generated for the main part of this
paper and data from dedicated runs in the quenched Schwinger model (QED with
massless fermions in 2D) to elucidate the effects that filtering and changing
$\be$ have on the spectral density of the hermitean Wilson operator
$\HW\!=\!\gaf D_{\mr{W},-1}$.

\bigskip

\begin{figure}[!p]
\vspace{-3mm}
\begin{center}
\epsfig{file=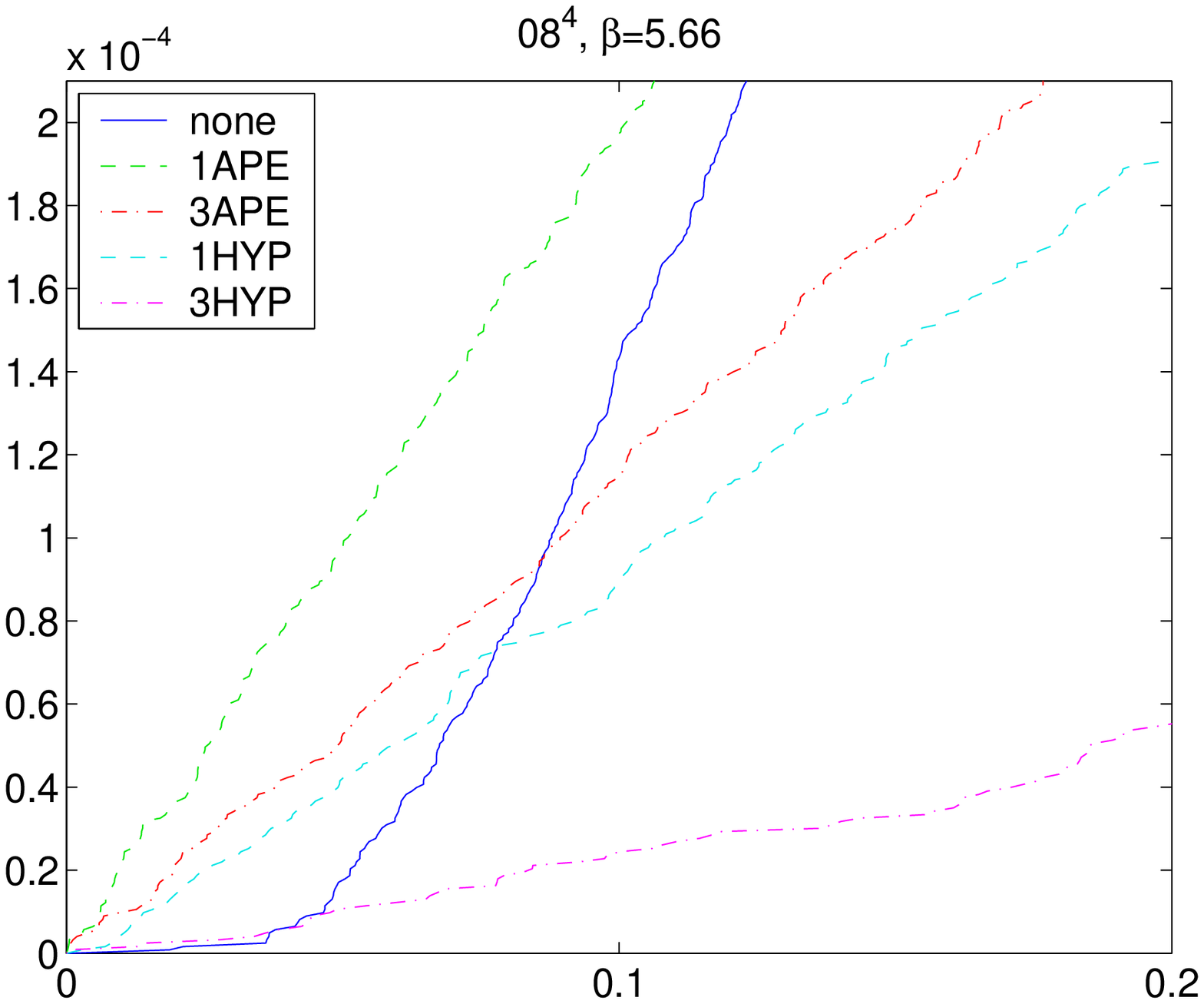,height=7.4cm}
\epsfig{file=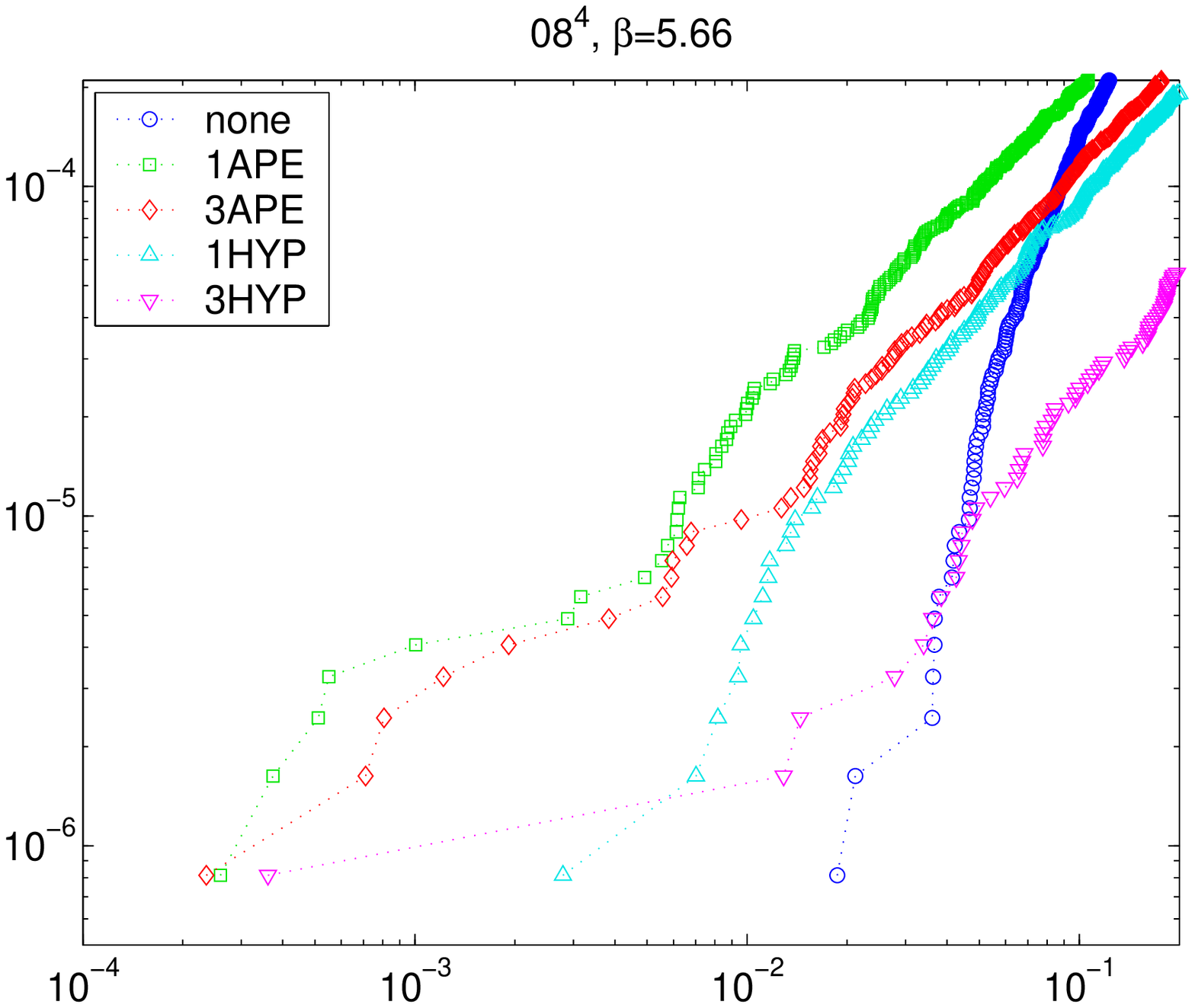,height=7.4cm}\\
\epsfig{file=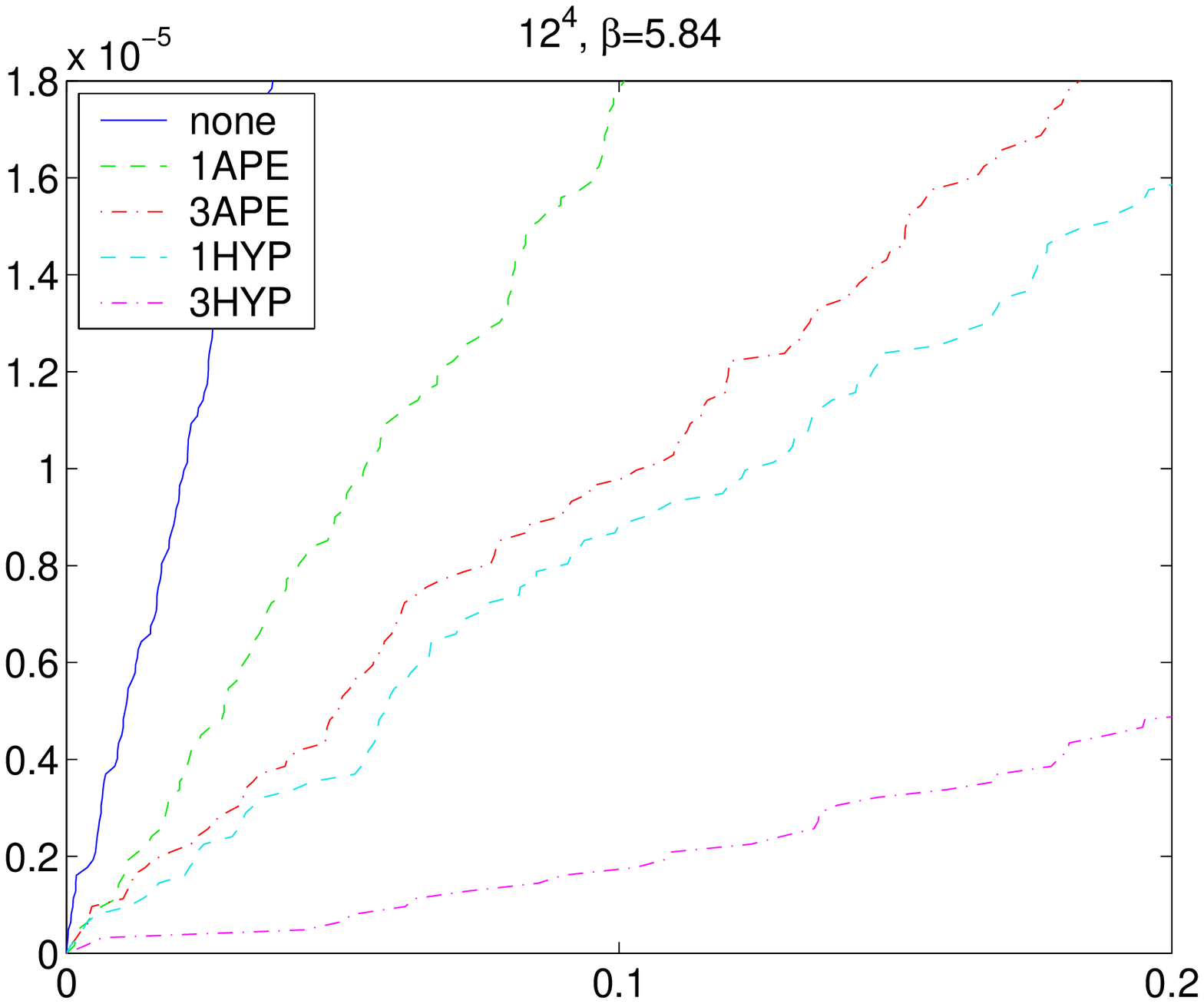,height=7.4cm}
\epsfig{file=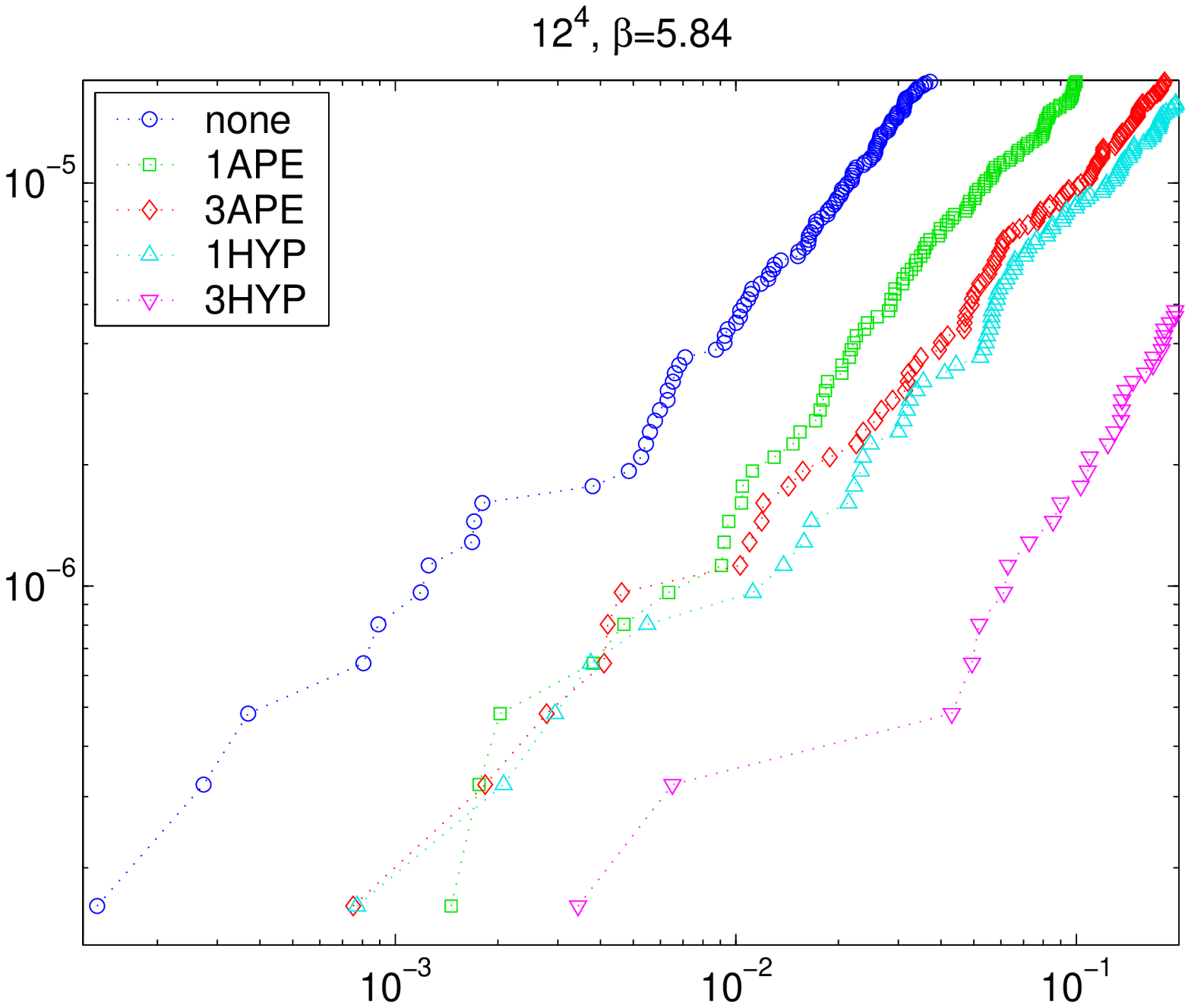,height=7.4cm}\\
\epsfig{file=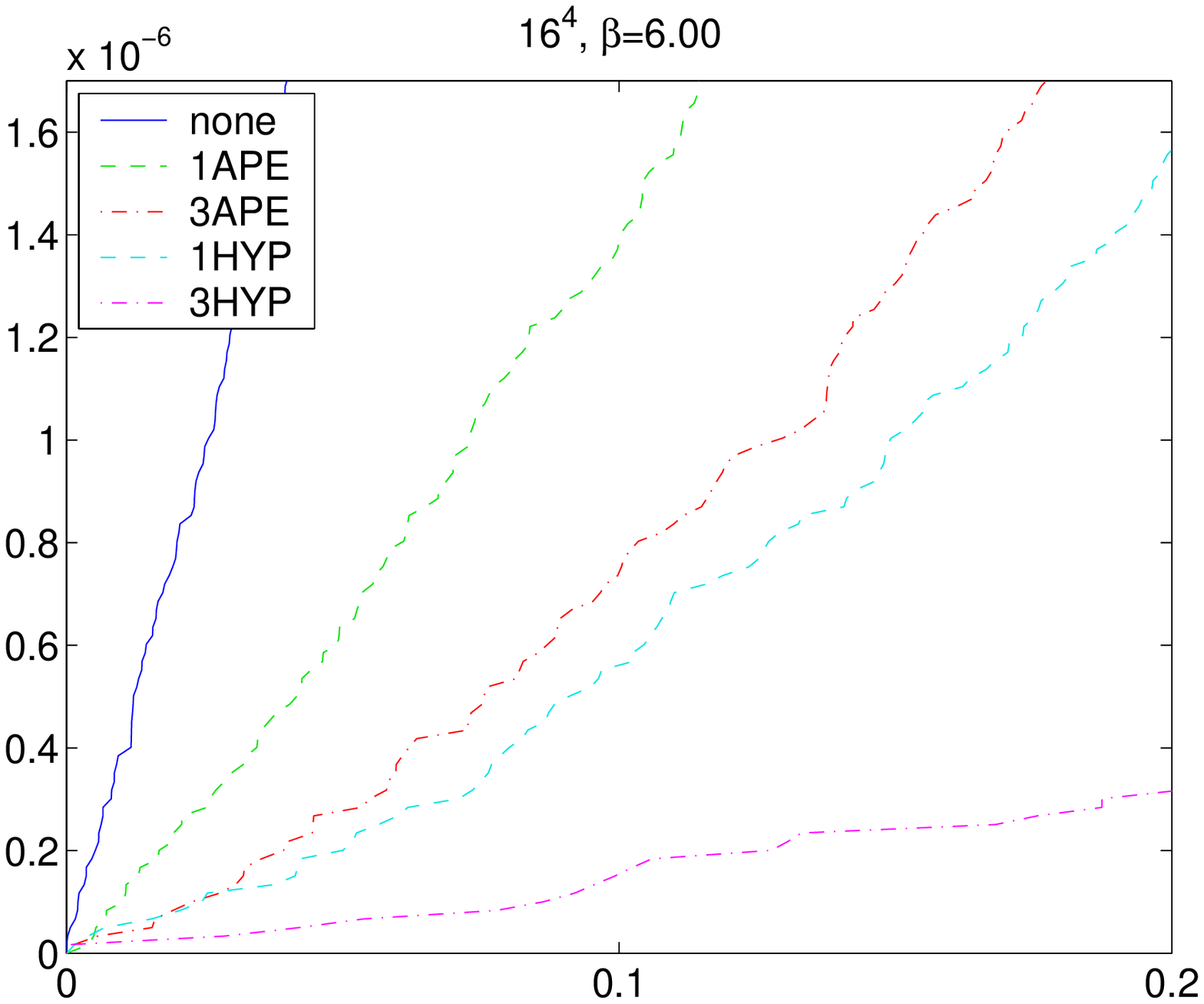,height=7.4cm}
\epsfig{file=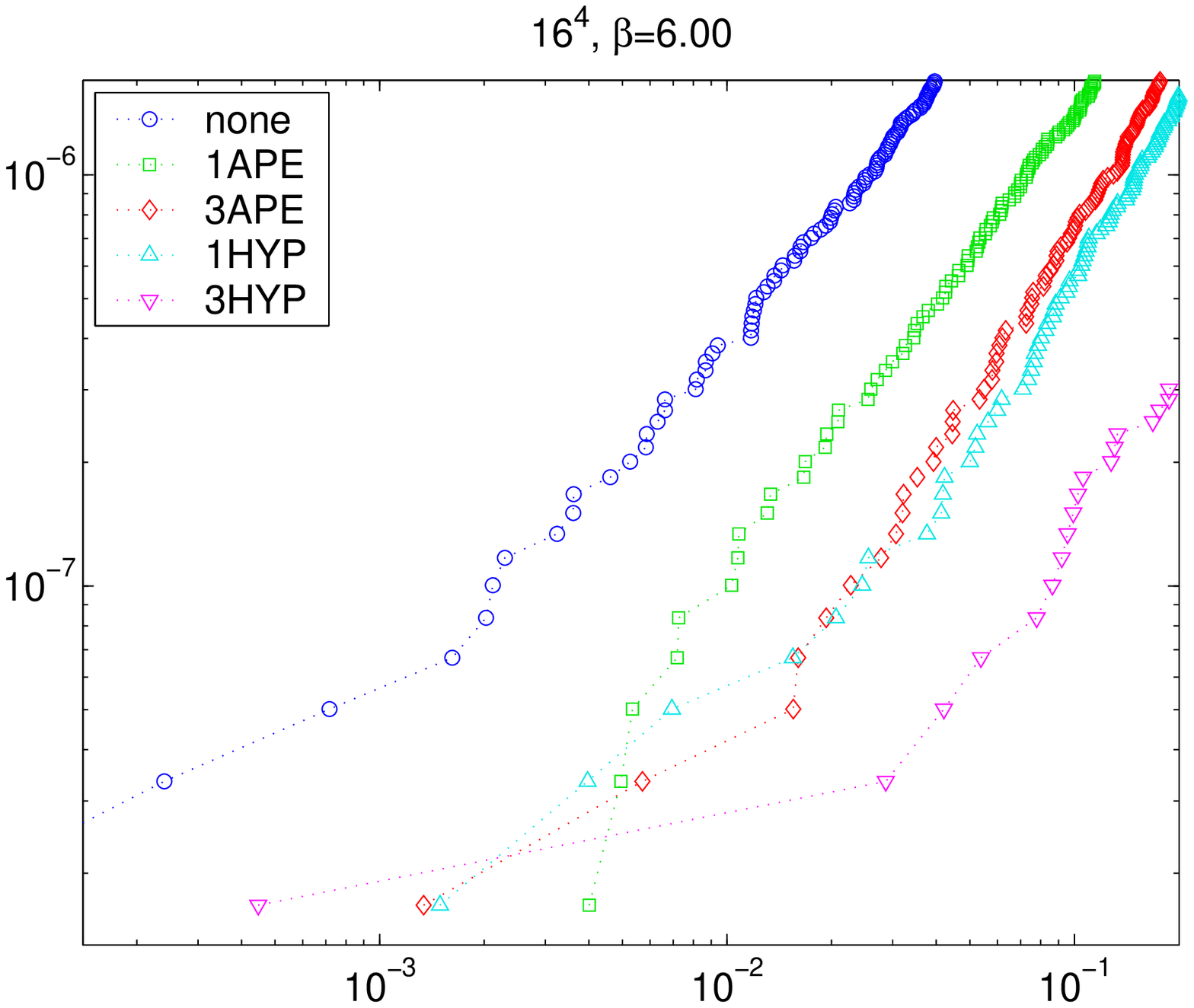,height=7.4cm}\\
\end{center}
\vspace{-6mm}
\caption{Cumulative eigenvalue distribution (CED) of $|\HW|$ in standard (left)
or double logarithmic (right) form for $\be\!=\!5.66,5.84,6.00$ (from top to
bottom) with 0,1,3 steps of APE or HYP filtering. At each $\be$ the upper cut
in the vertical direction is the same on the left and on the right and the
upper end in the horizontal direction is 0.2 throughout. Note the change in the
ordinate scale between different couplings. At $\be\!=\!6.26$ we definitely
lack the statistics needed to see a linearly dominated regime.}
\label{fig:ced_4D}
\end{figure}

\begin{figure}[!t]
\begin{center}
\epsfig{file=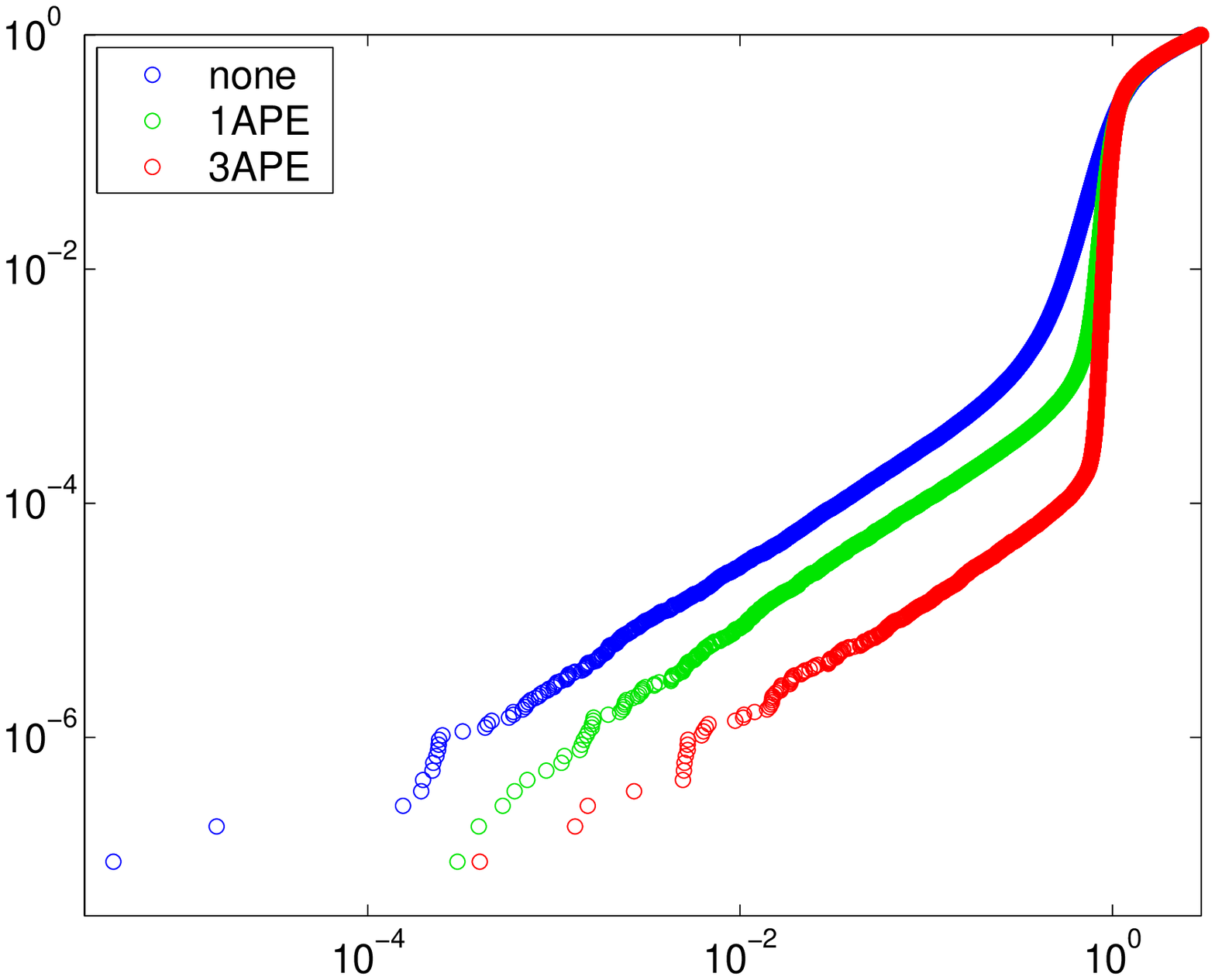,width=12cm}
\end{center}
\vspace{-6mm}
\caption{Log-log plot of the cumulative eigenvalue distribution (CED) of
$|\HW|$ in the Schwinger model ($\Nf\!=\!0$, $16^2$ geometry, $\be\!=\!3.2$,
$\rh\!=\!1$) without filtering (top curve) and after 1 or 3 filtering steps.
In all cases the CED starts out linearly, raises sharply somewhere near
$\la\!\simeq\!1$ and reaches 1 at $\la\!\simeq\!3$ (in 2D). This figure
includes all eigenvalues of all three operators on 22,500 decorrelated
configurations.}
\label{fig:ced_2D}
\end{figure}

Fig.\,\ref{fig:ced_4D} presents the cumulative eigenvalue distribution (CED) of
the 15 smallest eigenvalues of $|\HW|$ on the ensembles discussed before.
We show it both in standard form and in double logarithmic form, and the scale
on the ordinate follows from the requirement that it would extend up to 1, if
all eigenvalues were calculated (cf.\ Fig.\,\ref{fig:ced_2D} below).
For the two intermediate couplings ($\be\!=\!5.84,6.00$) we see the expected
linear rise of the CED near the origin, which soon gets complemented by a
higher order piece.
The coefficient of the linear part is a measure for the spectral density of
the hermitean Wilson operator at the origin, $\rh_{|\HW|}^{}(0)$.
That density being non-zero means that there is a finite probability to
encounter arbitrarily small eigenvalues.
The main effect of smearing is to reduce this spectral density, as is evident
from the double logarithmic plots -- here the initial slope 1 piece gets
shifted downwards, and this corresponds to a smaller coefficient in front
of the linear piece in the standard representation.
Our data at $\be\!=\!6.26$ are of lesser quality -- here we definitely cannot
identify a linearly dominated regime.
The situation is far more favorable in Fig.\,\ref{fig:ced_2D} where quenched
Schwinger model data are shown.
Apart from the statistics, the main difference is that all eigenvalues
(extending up to $\,\sim\!3$ in 2D) are included.
The higher the filtering level or $\be$, the more pronounced is the ``jump'' in
the CED at $\la\!\simeq\!1$.
Note that with a chiral kernel \emph{all} eigenvalues of $|\gaf D_\mr{kern}|$
would be there, i.e.\ the CED would be a step function at $\la\!=\!1$.
Finally, to come back to Fig.\,\ref{fig:ced_4D}, the situation at the strongest
coupling ($\be\!=\!5.66$) is different, since here the linear piece in the
unfiltered CED is \emph{not} larger than in the filtered versions.
This is, because our choice $\rh\!=\!1$ lets us ``loose'' the fermion -- at
this coupling our projection point is somewhere ``in'' the physical branch or
``to the left'' of it, while for the filtered version $\rh\!=\!1$ is still
appropriate.
One might avoid such a situation by choosing a larger $\rh$ with the unfiltered
kernel, but an even safer option might be to refrain from simulating unfiltered
overlap quarks on such coarse lattices.
It looks like this is a situation where the filtered overlap may help a lot,
since it allows simulations on coarser lattices than the unfiltered operator,
but in order to really be useful such simulations should be in the scaling
regime (and not just in the right universality class), and this is, of course,
not yet clear.

\bigskip

The spectral properties of $\HW$ play a role in the context of the physical
interpretation of the Aoki phase \cite{Aoki:1983qi}.
The latter is a conjectured phase, originally specific to $\Nf\!=\!2$ active
Wilson fermions at negative mass, in which after switching off an external
trigger term
\beq
S_\mr{source}=\pm h\psb\gaf\si_3\ps
\eeq
parity and flavor break spontaneously and a condensate ($\mr{const}\neq\!0$)
\beq
\lim_{h\to0^\pm}\<\psb\gaf\si_3\ps\>=\pm\mr{const}
\label{aoki_cond}
\eeq
forms.
Good numerical evidence for a non-zero condensate (\ref{aoki_cond}) in the
(dynamical) 2-flavor case for an appropriate choice of the negative mass
$-\rh(\be)$ is found in \cite{Ilgenfritz:2003gw}.
Ref.\,\cite{Bitar:1997ic} argues that in the massless limit of the continuum
theory a condensate of the form (\ref{aoki_cond}) is simply an axial rotation
of the usual (flavor diagonal) condensate and thus breaks neither parity nor
flavor.
They relate the spectral density of $\DW$ to that of $\HW$ and argue that the
absence of a gap (around the origin) of the latter is indicative of chiral
symmetry breaking and that $\rh_{|\HW|}(0)\!>\!0$ if and only if
(\ref{aoki_cond}) is non-zero.
This was later elucidated to be a continuum argument \cite{Sharpe:1998xm},
which --~in view of our Sec.\,4~-- might be an important point.

The next issue is whether there is an Aoki phase in the quenched theory
with 2 valence (but 0 sea) flavors \cite{Golterman:2003qe,Golterman:2005ie}.
The simplest expectation is that qualitatively the picture with the 5 Aoki
``fingers'' goes through, though the phase boundary is somewhat shifted
w.r.t.\ the $\Nf\!=\!2$ case.

Our 4D data in Fig.\,\ref{fig:ced_4D} clearly show the suppression of 
$\rh_{|\HW|}(0)$ as one approaches the continuum, but we cannot see any sign
that this distribution would vanish at some ``critical'' coupling.
Given the uniform pattern in the figures (apart from the scale on the $y$-axis
they seem qualitatively similar), it seems more likely to us that
$\rh_{|\HW|}(0)$ will stay non-zero for arbitrary couplings.

To test this view, we analyze the quenched Schwinger model where high
statistics can be reached.
The couplings and geometries are chosen such as to have a fixed physical
volume, with a box size about 5 times larger than the Compton wavelength of
the lightest degree of freedom in the chiral limit of the $\Nf\!=\!1$ theory.
A survey of the parameters is given in Tab.\,\ref{tab:sm_par} and for
technical details we refer to \cite{DuHo_schwinger}.

\begin{figure}[!p]
\epsfig{file=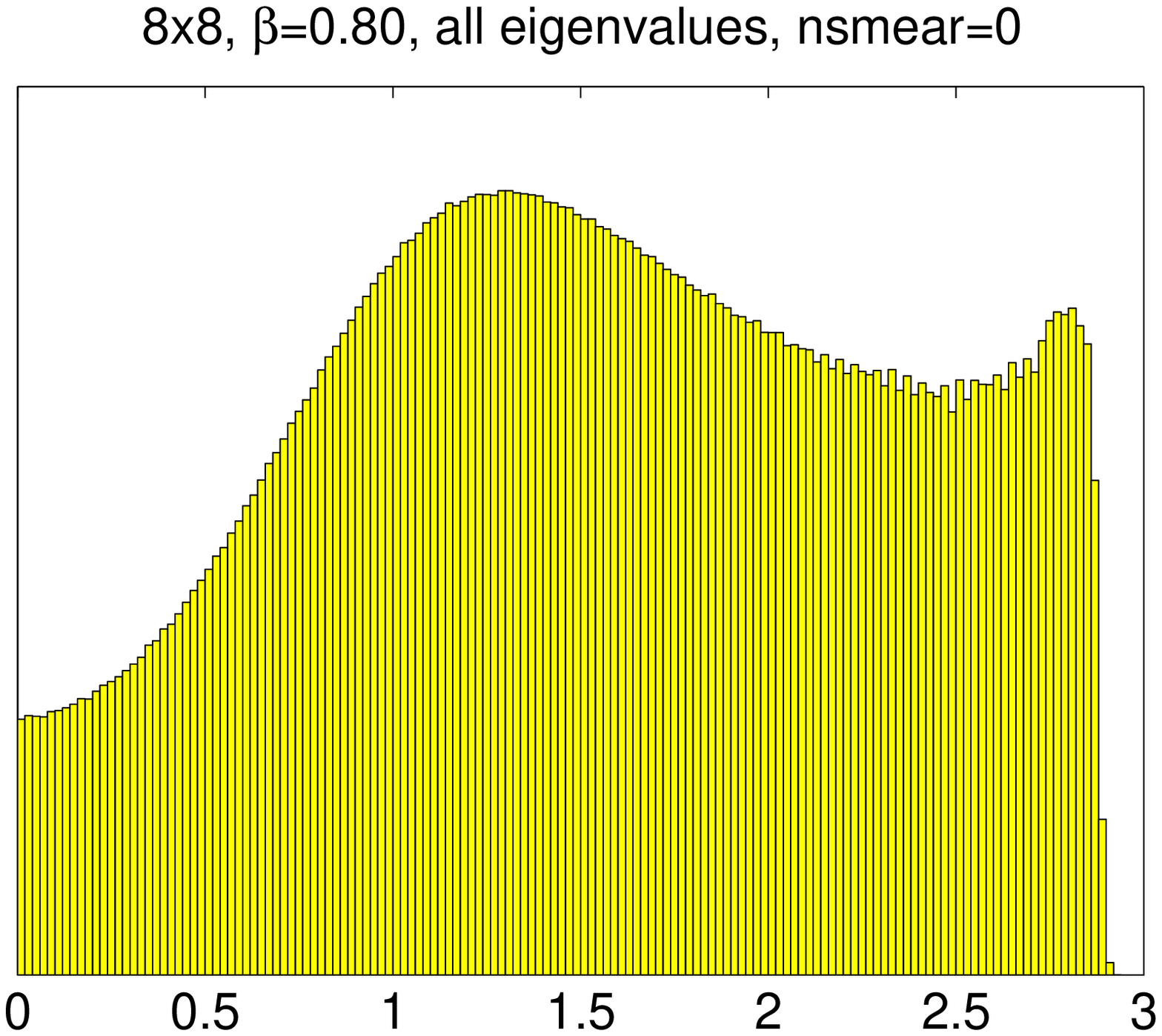,height=48mm,width=60mm}
\epsfig{file=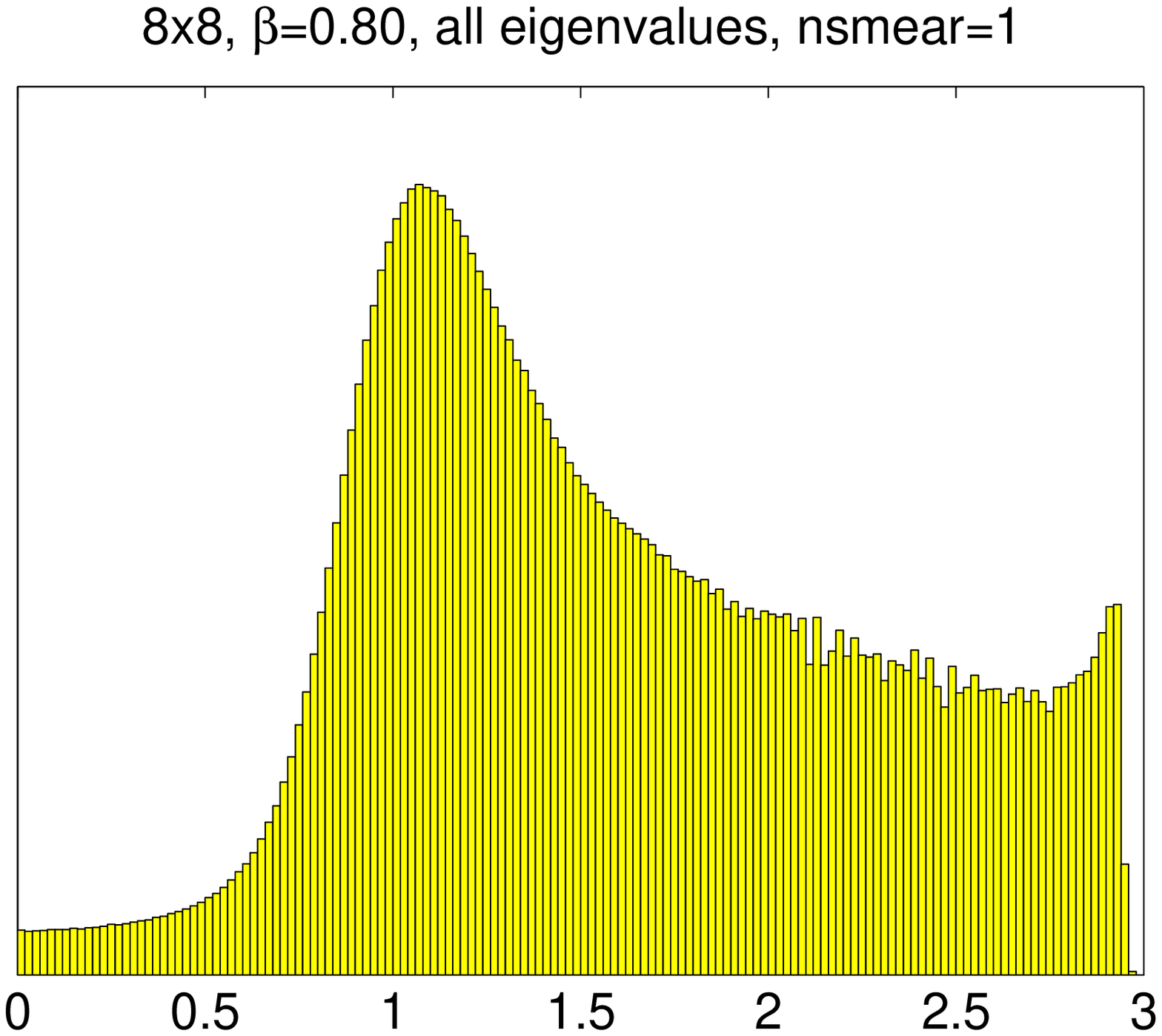,height=48mm,width=60mm}
\epsfig{file=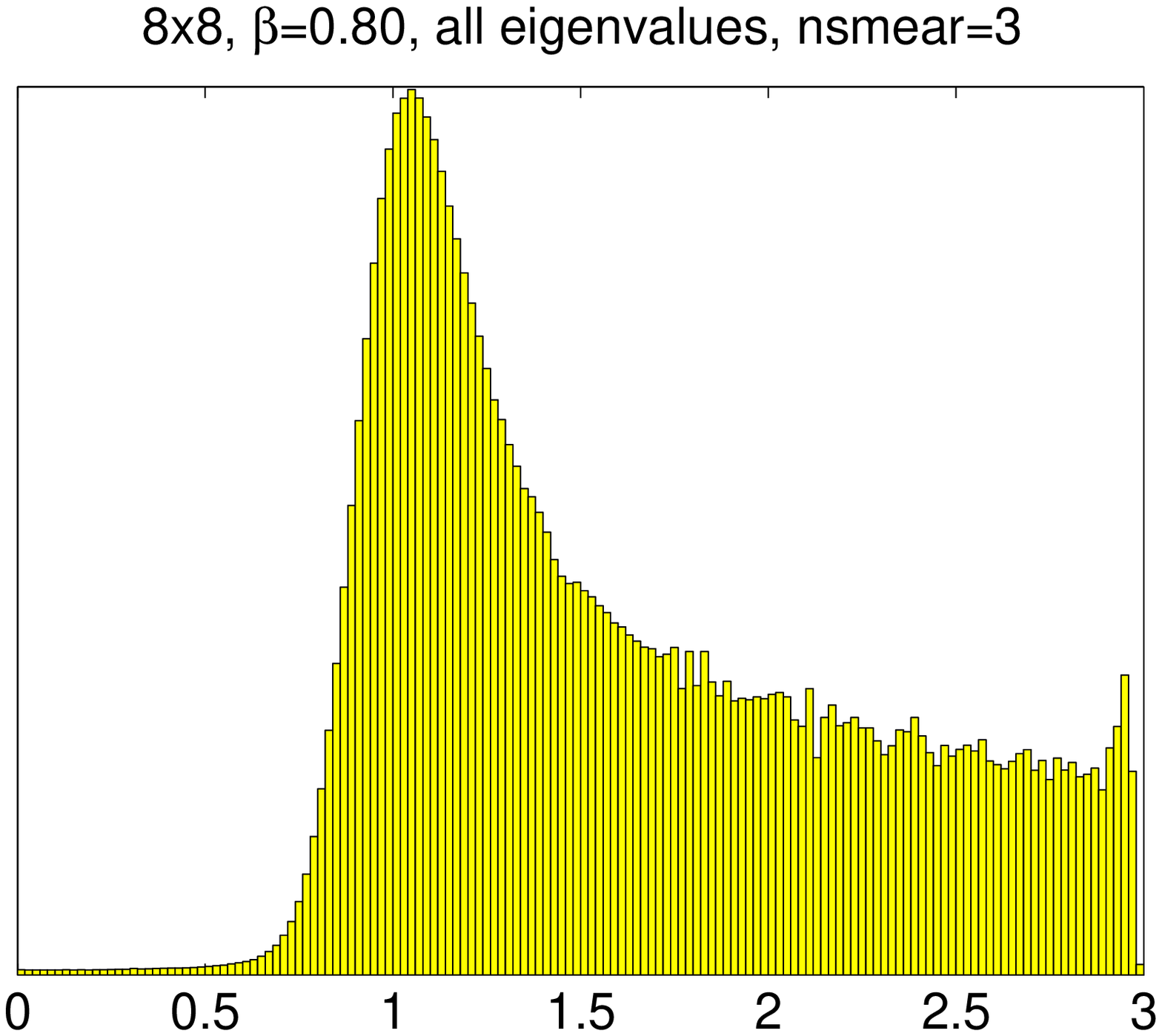,height=48mm,width=60mm}
\\
\epsfig{file=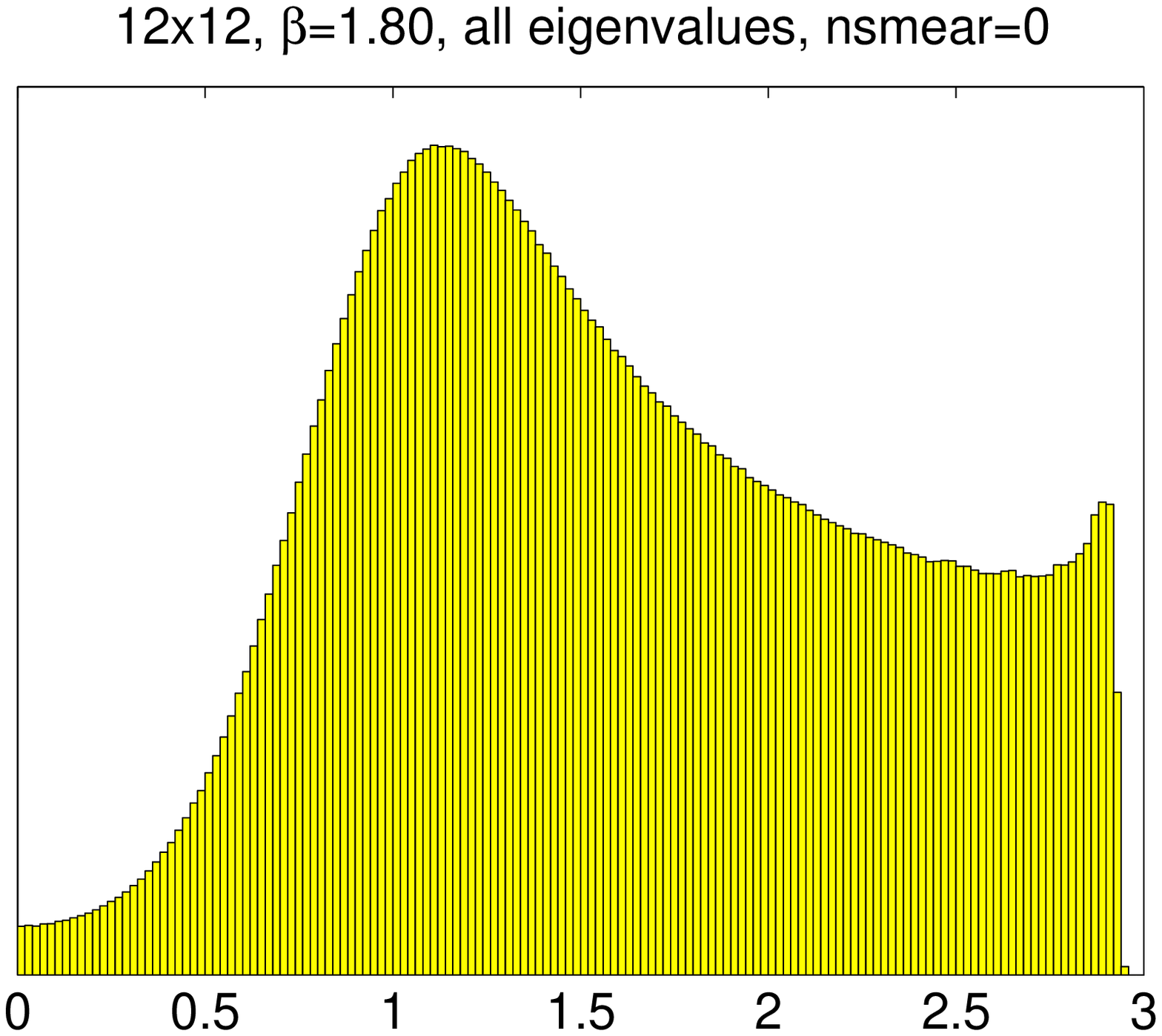,height=48mm,width=60mm}
\epsfig{file=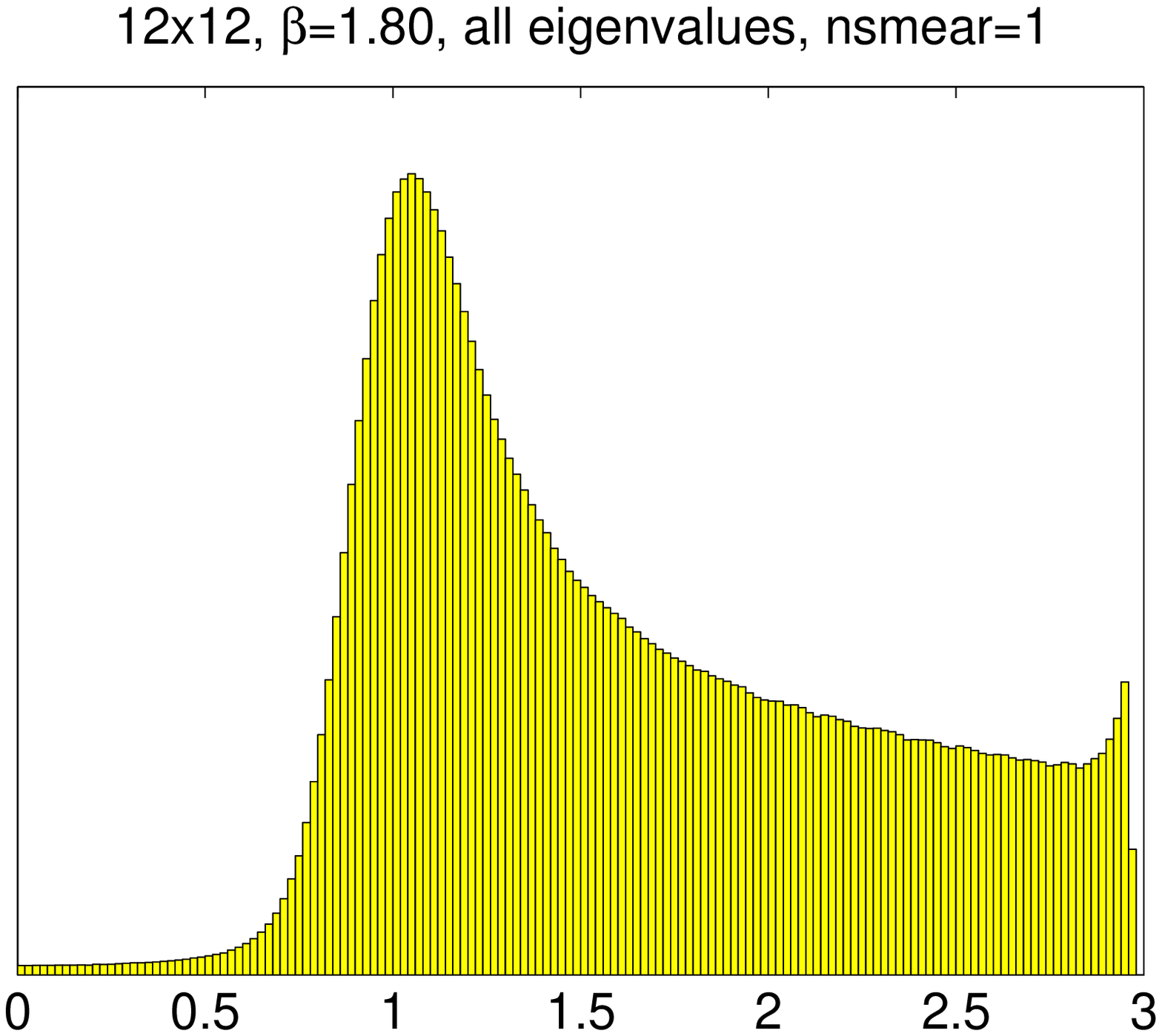,height=48mm,width=60mm}
\epsfig{file=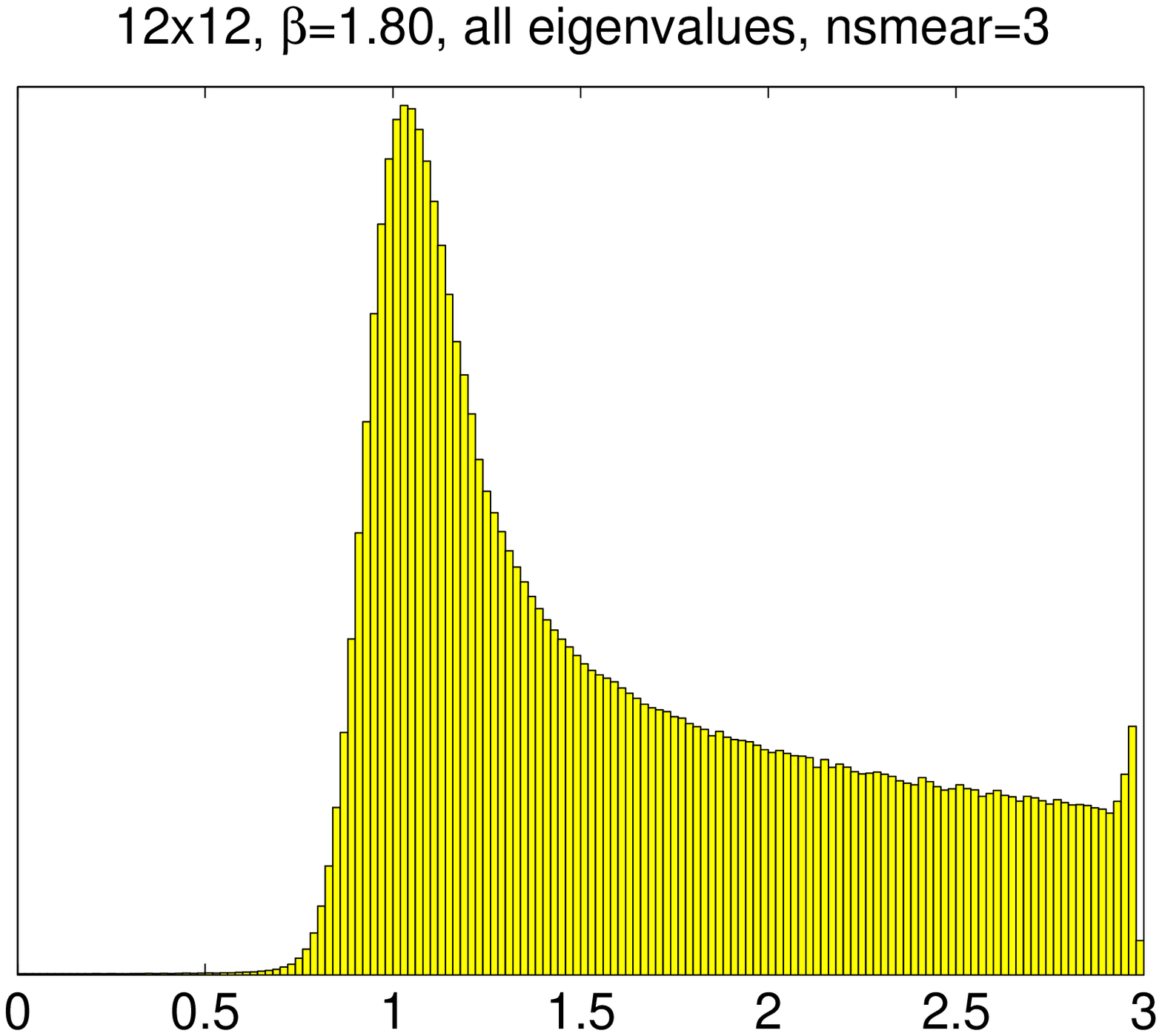,height=48mm,width=60mm}
\\
\epsfig{file=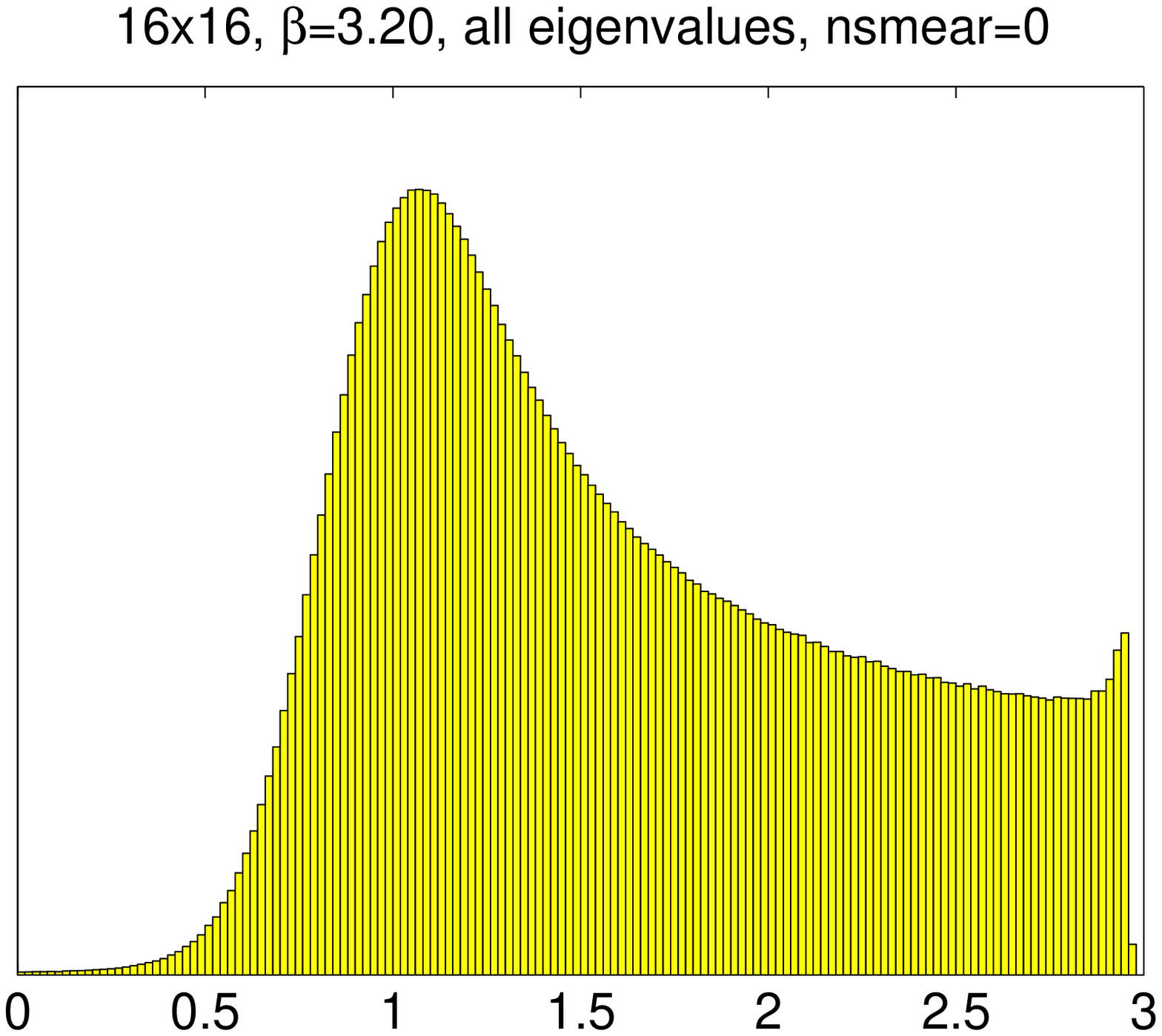,height=48mm,width=60mm}
\epsfig{file=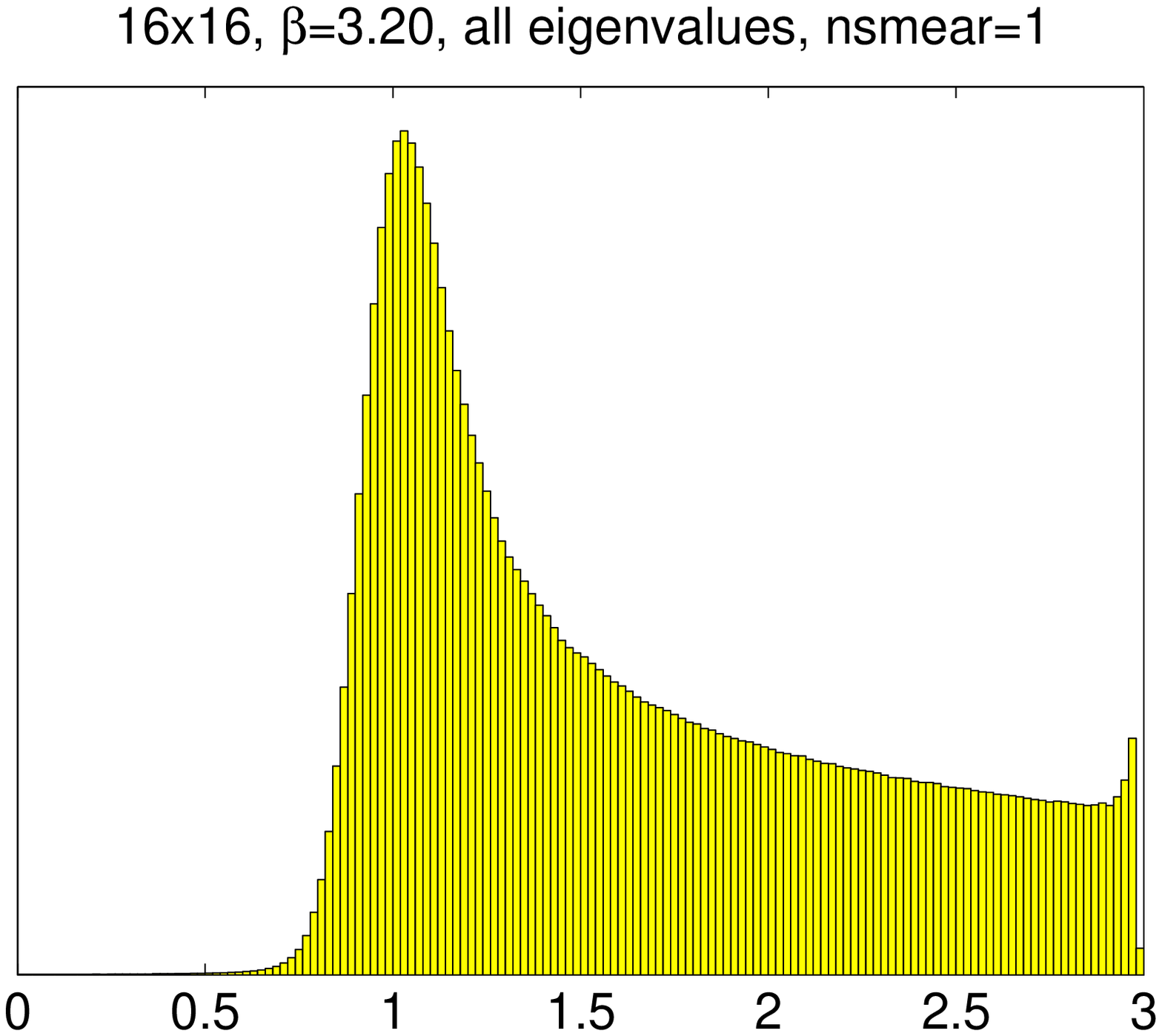,height=48mm,width=60mm}
\epsfig{file=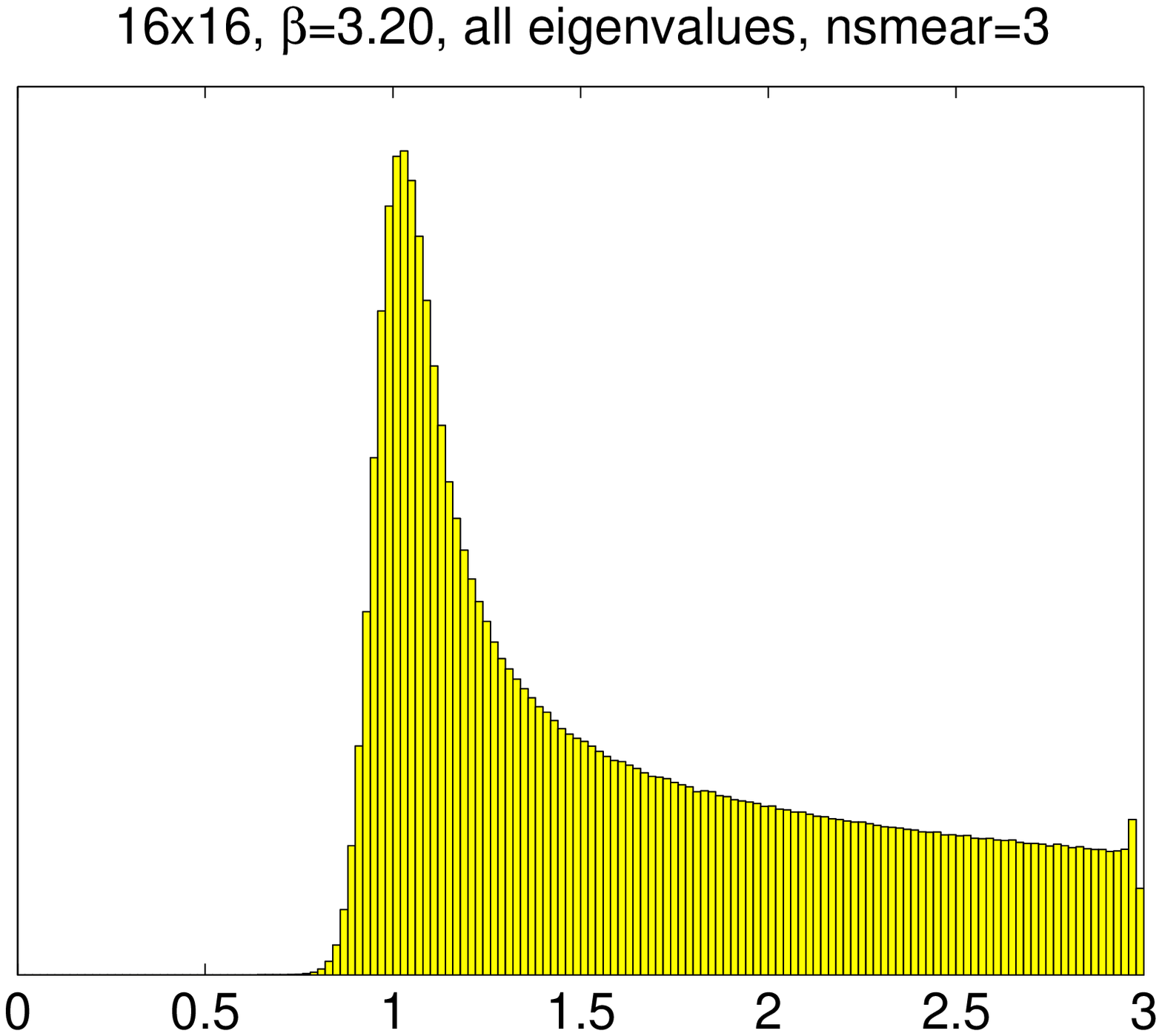,height=48mm,width=60mm}
\\
\epsfig{file=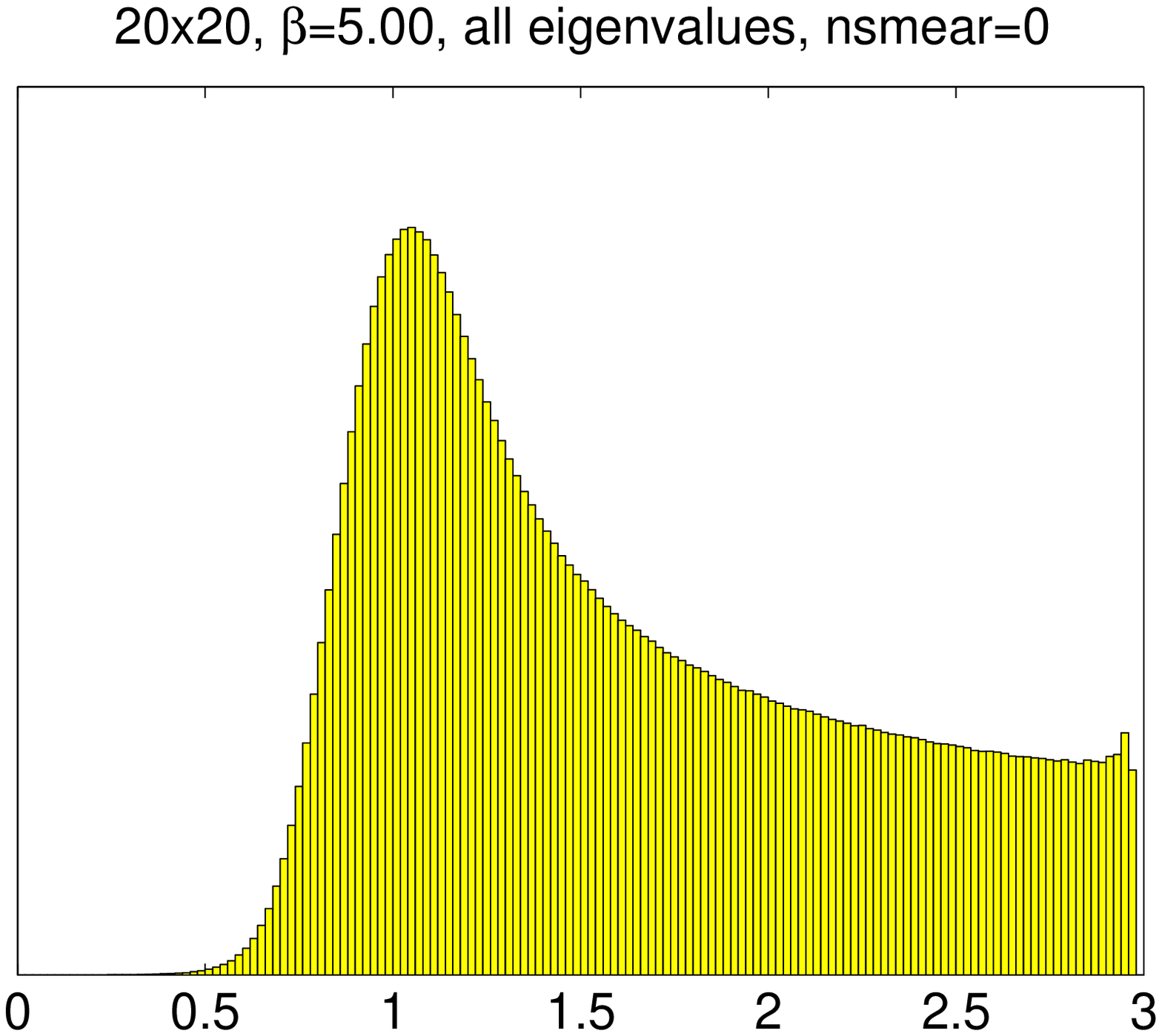,height=48mm,width=60mm}
\epsfig{file=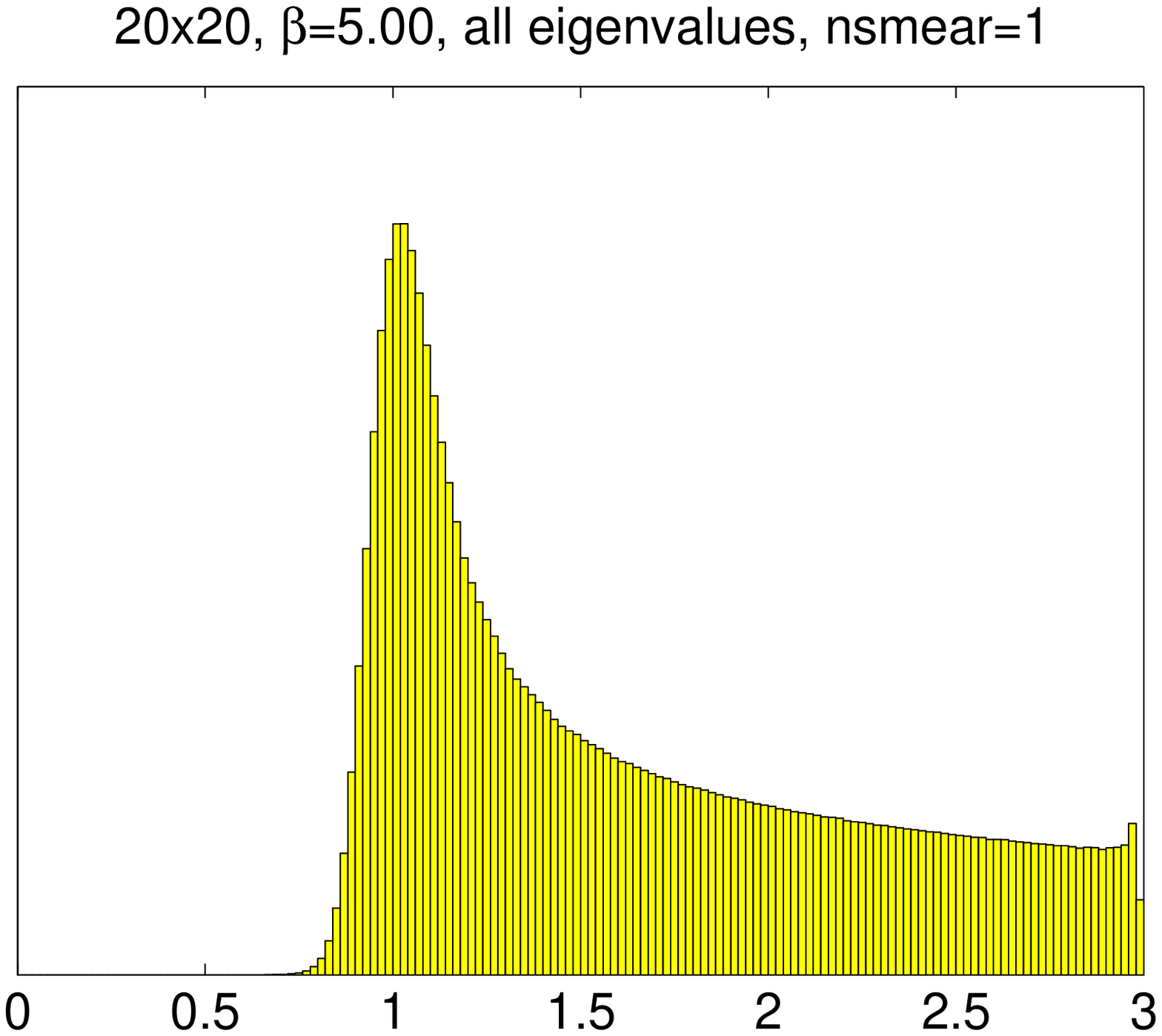,height=48mm,width=60mm}
\epsfig{file=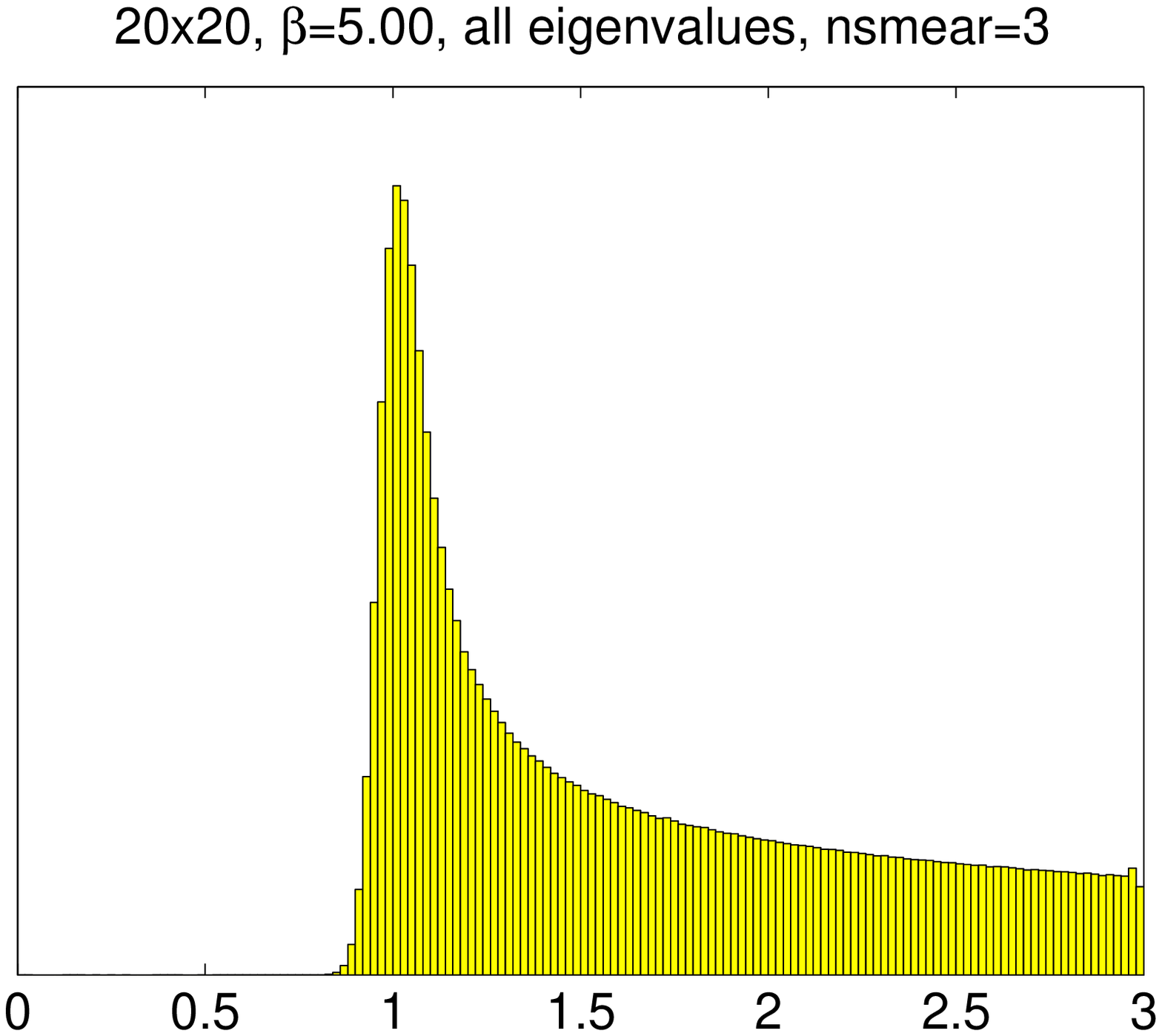,height=48mm,width=60mm}
\\
\epsfig{file=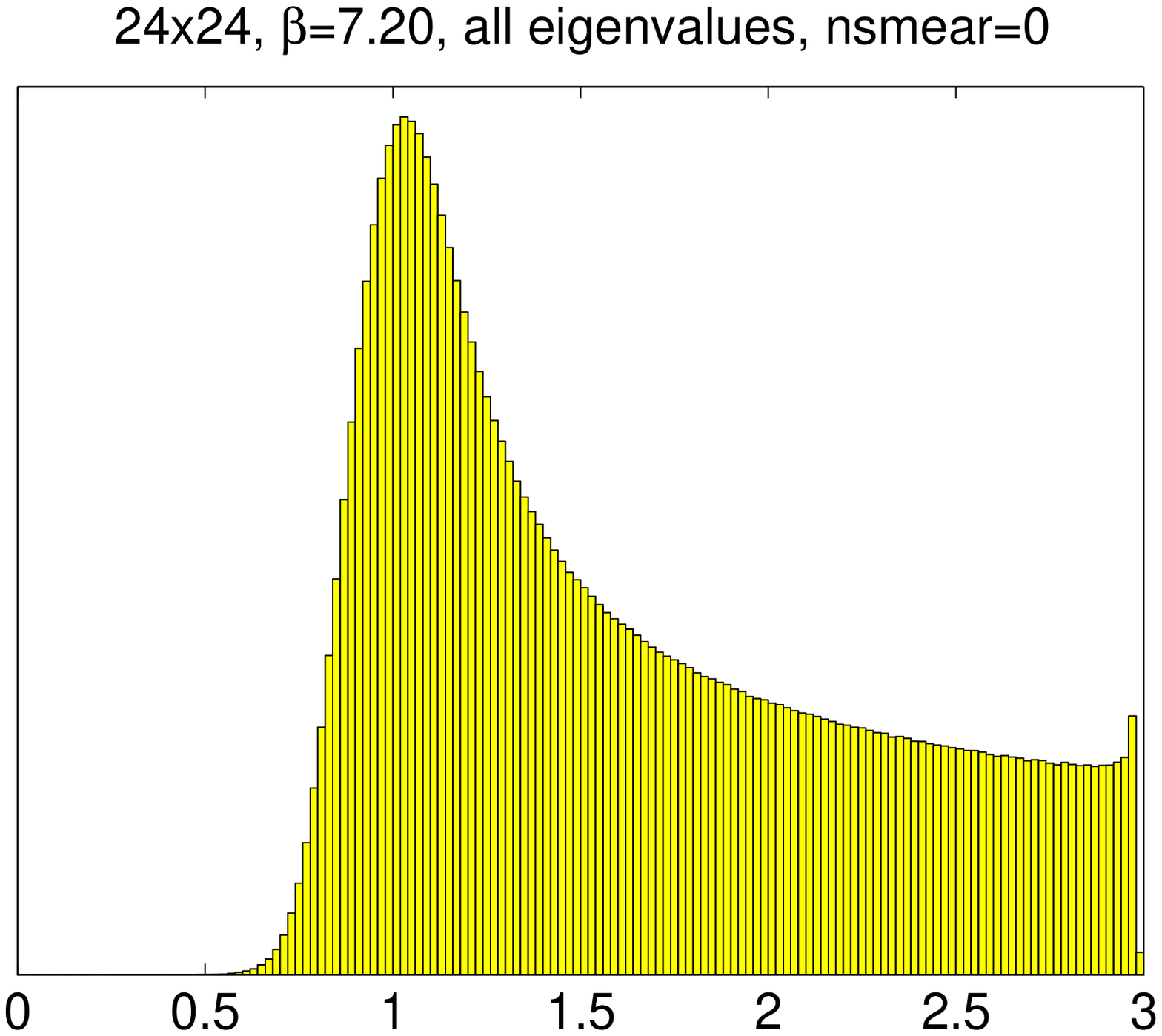,height=48mm,width=60mm}
\epsfig{file=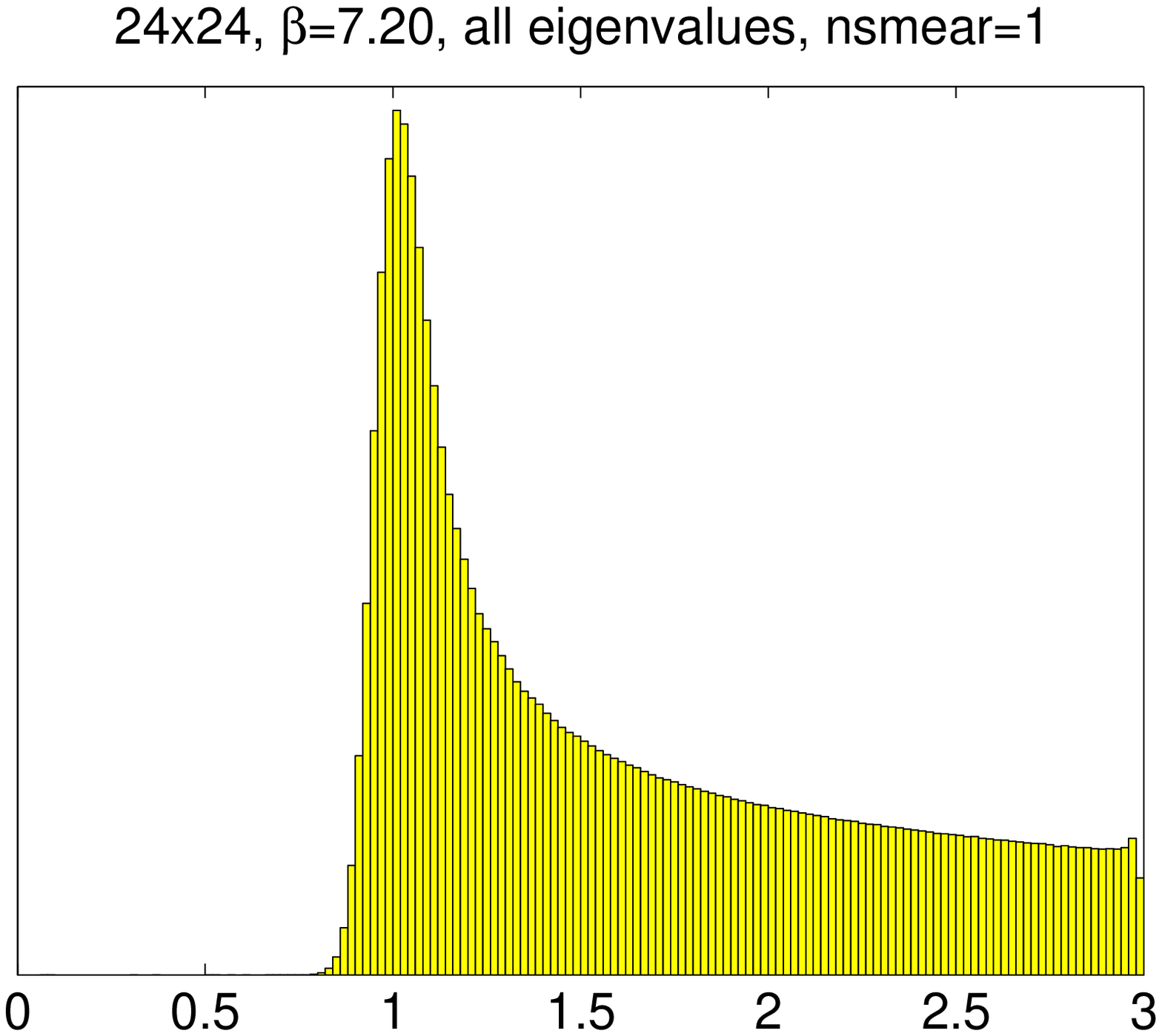,height=48mm,width=60mm}
\epsfig{file=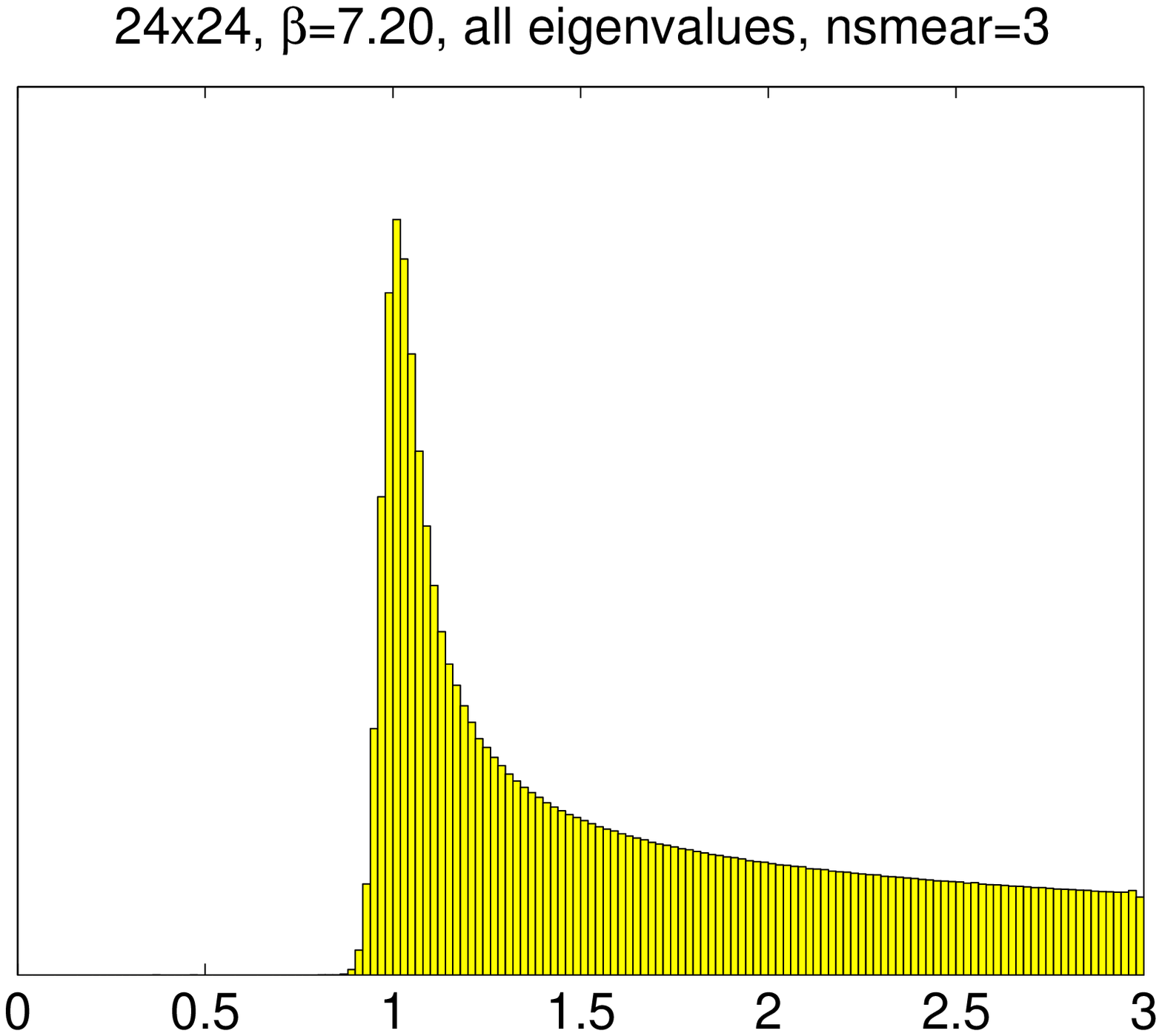,height=48mm,width=60mm}
\vspace{-2mm}
\caption{Distribution of all eigenvalues of $|\HW|\!=\!|\gaf D_{\mr{W},-1}|$ at
five $\be$ and 0,1,3 smearings.}
\label{fig:sm1}
\end{figure}

\begin{figure}[!p]
\epsfig{file=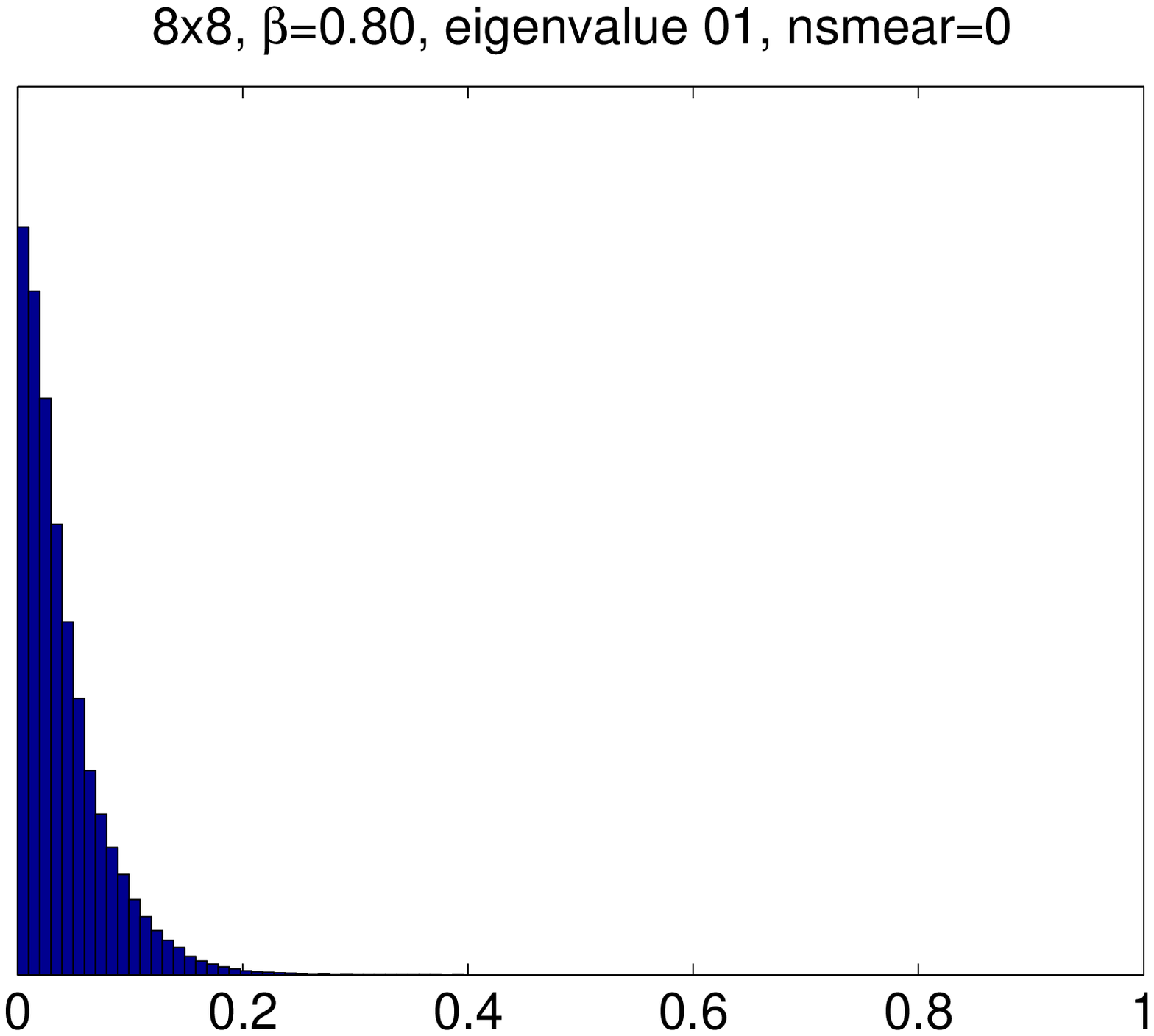,height=48mm,width=60mm}
\epsfig{file=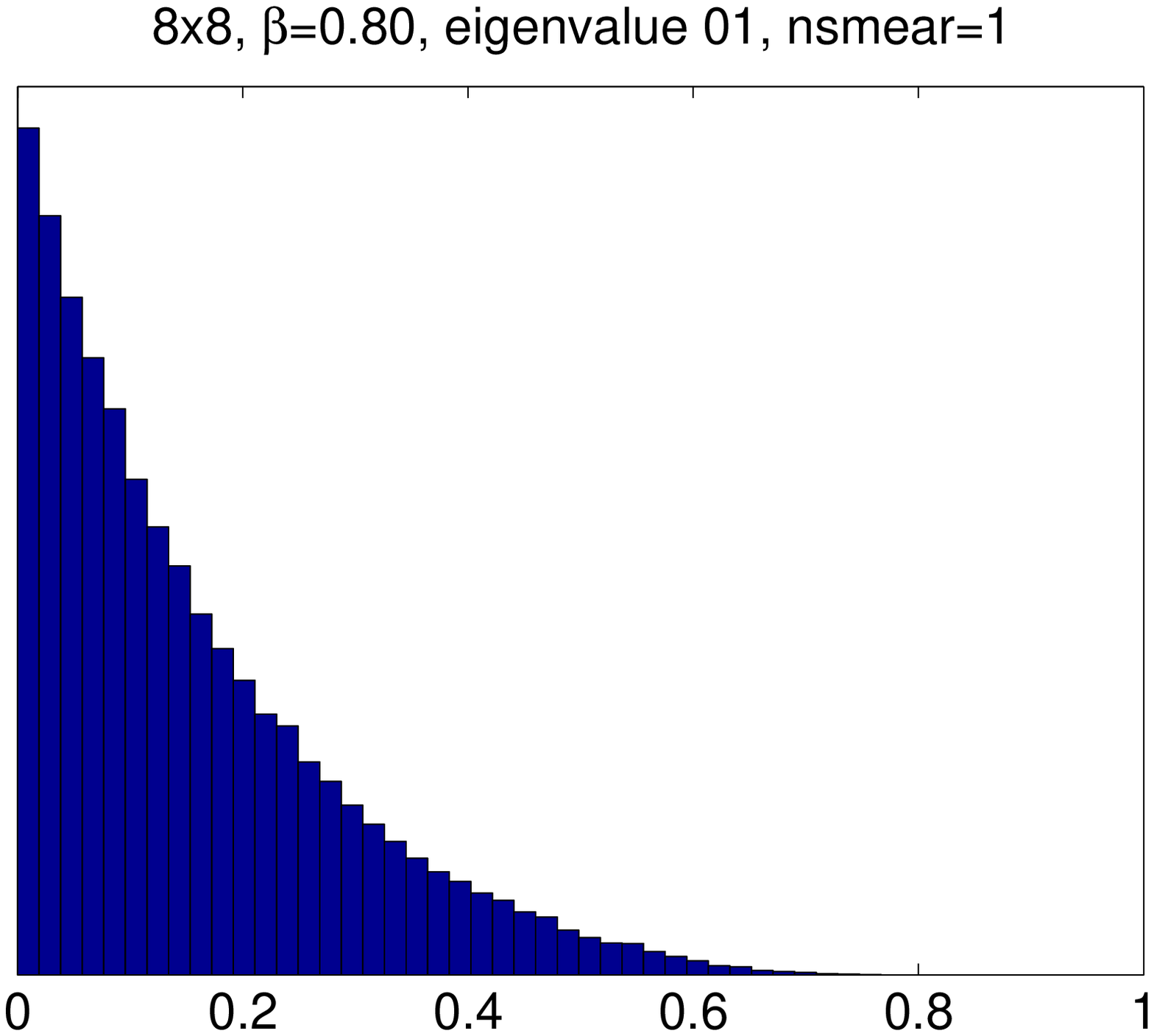,height=48mm,width=60mm}
\epsfig{file=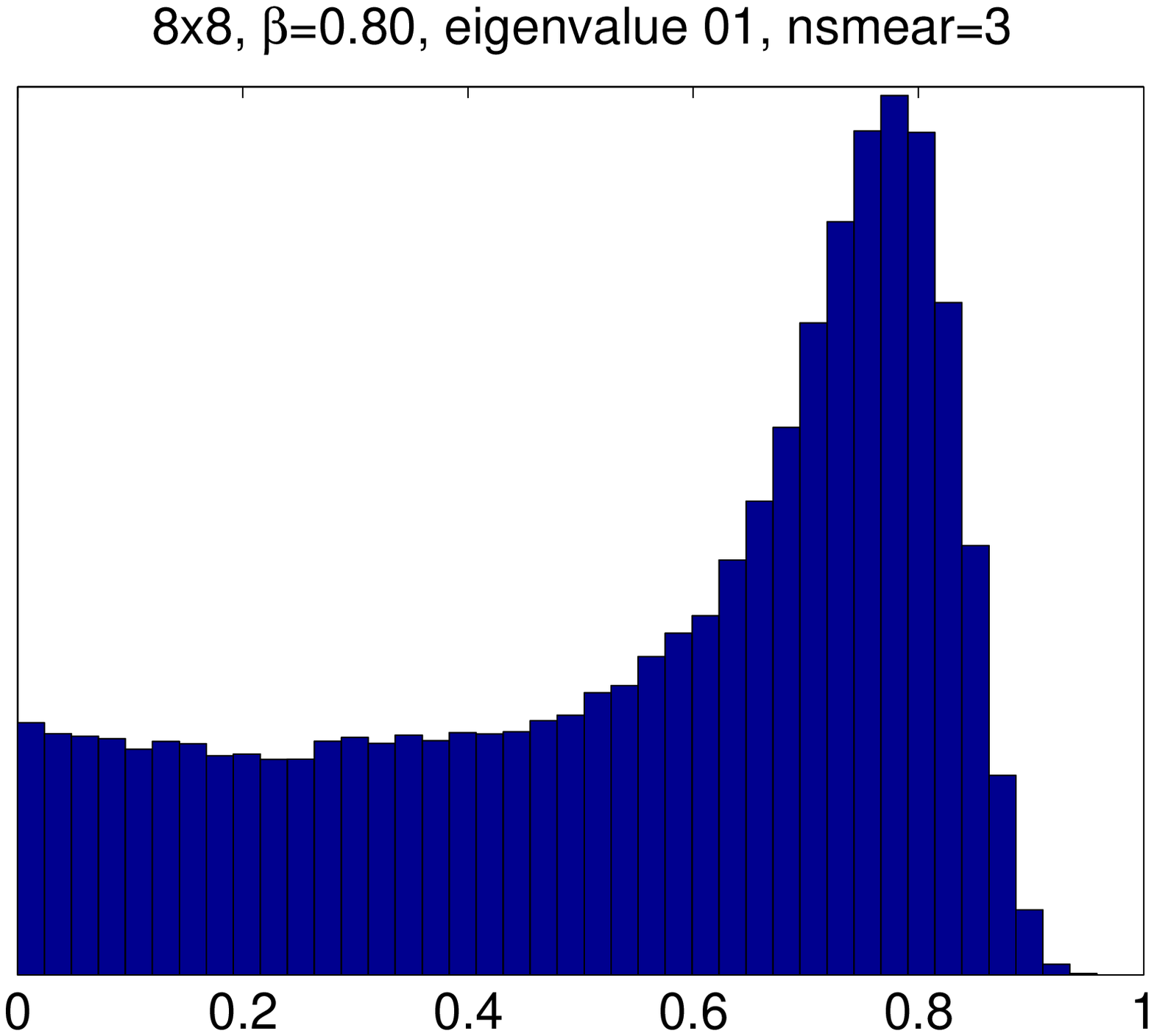,height=48mm,width=60mm}
\\
\epsfig{file=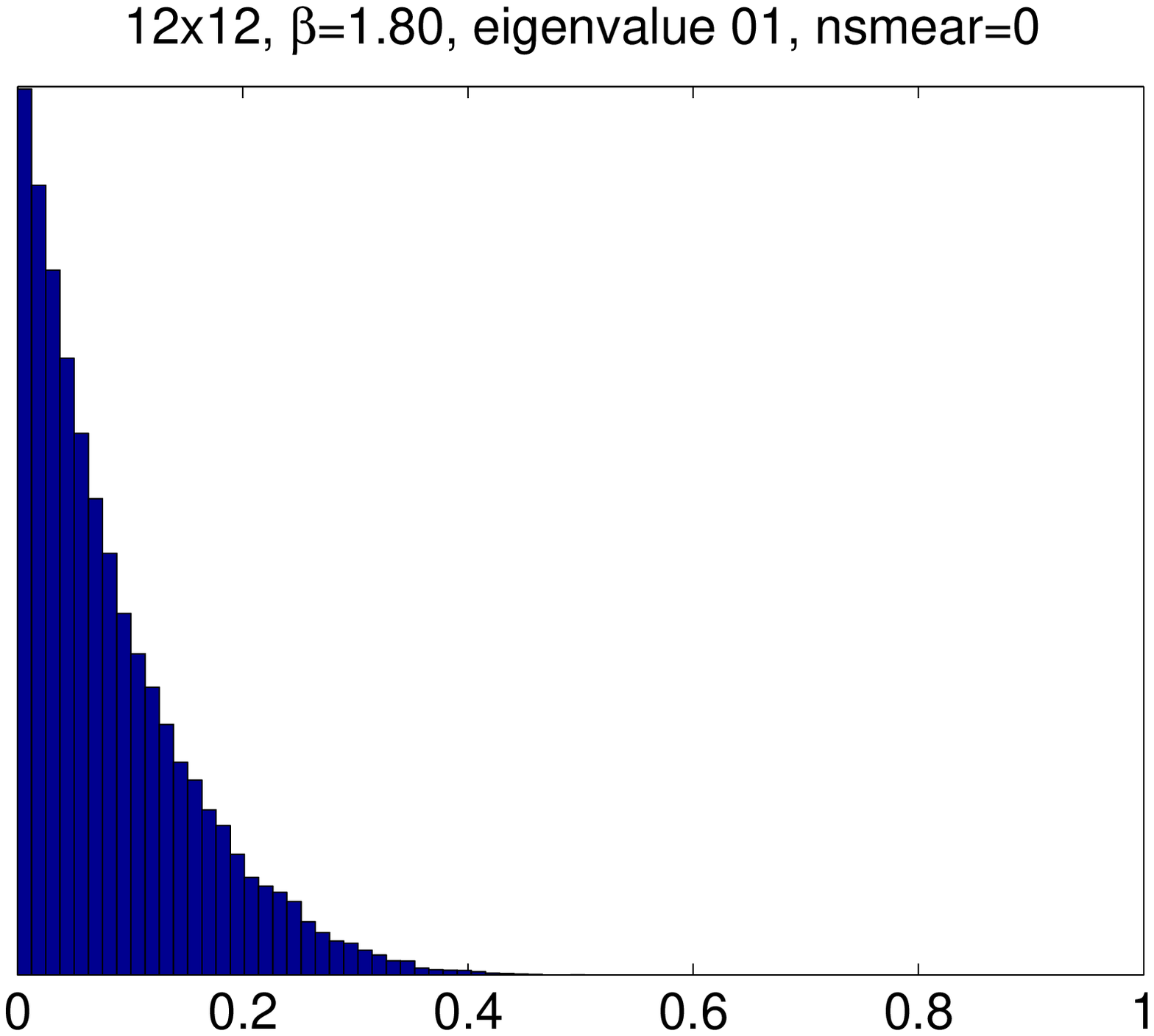,height=48mm,width=60mm}
\epsfig{file=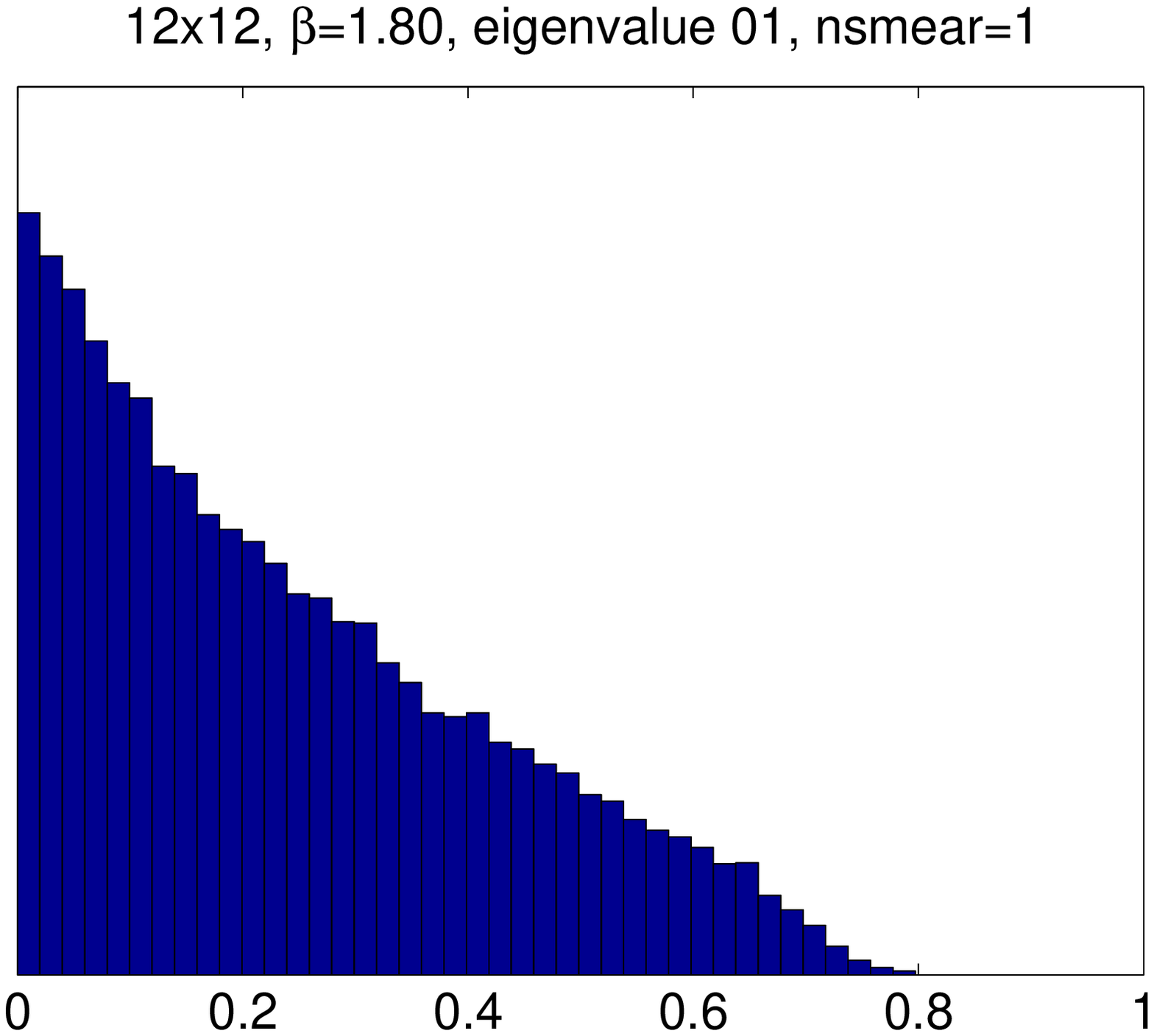,height=48mm,width=60mm}
\epsfig{file=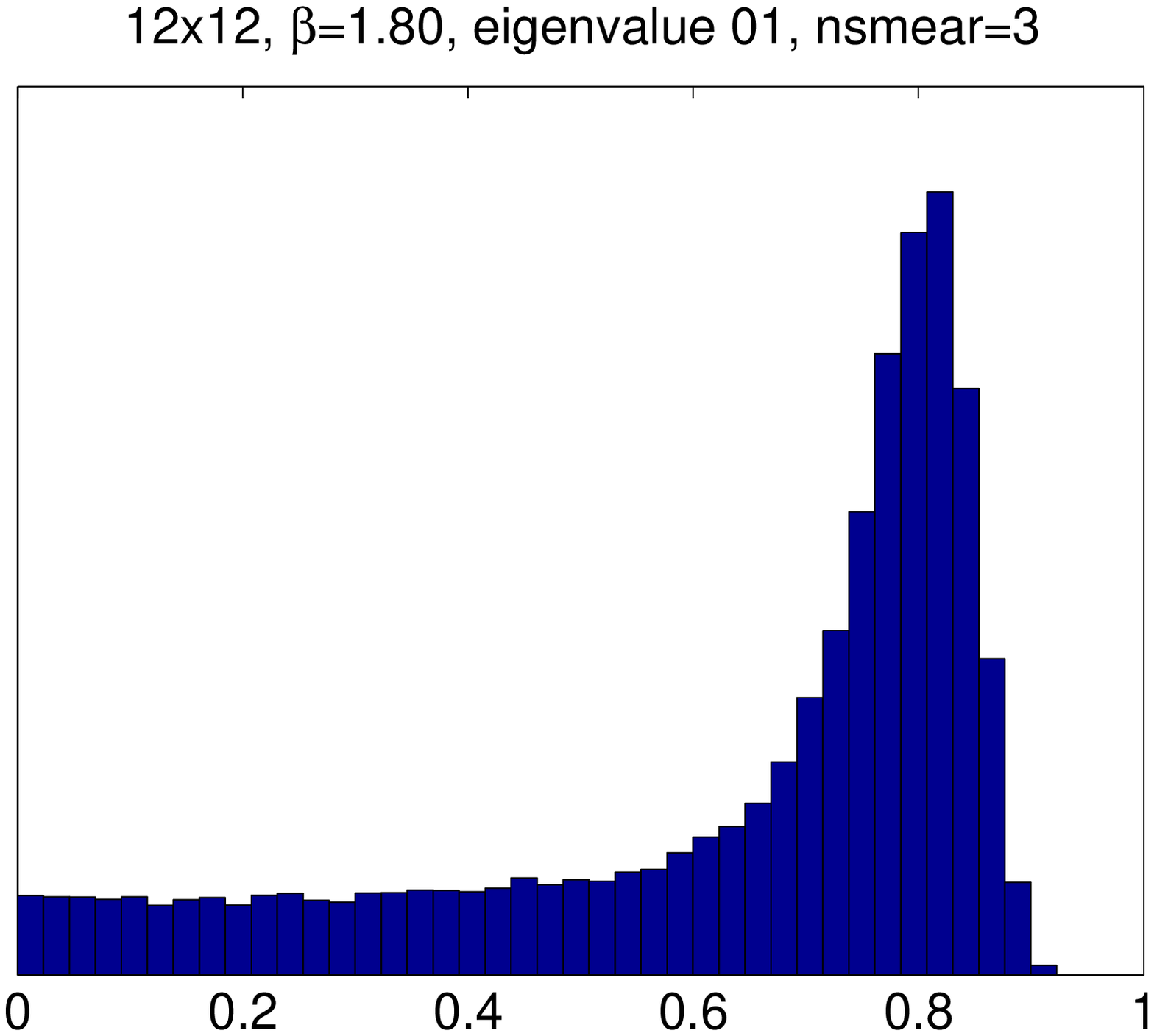,height=48mm,width=60mm}
\\
\epsfig{file=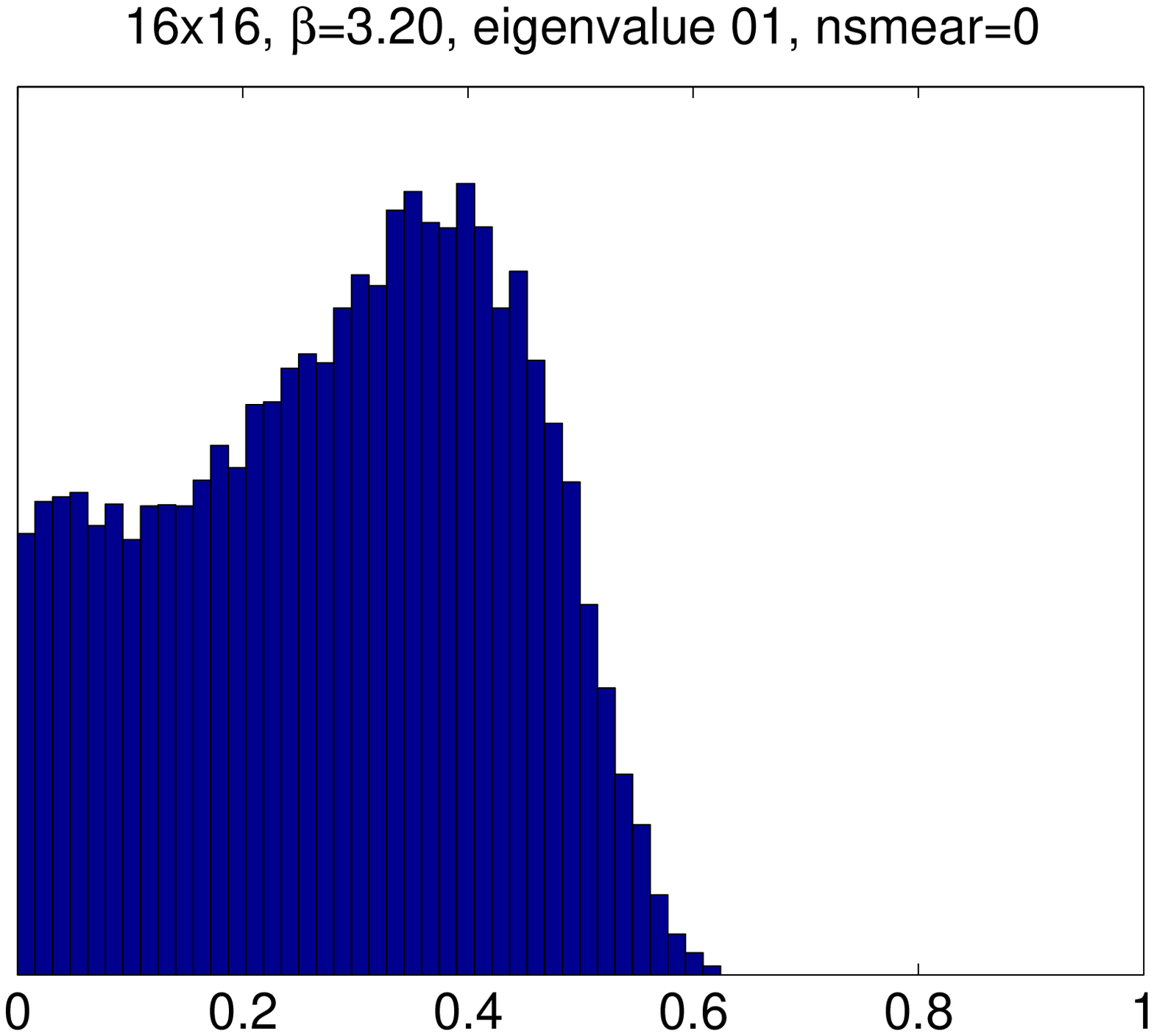,height=48mm,width=60mm}
\epsfig{file=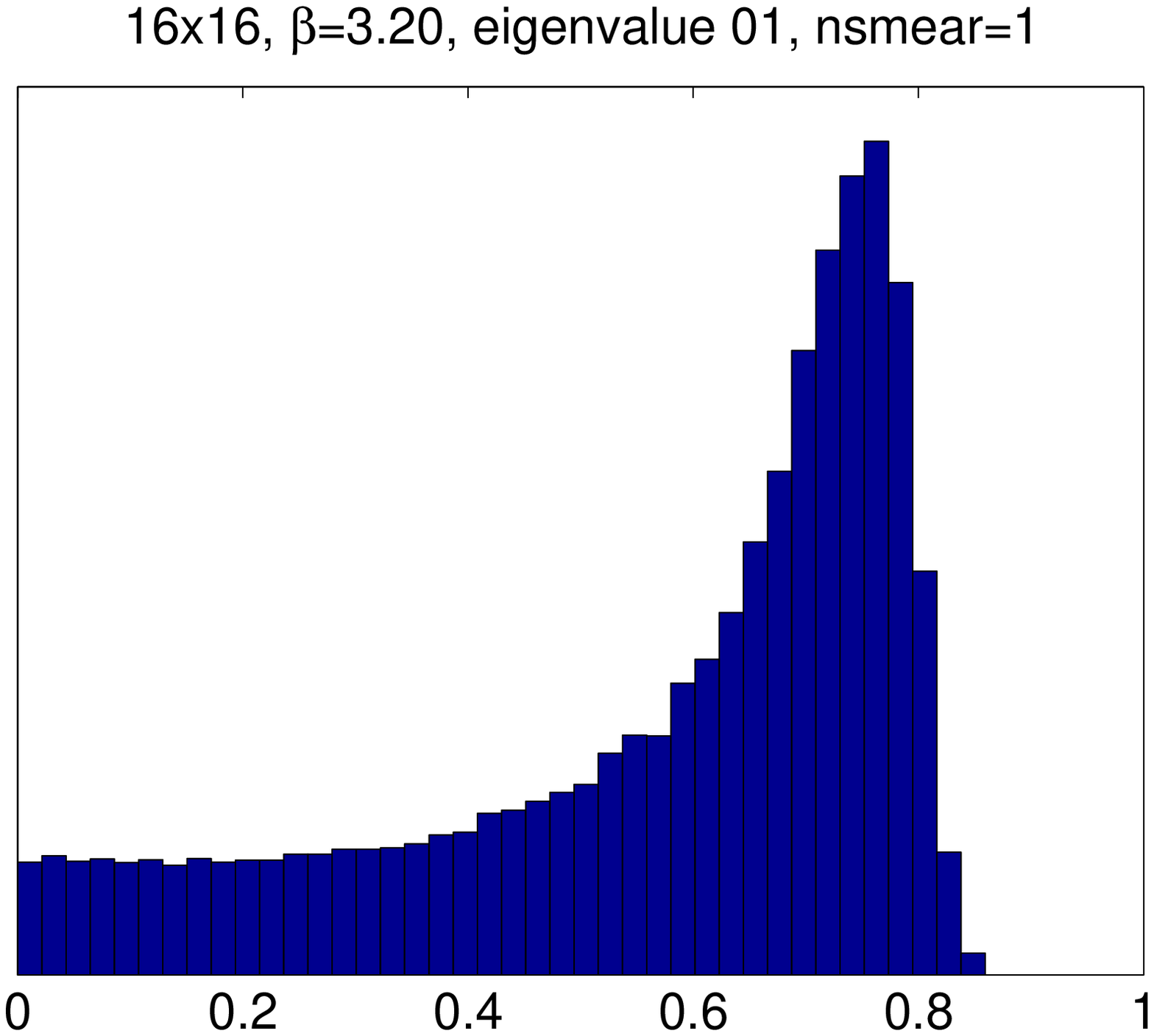,height=48mm,width=60mm}
\epsfig{file=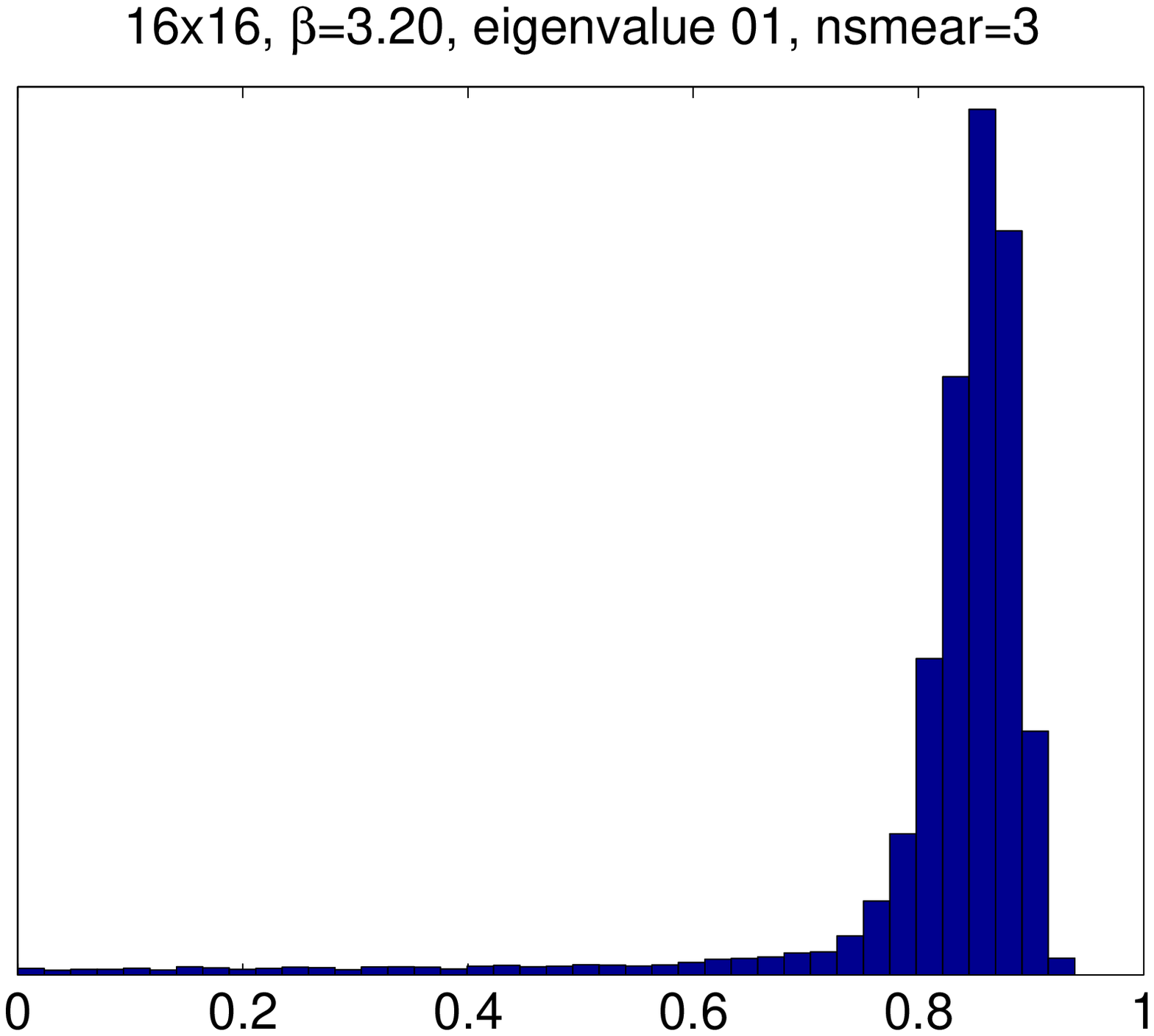,height=48mm,width=60mm}
\\
\epsfig{file=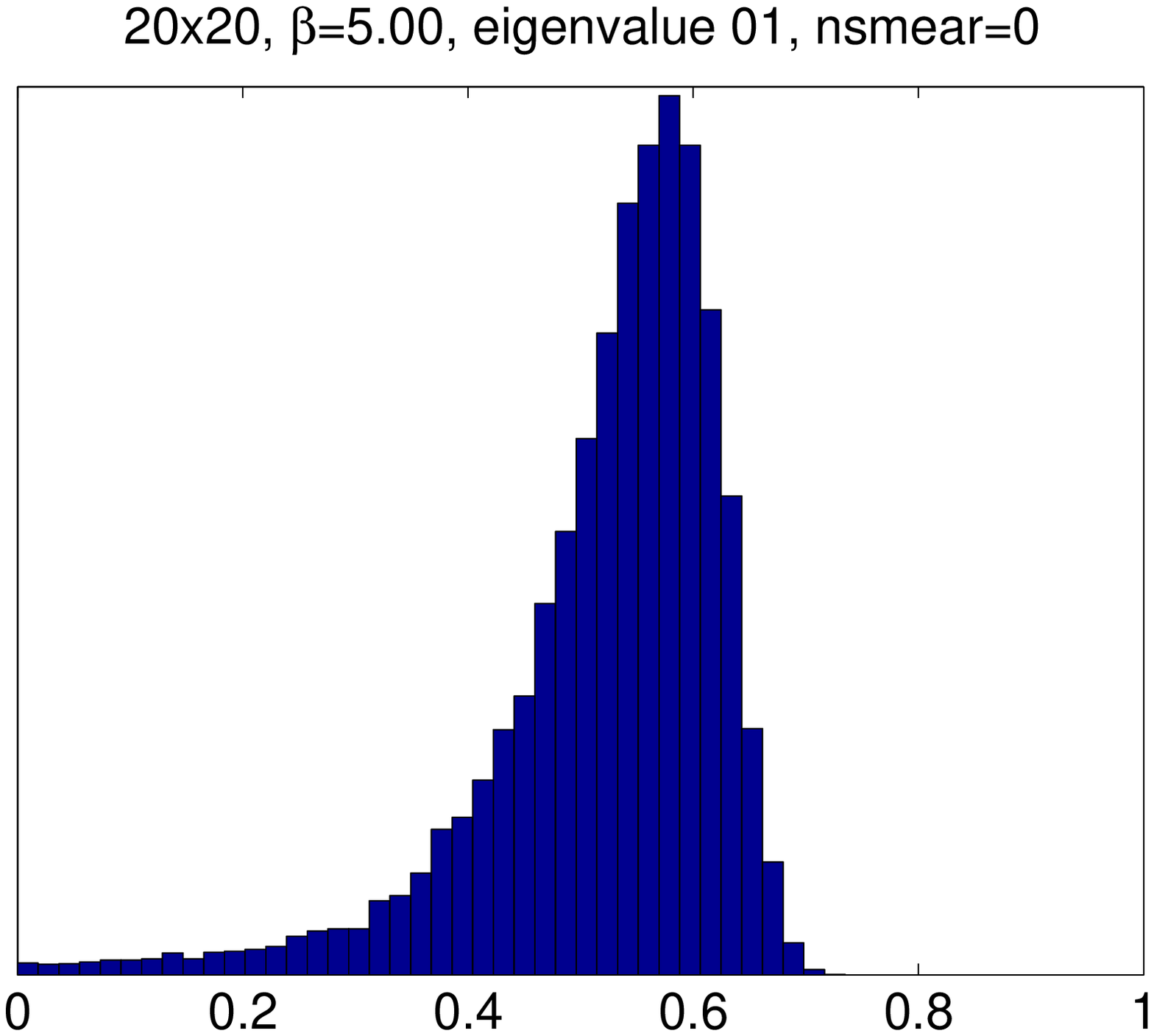,height=48mm,width=60mm}
\epsfig{file=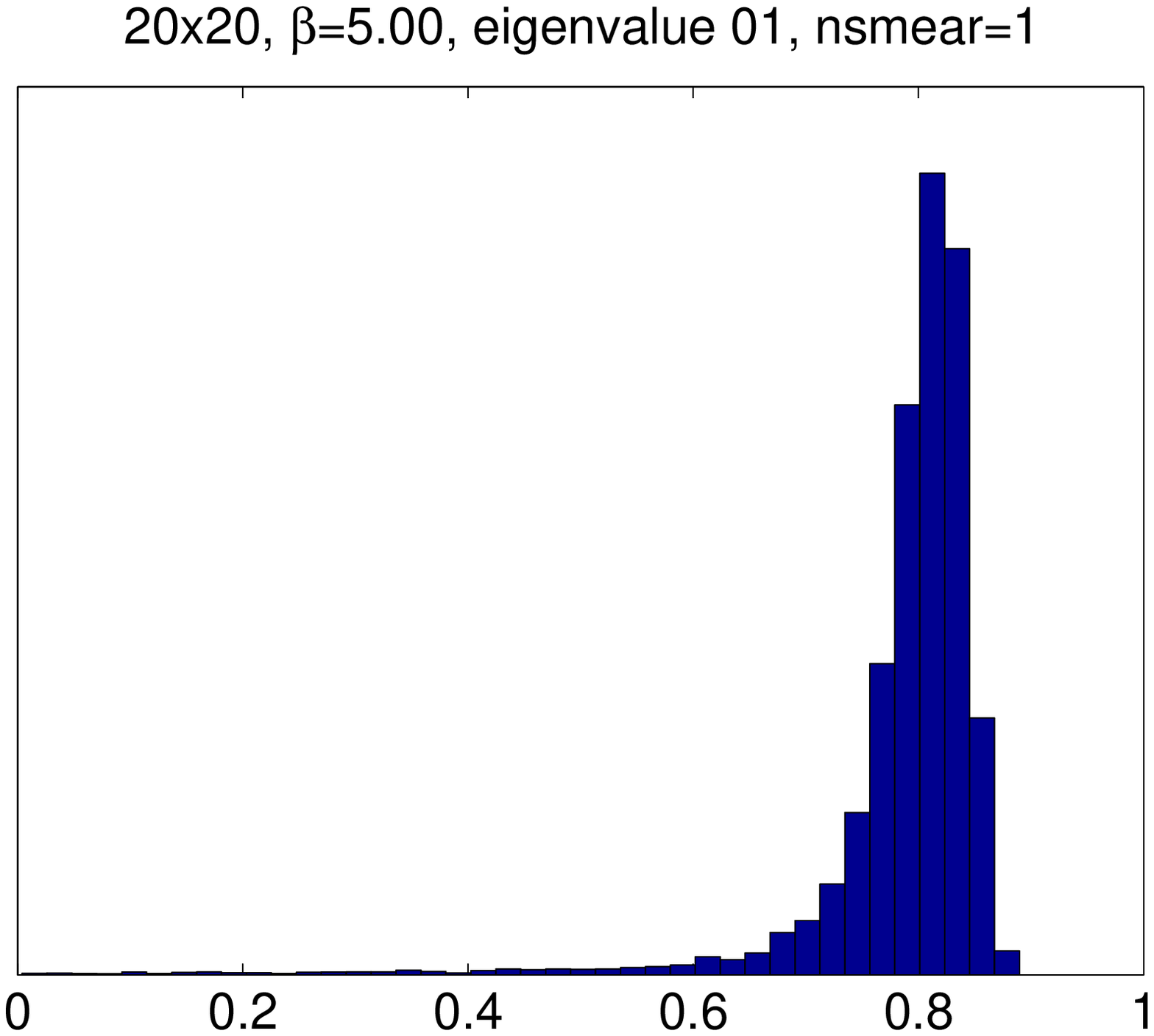,height=48mm,width=60mm}
\epsfig{file=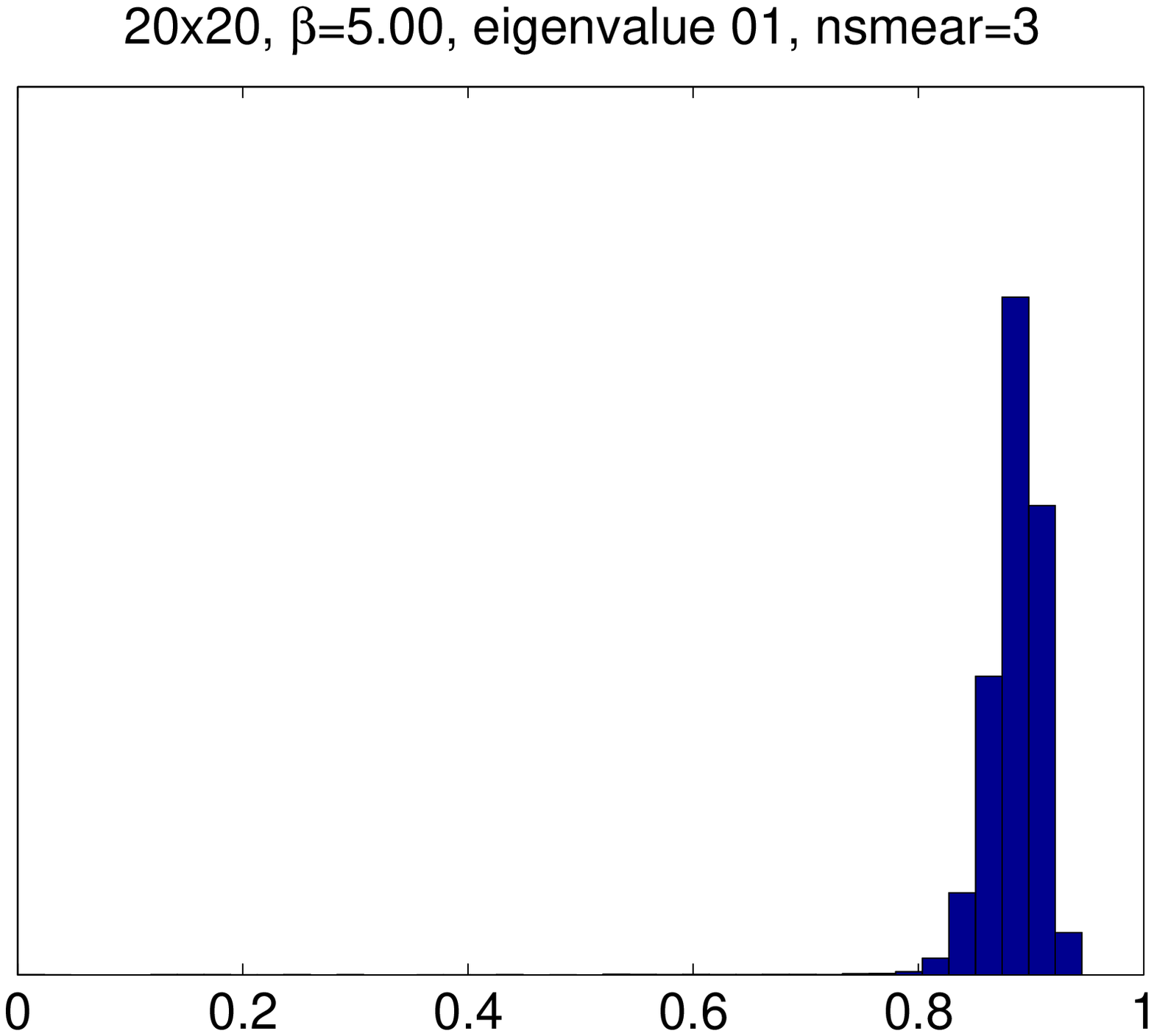,height=48mm,width=60mm}
\\
\epsfig{file=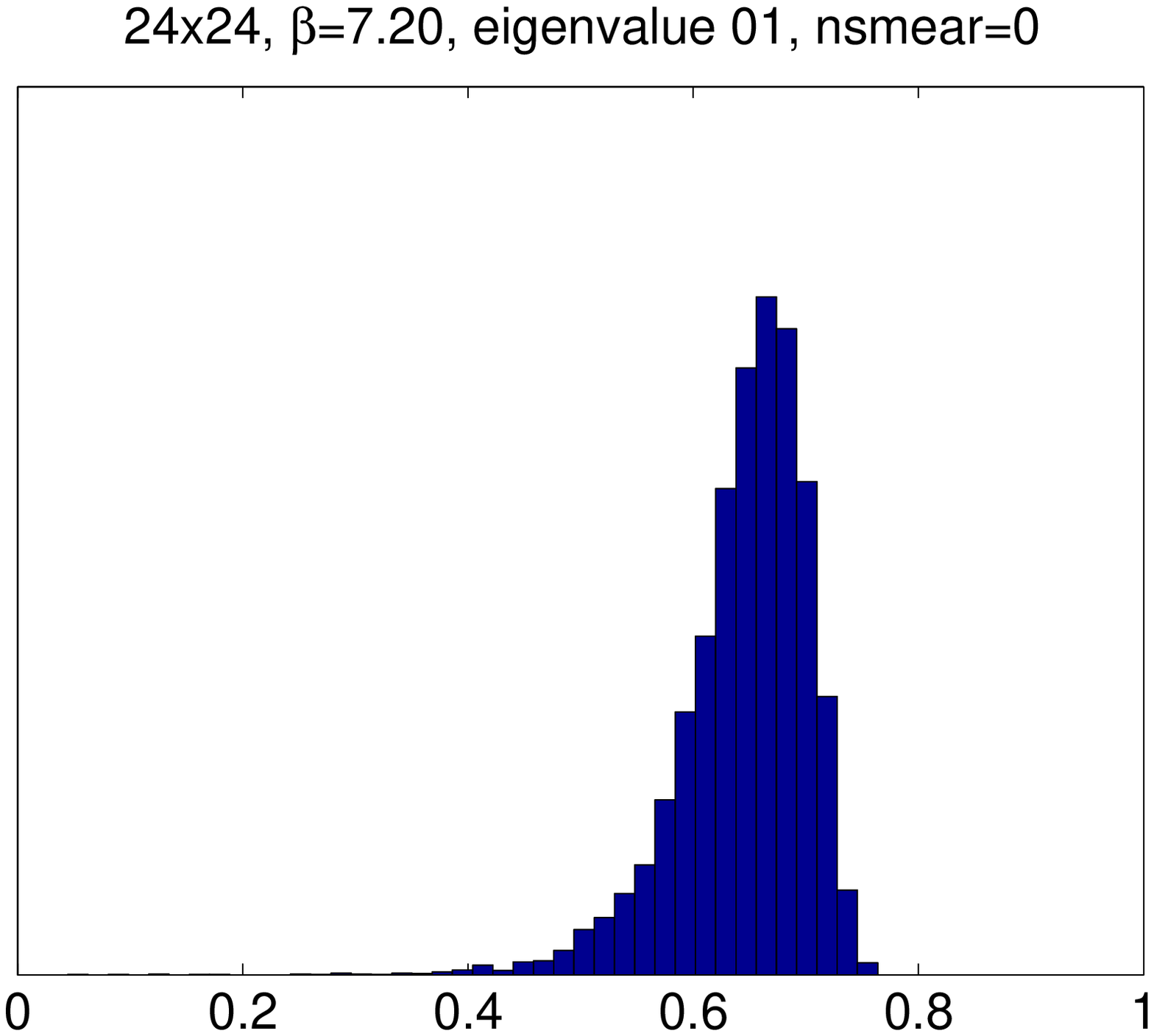,height=48mm,width=60mm}
\epsfig{file=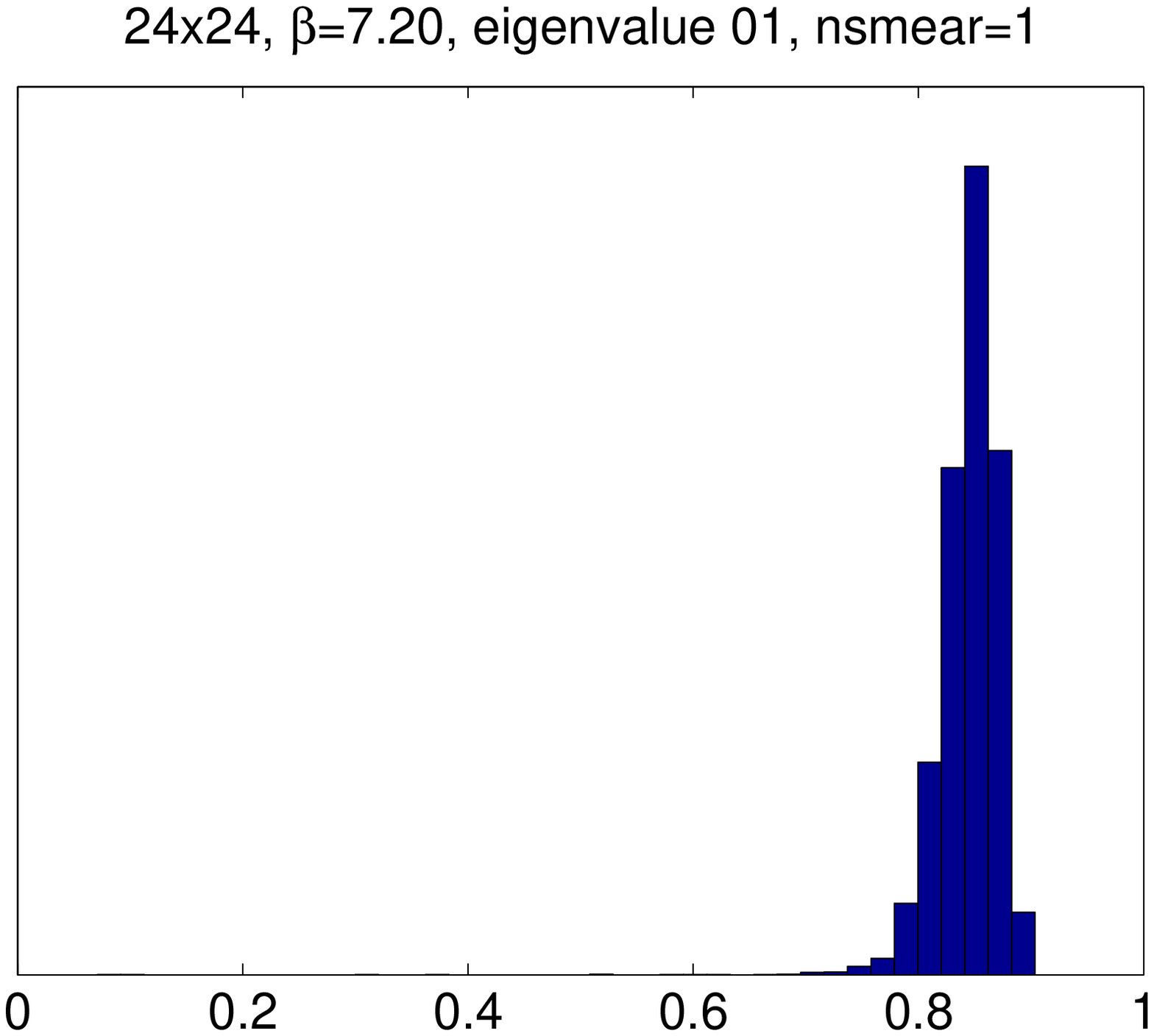,height=48mm,width=60mm}
\epsfig{file=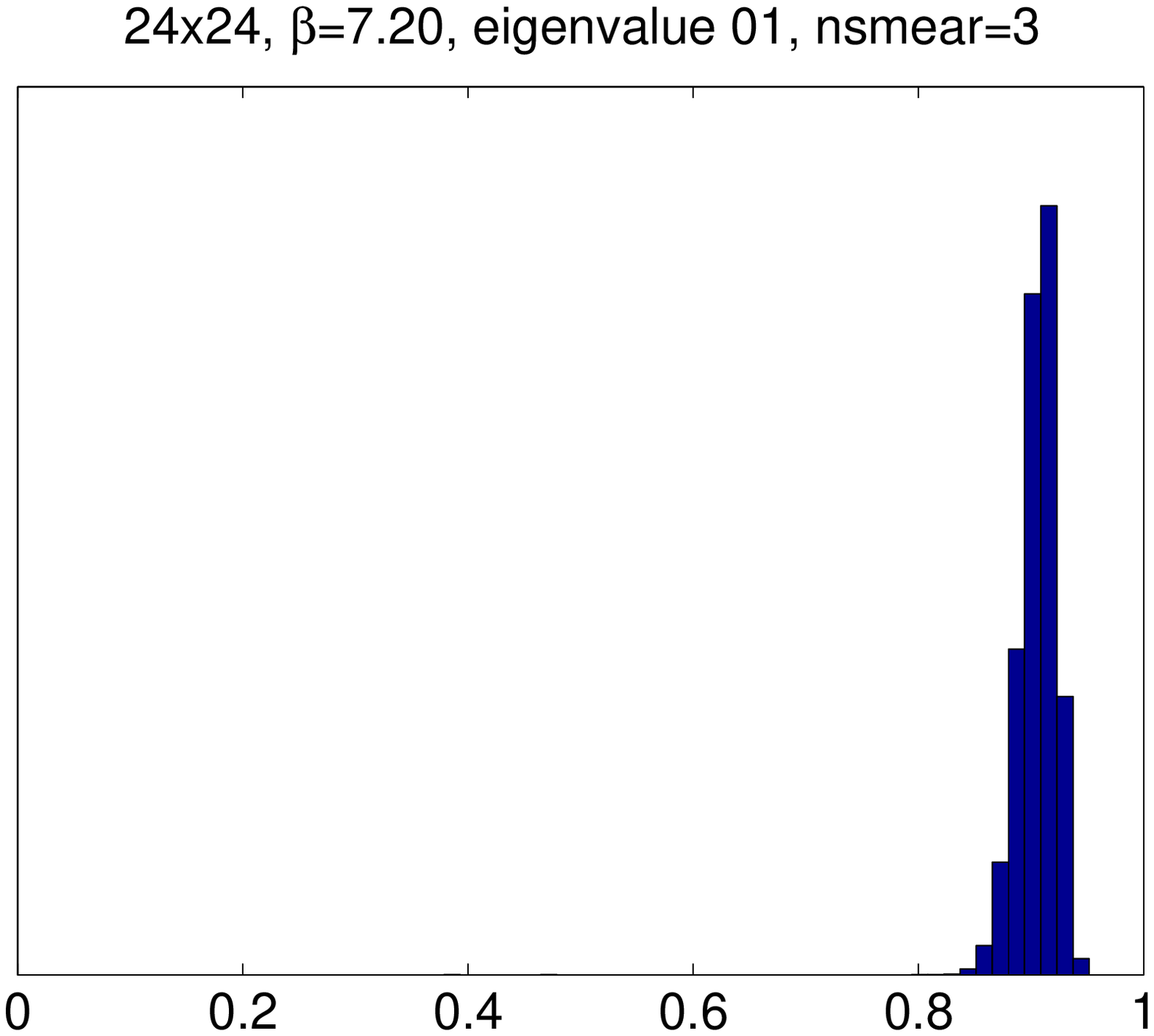,height=48mm,width=60mm}
\vspace{-2mm}
\caption{Distribution of the 1st eigenvalue of $|\HW|\!=\!|\gaf D_{\mr{W},-1}|$
at five $\be$, with 0,1,3 smearings.}
\label{fig:sm2}
\end{figure}

\begin{figure}[!p]
\begin{center}
\epsfig{file=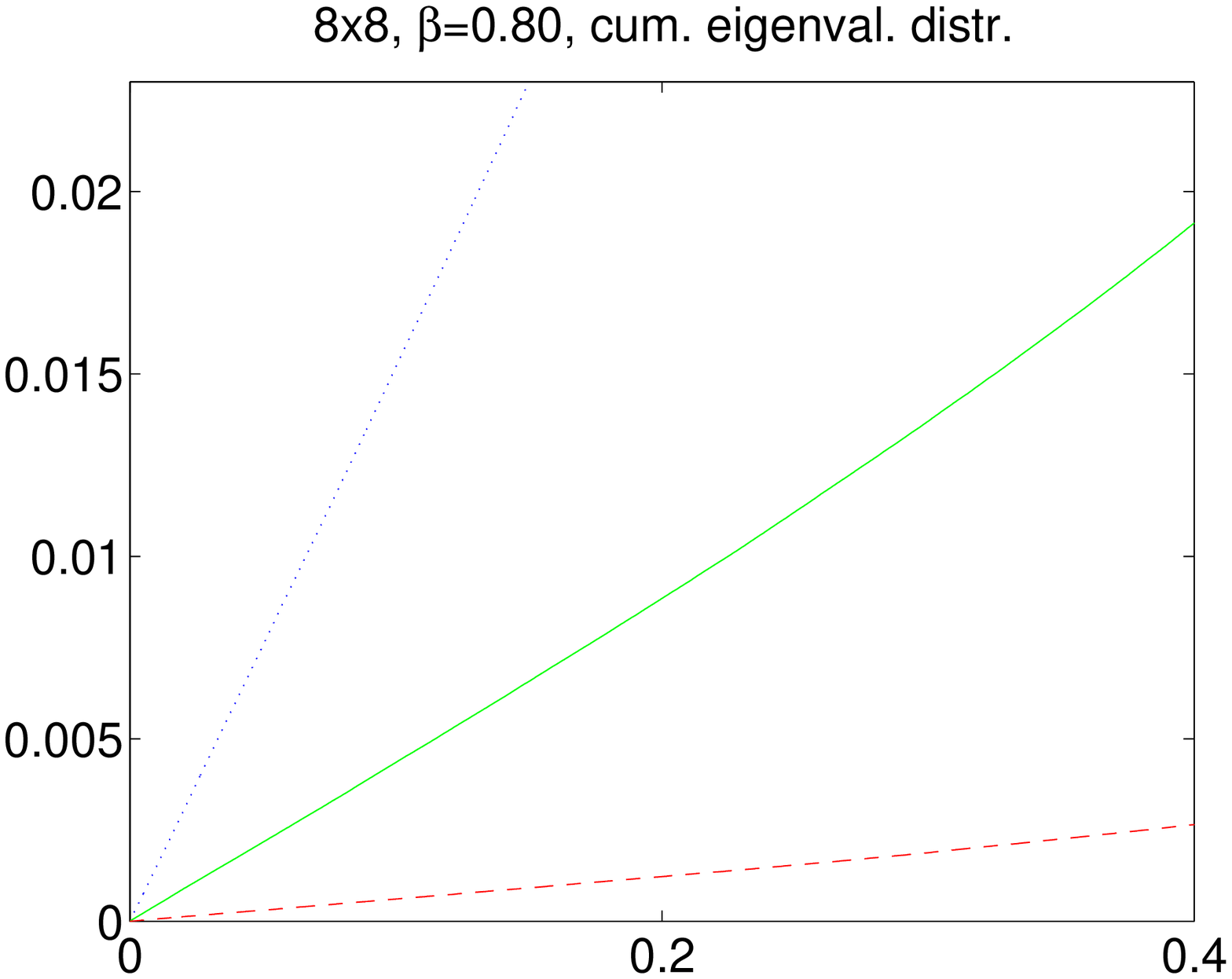,height=48mm,width=60mm}
\\
\epsfig{file=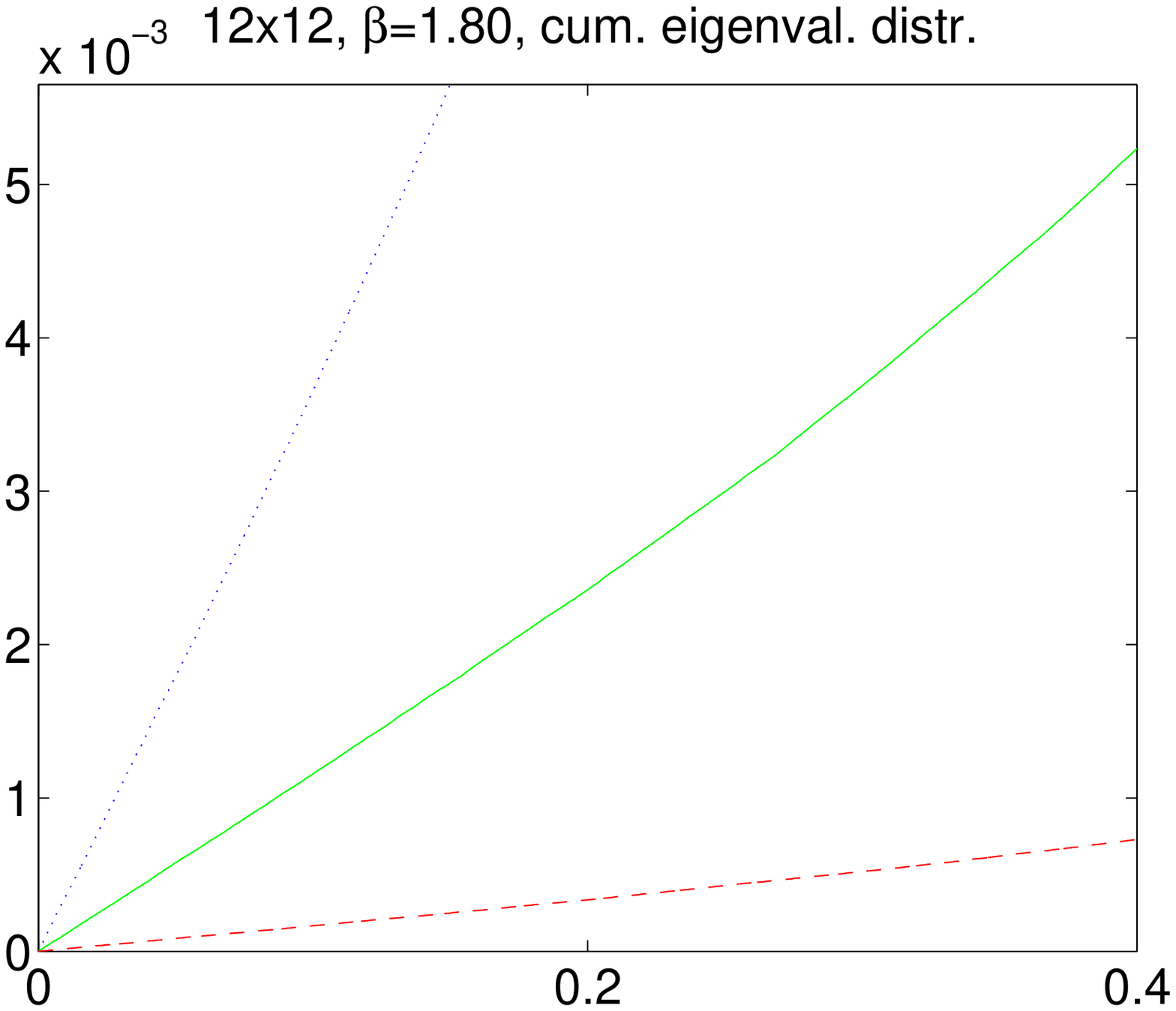,height=48mm,width=60mm}
\\
\epsfig{file=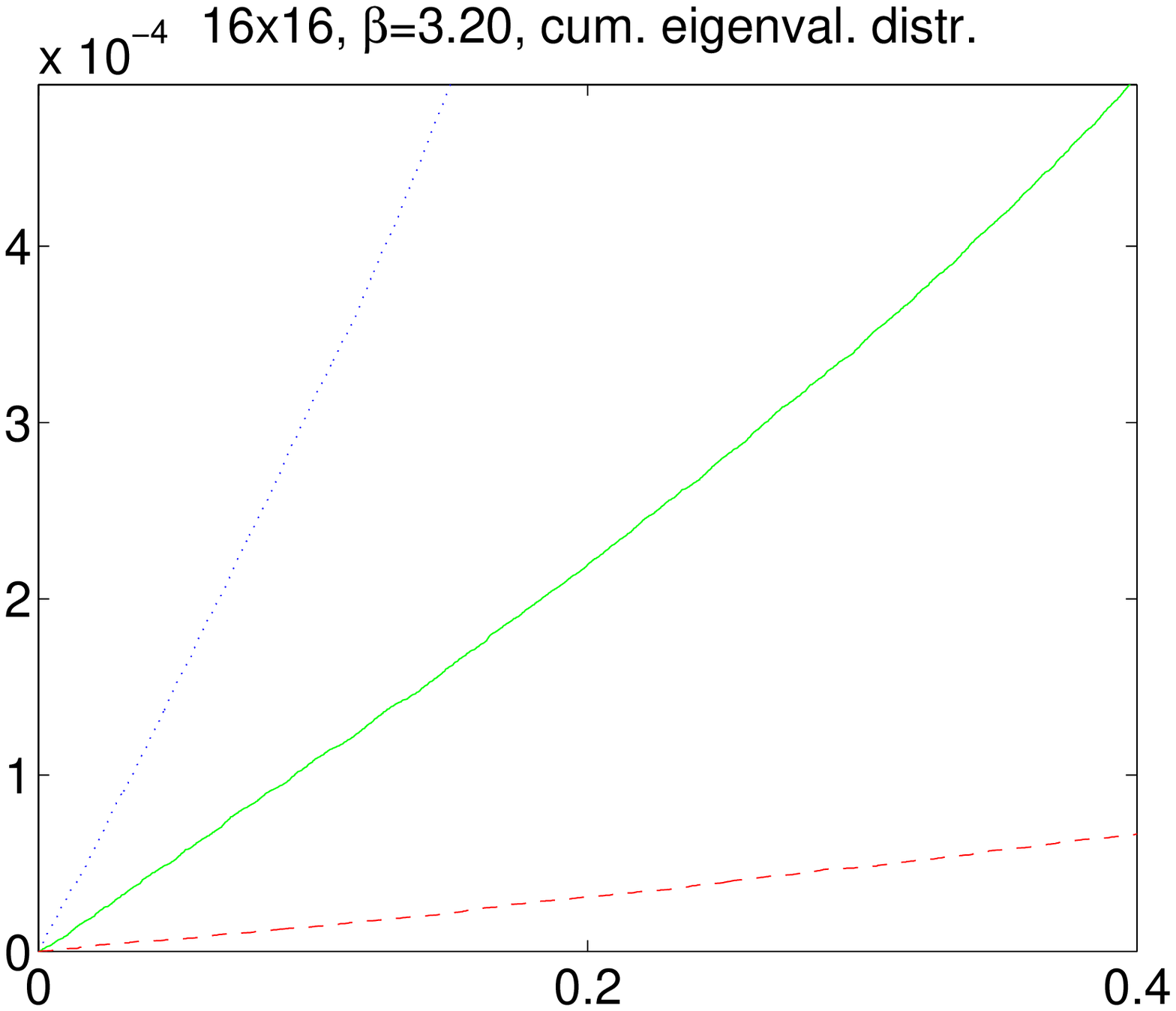,height=48mm,width=60mm}
\\
\epsfig{file=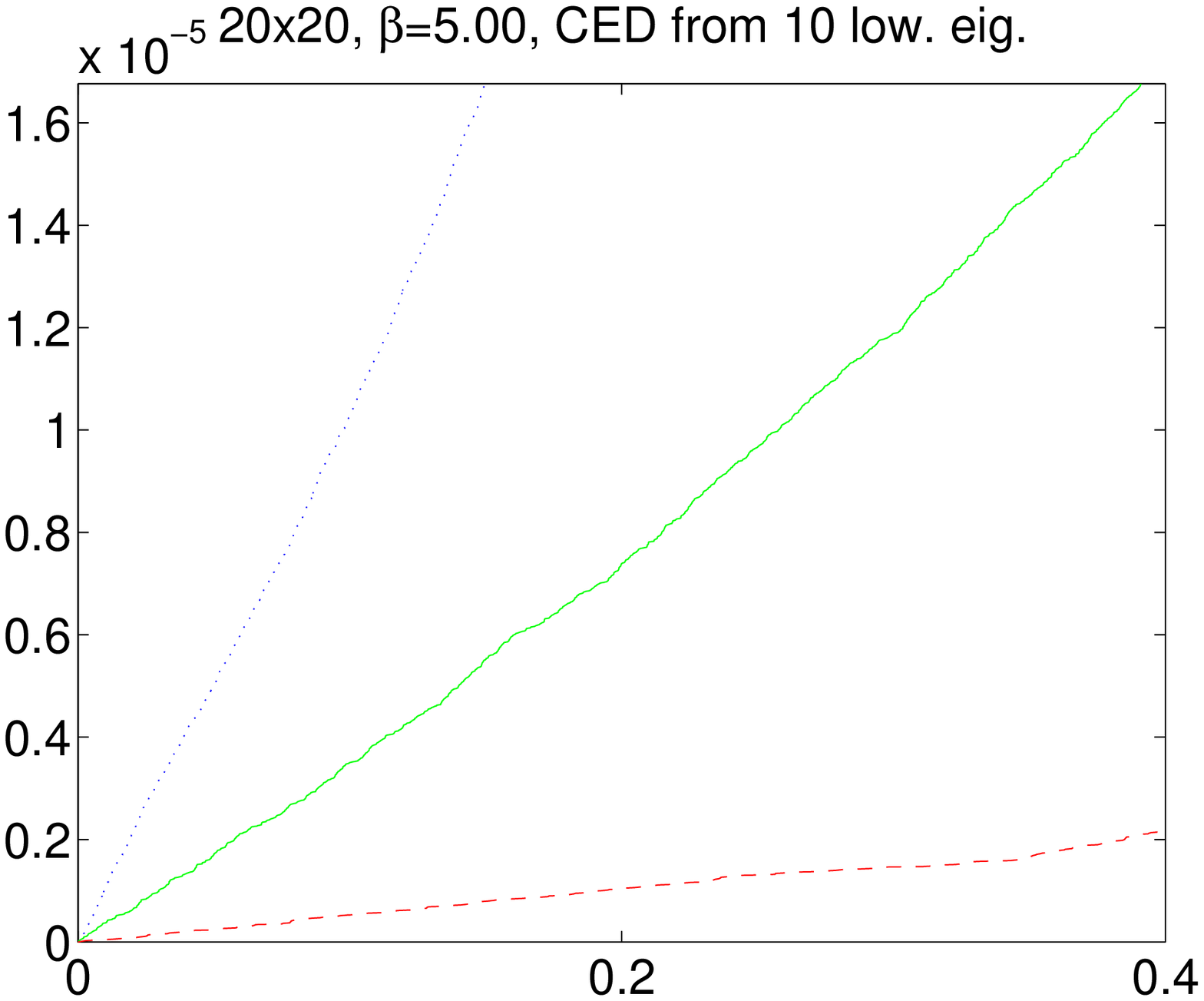,height=48mm,width=60mm}
\\
\epsfig{file=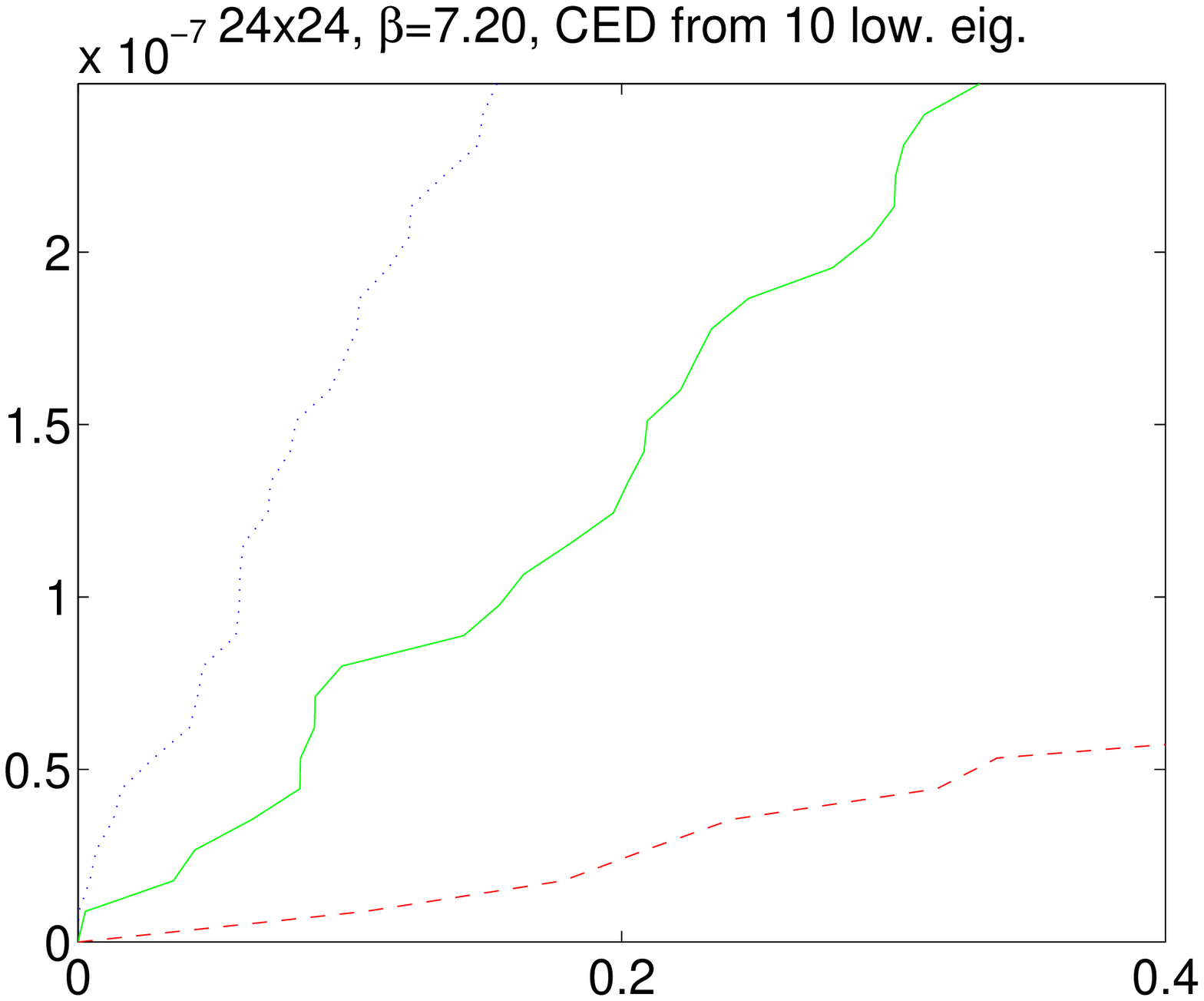,height=48mm,width=60mm}
\end{center}
\vspace{-8mm}
\caption{Cumulative eigenvalue distribution of $|\HW|\!=\!|\gaf D_{\mr{W},-1}|$
at five $\be$, with 0,1,3 smearings.}
\label{fig:sm3}
\end{figure}

Fig.\,\ref{fig:sm1} provides an overview over the complete $|\HW|$ eigenvalue
distribution; one sees a ``peak'' at $\la\!=\!1$ forming that gets more
pronounced with higher $\be$ and higher filtering level.
This ``peak'' corresponds to the
``jump'' at $\la\!=\!1$ in the CED of $|\HW|$ in Fig.\,\ref{fig:ced_2D}.

Fig.\,\ref{fig:sm2} presents the distribution of the lowest eigenvalue of
$|\HW|$.
At low $\be$ this distribution accumulates at zero, at intermediate values of
the coupling there is a horizontal band of eigenvalues connecting down to zero,
and at the largest $\be$ there are just scattered eigenvalues.
Evidently, one cannot draw a final conclusion whether these scattered
eigenvalues really make up for a non-zero $\rh(0)$, but it seems worth while to
study this band in the region of $\be$ values where it is clearly visible and
see whether changing $\be$ implies some structure, or whether it just stays
flat, regardless of $\be$.

Fig.\,\ref{fig:sm3} presents the CED in the area of interest, the very-low
$\la$ region.
At $\be\!=\!5.0,7.2$ we show the data from the high-statistics run with
10 eigenvalues per configuration (bottom line of Tab.\,\ref{tab:sm_par}),
but we checked that the results are consistent with what we get from the runs
where all eigenvalues were determined.
At a given coupling filtering clearly reduces $\rh_{|\HW|}(0)$.
The overall impression is that changing $\be$ merely rescales the $y$-axis,
in striking analogy with what we have seen in 4D (Fig.\,\ref{fig:ced_4D}).
If this is indeed true, the natural conclusion is that $\rh_{|\HW|}(0)\!>\!0$
at any finite $\be$ in the quenched theory.

\begin{figure}[!t]
\begin{center}
\epsfig{file=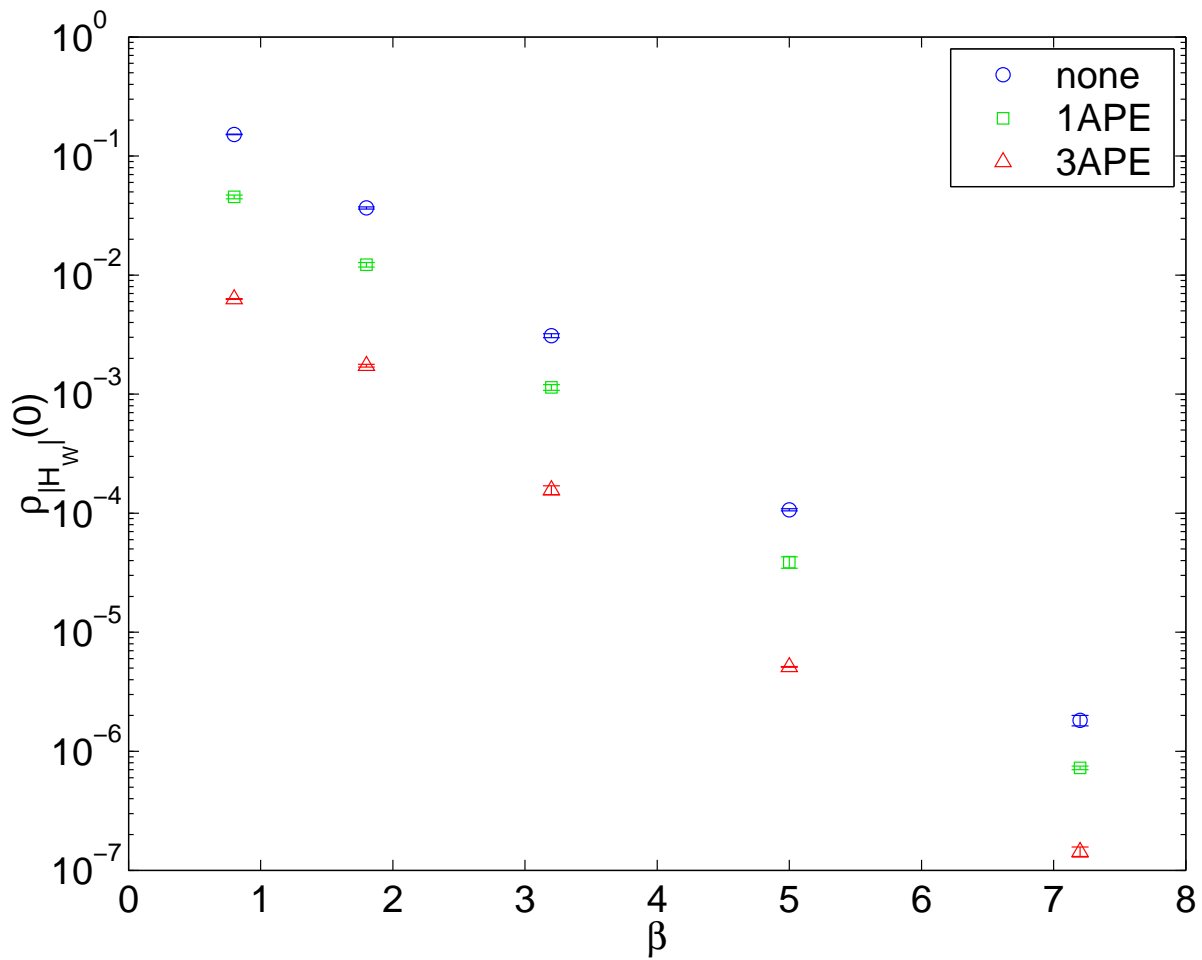,width=16cm}
\end{center}
\vspace{-6mm}
\caption{$\rh_{|\HW|}^{}(0)$ versus $\be$ in the quenched Schwinger model.
Filtering seems to reduce, for all $\be$, the eigenvalue density by a
near-universal factor. The data suggest an exponential fall-off at large
$\be$.}
\label{fig:sm6}
\end{figure}

Fig.\,\ref{fig:sm6} contains a summary of our determinations of the
spectral densities $\rh_{|\HW|}(0)$ for various $\be$ and filtering levels,
extracted from the initial slopes in the CED shown in Fig.\,\ref{fig:sm3}.
It looks like eventually the density decreases exponentially in $\be$ and
changing the filtering level amounts to an overall rescaling factor which is,
in good approximation, independent of $\be$.
Obviously, this is just numerical evidence, but the message seems to be as
clear as one can possibly hope for from a numerical experiment.
Note that for practical reasons we cannot take the infinite volume limit,
but given our physical box size we expect finite volume effects to be
exponentially small.

\begin{table}
\begin{center}
\begin{tabular}{|l|ccccc|}
\hline
$\be$           & $0.8$& $1.8$& $3.2$& $5.0$& $7.2$\\
geometry        &$ 8\!\times\! 8$&$12\!\times\!12$&$16\!\times\!16$&
                 $20\!\times\!20$&$24\!\times\!24$ \\
statistics      &90,000&40,000&22,500&14,400&10,000\\
statistics      & --- &  --- &  --- &275,000&75,000\\
\hline
\end{tabular}
\end{center}
\vspace{-4mm}
\caption{Matched quenched 2D lattices; reweighting to $\Nf\!=\!1$ the lightest
particle in the chiral limit would fit $5$ times into the box. In the first
set all $|\HW|$ eigenvalues are determined with statistics such as to have an
equal number of eigenvalues. In the second set only the lowest 10 eigenvalues
are computed.}
\label{tab:sm_par}
\end{table}

We remind the reader that (both in 2D and in 4D) we were working at fixed
negative mass $m_0\!=\!-1$ and pushed towards the continuum line.
In other words, we were trying to stay as far outside the Aoki phase as one
can, if one wants to be in the supercritical region with one overlap fermion.
Of course, our data do not exclude the existence of a critical $\be$, but they
favor the view that there is no $\be_\mr{crit}$ that makes the
$|H_{\mr{W},-1}|$ spectral density strictly zero and therefore we conjecture
that $\rh_{|\HW|}(0)\!>\!0$ throughout the supercritical region.
Still, we do not see why this would create a problem for the localization
of the overlap operator, since the two seem not one-to-one inversely connected.

Finally, there is a simple argument that $\rh_{|\HW|}(0)\!>\!0$
holds in all quenched or unquenched theories with a massive overlap determinant
at all couplings.
Acquire infinite statistics at $\be\!=\!0, \Nf\!=\!0$.
When integrating over the full configuration space the spectral density is
certainly non-zero.
Results for the case of interest at finite $\be$ and maybe finite $\Nf$ can be
obtained through reweighting.
As long as one can guarantee that there is no configuration where the
reweighting factor vanishes,
the spectral density will be modified, but it cannot be made strictly zero.
This holds true in the quenched case and in the dynamical theory with an
overlap determinant ($m\!\neq\!0,-\!2\rh$), but it would not be true with
Wilson fermions at a negative mass.


\section*{App.\,B: Overlap operator locality in the free case}


In this appendix we collect some technical points to make sure that a
numerical investigation of the localization $\nu$ versus $\rh$ as shown
in Fig.\,\ref{fig:localization_free_summary48} for a $48^4$ lattice is
not overwhelmed with finite size effects.

\begin{figure}[p!]
\begin{center}
\epsfig{file=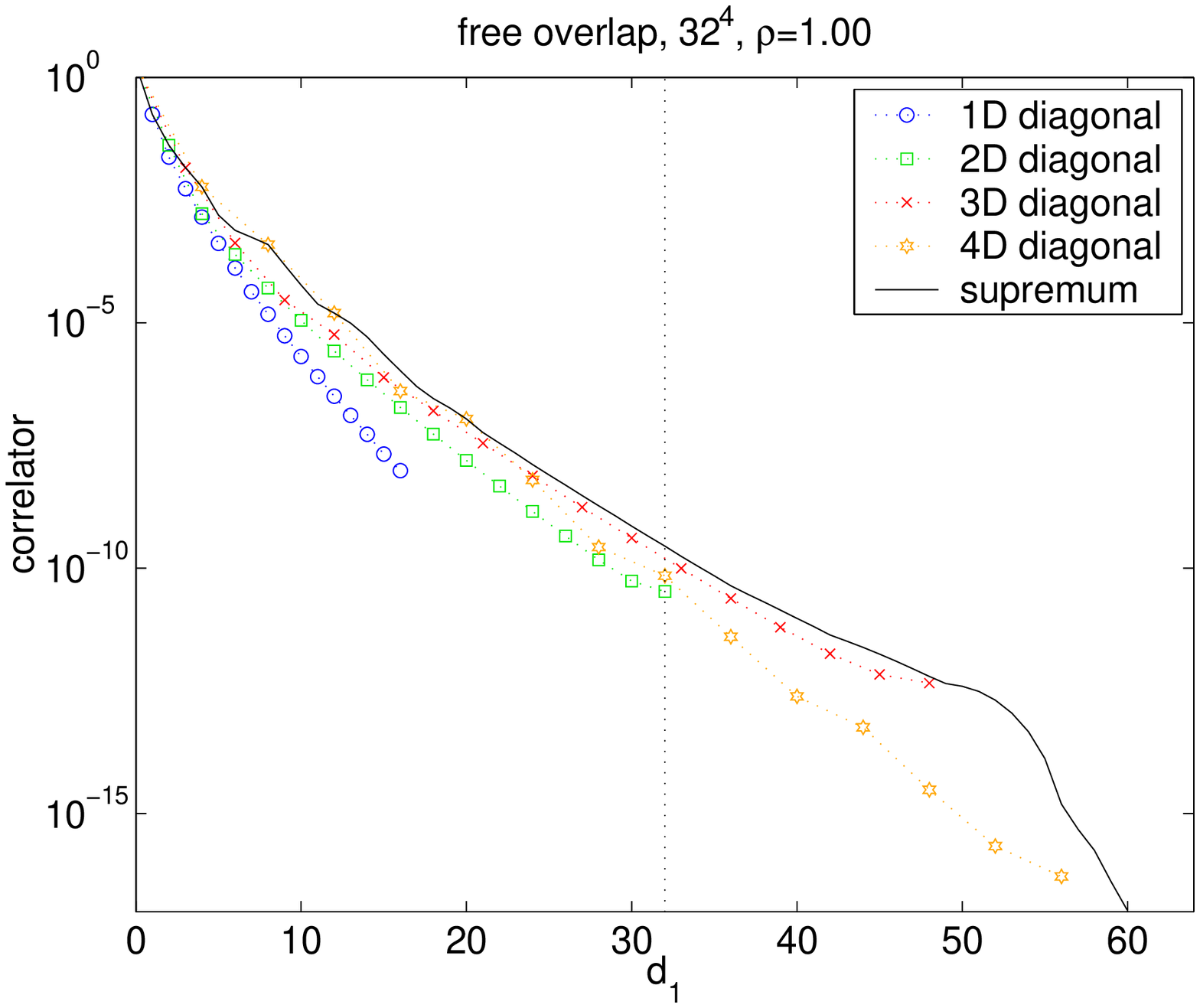,height=7.4cm}
\epsfig{file=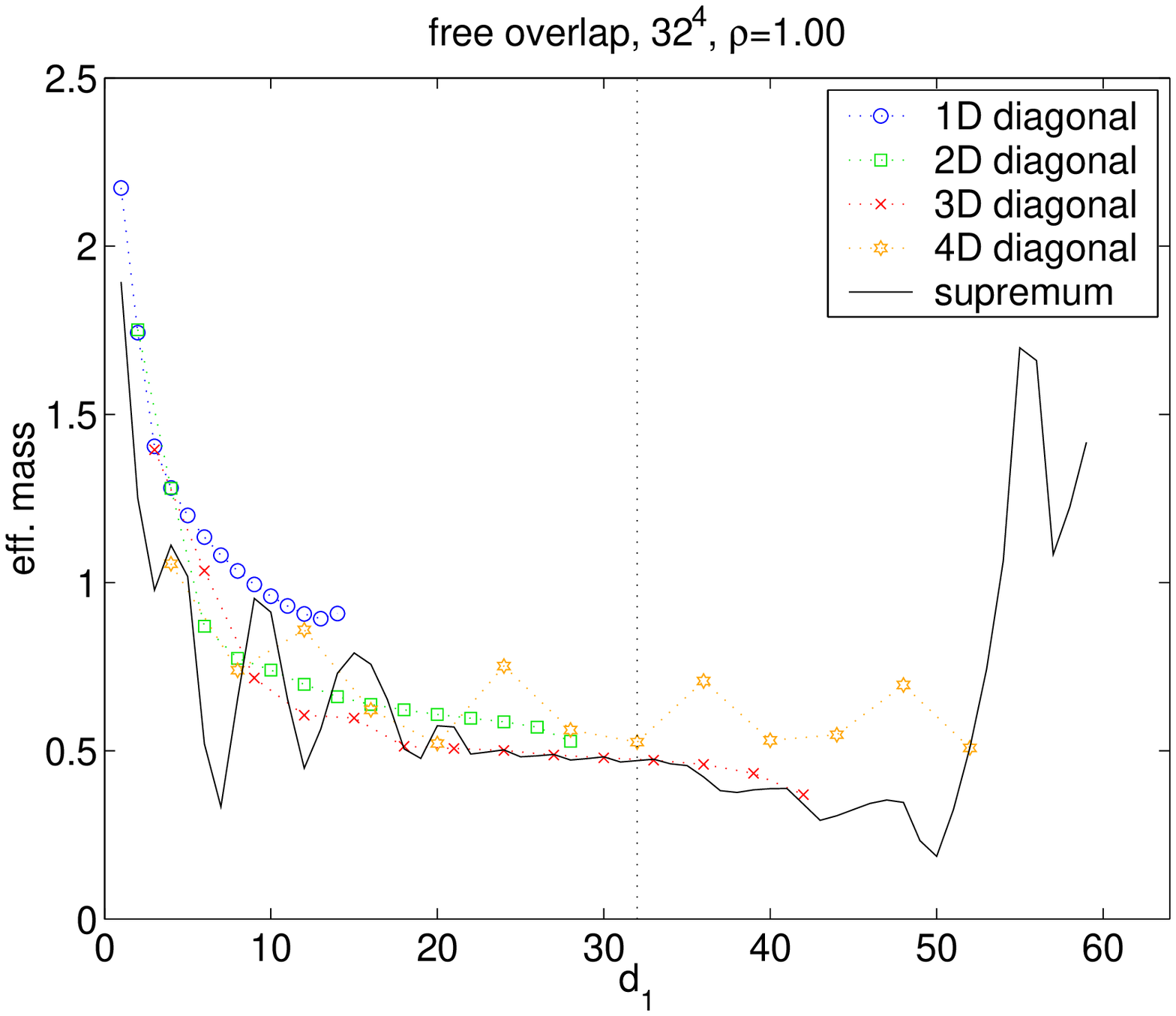,height=7.4cm}\\
\epsfig{file=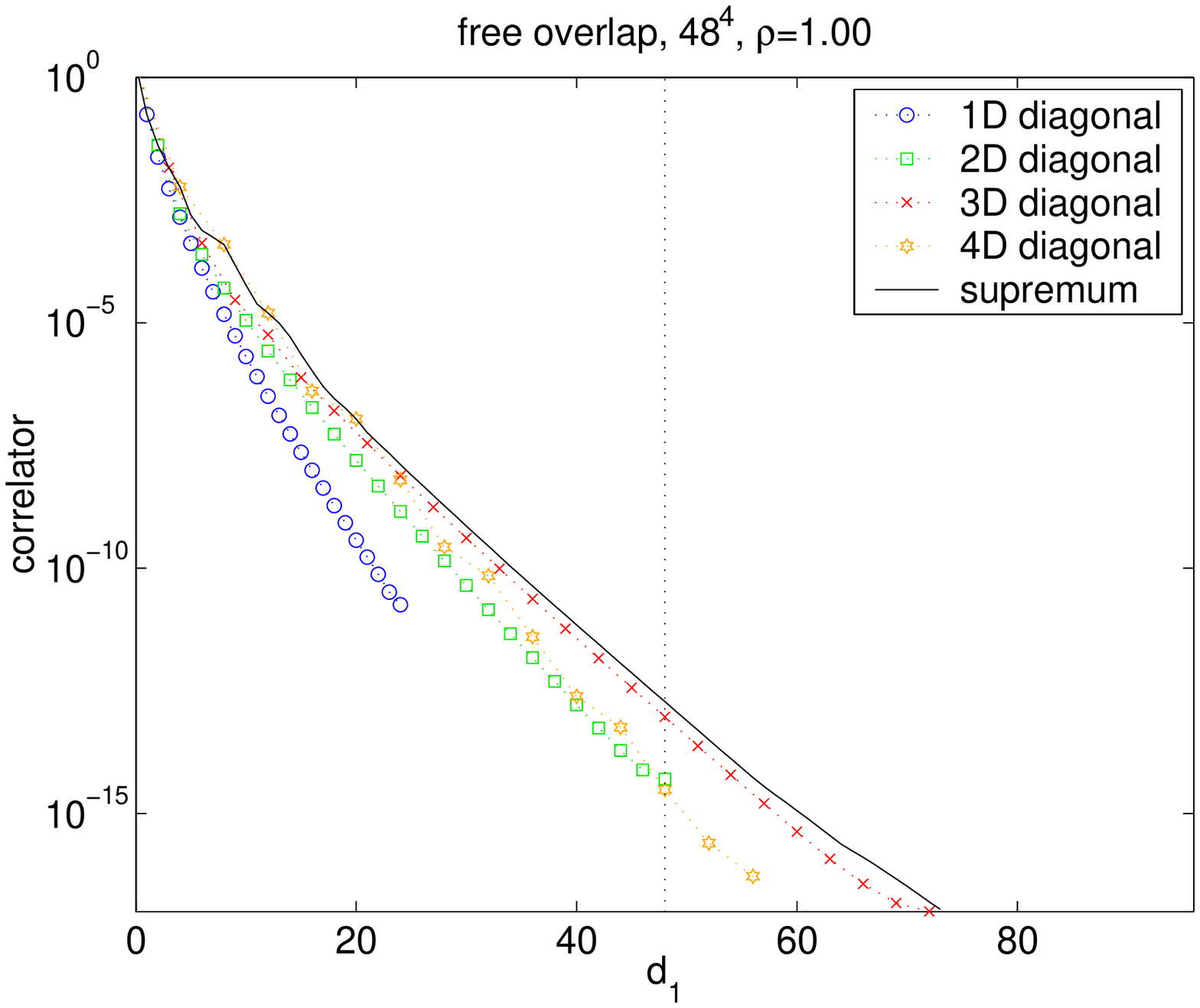,height=7.4cm}
\epsfig{file=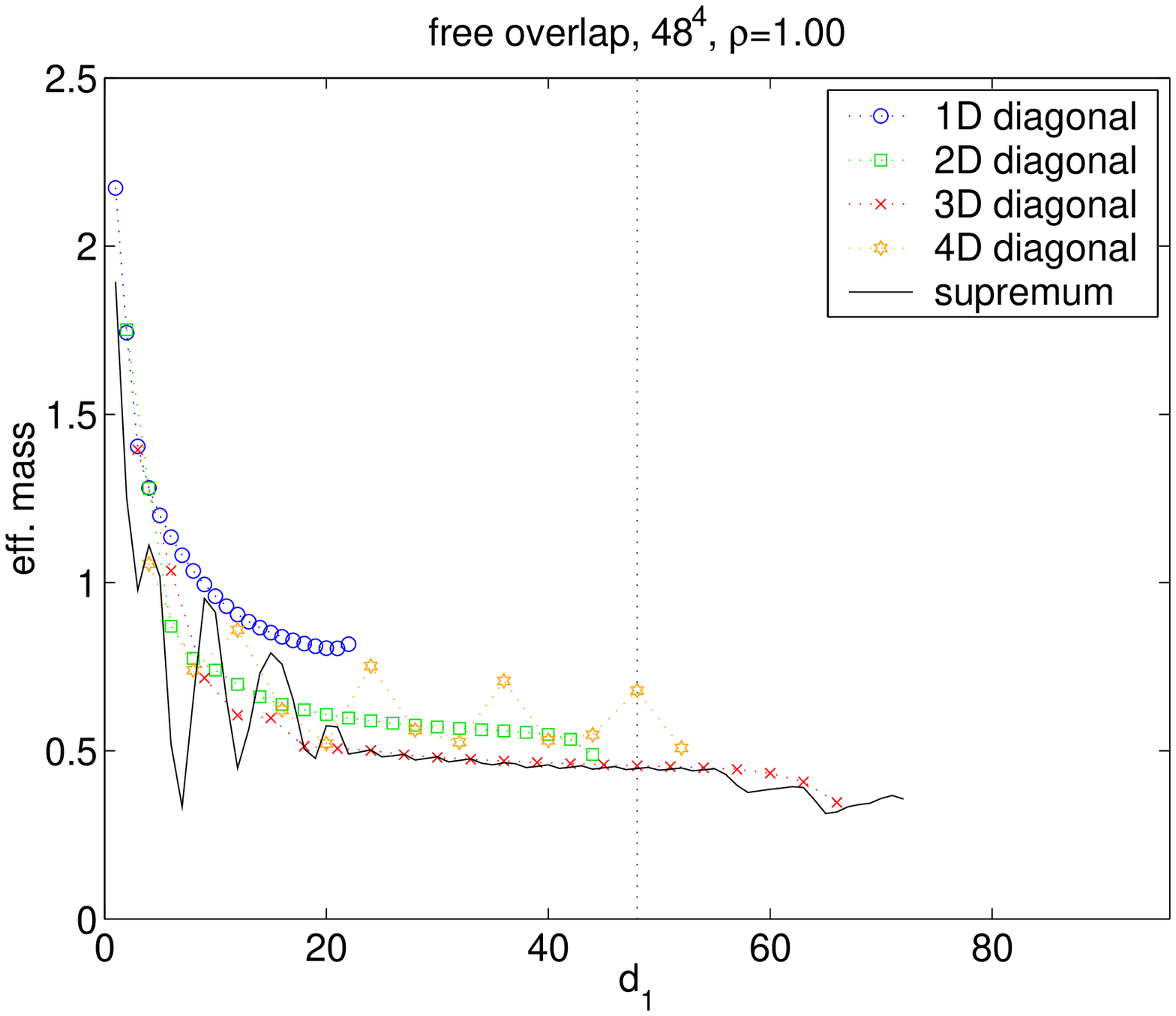,height=7.4cm}\\
\epsfig{file=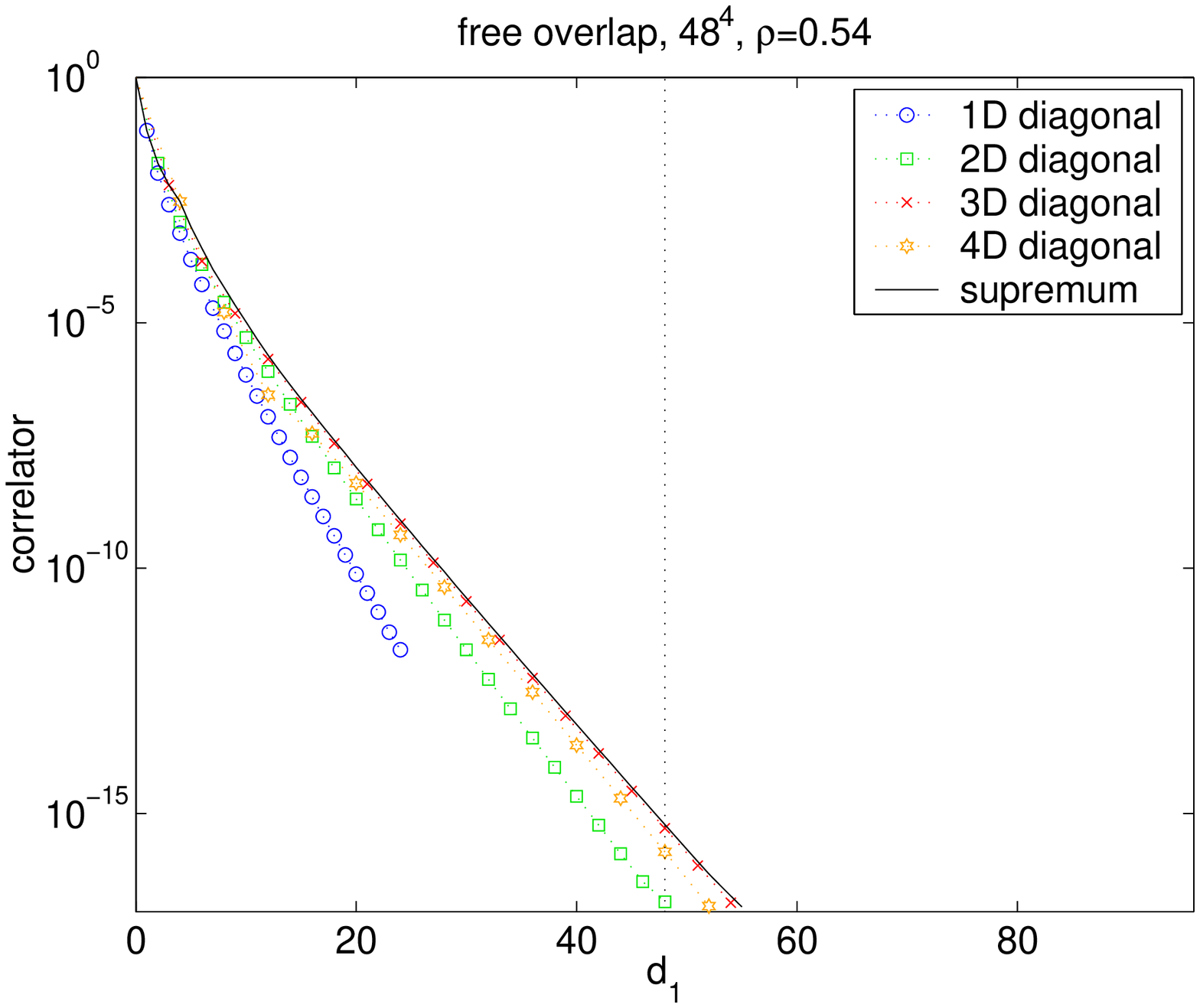,height=7.4cm}
\epsfig{file=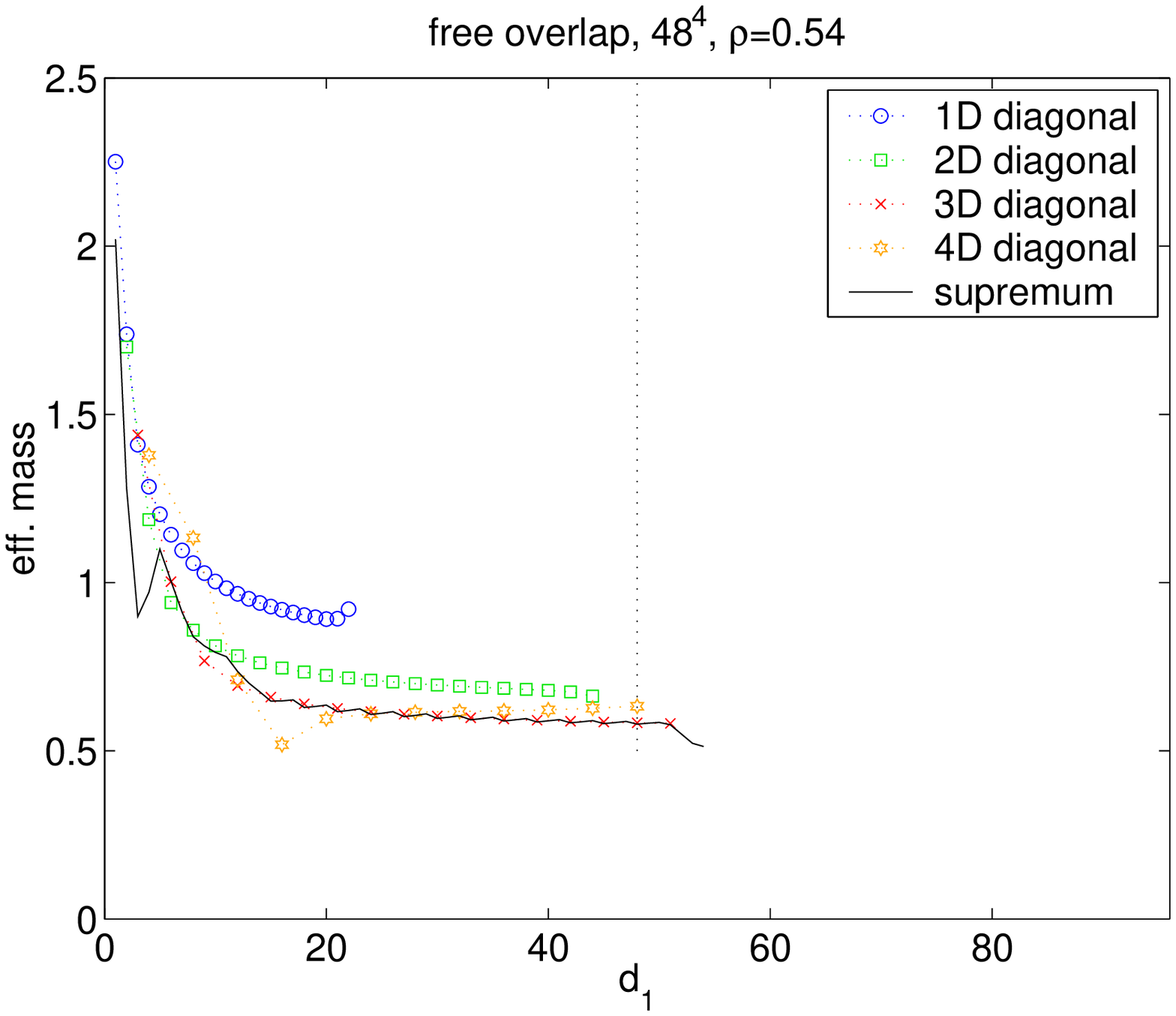,height=7.4cm}\\
\end{center}
\vspace{-8mm}
\caption{Free overlap couplings in four directions and the function
(\ref{supremum}) together with their ``effective masses'' for
($32^4$, $\rh\!=\!1$) and ($48^4$, $\rh\!=\!1$). Some effective masses are
missing, since correlator values below $10^{-17}$ have been cut off (double
precision limit). The good agreement of the $32^4$ and $48^4$ data with
evaluation point $d_1\!=\!L$ (dotted vertical lines) suggests that these data
are much less affected by finite size effects than those near the maximal
$d_1\!=\!2L$. The correlator with $\rh\!=\!0.54$ is visibly steeper.}
\label{fig:localization_free_1.00}
\end{figure}

Considering $f(d_1)$ as given in (\ref{supremum}) \cite{Hernandez:1998et} on a
finite lattice one encounters a technical problem that is evident in
Fig.\,\ref{fig:localization_free_1.00}.
The free field case is far from showing rotational symmetry and the supremum
function (\ref{supremum}) has a couple of initial bumps (in particular at
$d_1\!=\!4,8,12$), and this means that one needs to go to sufficiently large
distances to measure the slope in a logarithmic representation.
On the other hand, the choice to measure the distance in the 1-norm leads to
rather large finite size effects for $d_1\!>\!L$, in particular the region near
the maximal distance $d_1\!=\!2L$ is heavily contaminated.
Therefore, we tried
\beq
\nu={1\ovr2}\log(f(L\!-\!1)/f(L\!+\!1))
\label{def_nu}
\eeq
as a technical definition of the localization $\nu$ in
(\ref{overlap_localization}).
The comparison between the $32^4$ and $48^4$ geometries shows that our choice
to evaluate the logarithmic derivative at $d_1\!=\!L$ produces rather
consistent $\nu$ values, and we take this as a sign that they cannot be far
from the asymptotic exponent.
For the (two-digit-precision) projection parameter that we find to be optimal
w.r.t.\ locality in the free case, $\rh_\mr{opt}^\mr{free}\!\simeq\!0.54$,
the correlator is explicitly shown to be steeper than in the $\rh\!=\!1$ case.

\clearpage



\end{document}